\documentclass[longauth]{aa}  
\usepackage{graphicx}
\usepackage{txfonts}
\usepackage{xcolor}

\newcommand{\orcid}[1]{\href{https://orcid.org/#1}{\includegraphics[width=0.90pt]{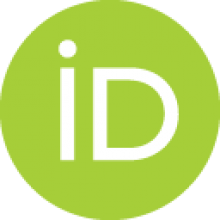}}}
\usepackage{hyperref}
\hypersetup{colorlinks, 
 linkcolor = blue,
 urlcolor = blue,
 citecolor = blue,
 anchorcolor = blue
}

\begin{document}

   \title{Mechanisms Affecting Galaxies Nearby and Environmental Trends (MAGNET) 
   }
   \titlerunning{MAGNET. Survey goals and theoretical predictions}

   \authorrunning{B. Vulcani et al.}
   \author{Benedetta Vulcani\orcid{0000-0003-0980-1499}\inst{1} \and
   Gabriella De Lucia\orcid{}\inst{2,3} \and
   Daria Zakharova\orcid{0009-0001-1809-4821}\inst{2} \and
   Paolo Serra\orcid{}\inst{4} \and Lizhi Xie\orcid{0000-0003-3864-068X}\inst{5,2} \and 
   Stefania Barsanti\orcid{}\inst{6} \and 
   Bianca Maria Poggianti\orcid{}\inst{1} \and
   Alessia Moretti \orcid{0000-0002-1688-482X}\inst{1} \and 
   Marco Gullieuszik\orcid{}\inst{1} \and 
          Yannick Bahé\orcid{}\inst{7} \and  Fabio Fontanot\orcid{}\inst{2} \and Jacopo Fritz\orcid{}\inst{8} \and Fabio Gastaldello\orcid{0000-0002-9112-0184}\inst{9} \and Massimo Gaspari\orcid{}\inst{10} \and Michaela Hirschmann\orcid{}\inst{11,2}
          \and Yara Jaffé\orcid{}\inst{12,13} \and Konstantinos Kolokythas\orcid{0000-0002-3104-6154}\inst{14, 15} \and Alessandro Ignesti\orcid{0000-0003-1581-0092}\inst{1} \and Augusto Lassen\orcid{0000-0003-3575-8316}\inst{1} \and Alessandro Loni\orcid{0000-0001-9556-3786}\inst{16} \and Lorenzo Lovisari\orcid{0000-0002-3754-2415}\inst{9,17} \and Antonino Marasco\orcid{0000-0002-5655-6054}\inst{1} \and Sphesihle Makhathini\orcid{}\inst{18,15,6} \and Sean McGee\orcid{0000-0003-3255-3139}\inst{19} \and Moses Mogotsi\orcid{}\inst{15}  \and D.J. Pisano\orcid{} \inst{20}\and Mpati Ramatsoku\orcid{}\inst{14,6} \and Oleg Smirnov\orcid{}\inst{14,15} \and Rory Smith\orcid{}\inst{12,13} \and  Stephanie Tonnesen\orcid{}\inst{21} \and Marc Verheijen\orcid{}\inst{22} 
          }
   \institute{INAF- Osservatorio astronomico di Padova, Vicolo Osservatorio 5, I-35122 Padova, Italy\\
              \email{benedetta.vulcani@inaf.it}
         \and
   INAF – Osservatorio Astronomico di Trieste, Via Tiepolo 11, I-34131 Trieste, Italy
   \and
   IFPU - Institute for Fundamental Physics of the Universe, via Beirut 2, 34151, Trieste, Italy  
    \and 
    INAF – Osservatorio Astronomico di Cagliari, Via della Scienza 5, I-09047 Selargius, (CA), Italy
    \and
    Tianjin Normal University, Binshuixidao 393, 300387, Tianjin, China
    \and
    Sydney Institute for Astronomy (SIfA), School of Physics, The University of Sydney, NSW 2006, Australia
    \and
    School of Physics and Astronomy, University of Nottingham, University Park, Nottingham NG7 2RD, UK
    \and
    Instituto de Radioastronomía y Astrofísica, UNAM, Campus Morelia, A.P. 3-72, C.P. 58089, Mexico
    \and
INAF - Istituto di Astrofisica Spaziale e Fisica Cosmica, Via Alfonso Corti 12, 20133 Milano, Italy
    \and
    Department of Physics, Informatics and Mathematics, University of Modena and Reggio Emilia, 41125 Modena, Italy
    \and
    Institute of Physics, GalSpec Laboratory, EPFL, Observatoire de Sauverny, Chemin Pegasi 51, CH-1290 Versoix, Switzerland
    \and Departamento de Física, Universidad Técnica Federico Santa María, Avenida España 1680, Valparaíso, Chile
    \and
    Millennium Nucleus for Galaxies (MINGAL), Chile
    \and Centre for Radio Astronomy Techniques and Technologies (RATT), Department of Physics and Electronics, Rhodes University, Makhanda 6140, South Africa
    \and 
    South African Radio Astronomy Observatory, Cape Town, 7925, South Africa
    \and 
    INAF – Astronomical Observatory of Capodimonte, Salita Moiariello 16 I-80131, Naples, Italy
    \and 
Center for Astrophysics | Harvard \& Smithsonian, 60 Garden St., Cambridge, MA 02138, USA         
    \and
    School of Physics, University of the Witwatersrand, Johannesburg, Gauteng, South Africa
    \and 
School of Physics and Astronomy, University of Birmingham, Birmingham, UK, B15 2TT
    \and
    Department of Astronomy, University of Cape Town, Private Bag X3, Rondebosch 7701, South Africa
    \and
    Flatiron Institute, Center for Computational Astrophysics, 162 5th Avenue, New York, NY 10010, USA
    \and
    Kapteyn Astronomical Institute, University of Groningen, Postbus 800, 9700 AV Groningen, The Netherlands
    }


 
  \abstract
{Galaxy evolution is profoundly shaped by intricate internal and external mechanisms that regulate the baryon cycle and star formation activity. To characterize the role of these processes as a function of galaxy environment, we present a theoretical framework based on the GAlaxy Evolution and Assembly (GAEA) semi-analytic model. We extracted portions of simulated volumes that include isolated galaxies, pairs, group, and filament members at $z \sim 0$, specifically avoiding massive clusters. Galaxies were classified using both intrinsic (halo-based) and observational (2D projected) parameterizations, reconstructing their environmental histories from $z = 2$ and identifying mergers, tidal interactions, ram pressure stripping (RPS), and starvation. 
GAEA predictions show that 2D information biases environment definitions, decreasing isolated and group fractions while doubling pairs. More than half of galaxies remain unaffected by the investigated processes since $z = 2$. Among the galaxies affected by external mechanisms, mergers dominate at high stellar masses (40--60\% at $\log (M_\ast/M_\sun)>10.5$). Tidal interactions are less frequent, and their incidence increases with stellar mass (up to 20\%). RPS dominates in groups and filaments at intermediate masses ($\sim$50\%), while starvation ranges from 20 to 30\%. 
The incidence of the different mechanisms depends strongly on both mass and environment, though their imprints on global properties (e.g., colors, gas fractions, sizes) are often subtle. Quenched fractions rise steadily from isolated galaxies to groups. Distinct evolutionary pathways emerge: at low masses ($\log (M_\ast/M_\sun)<9.5$), galaxies in groups and filaments have a faster mass growth than galaxies in the other environments, especially those undergoing starvation,  mergers and -- to less extent -- RPS. Differences are reduced moving to higher masses, where no clear dependence on physical mechanism emerge, even though at these masses a clear star formation suppression is evident in mergers and starved galaxies. This theoretical investigation provides essential context for the recently started multi-wavelength program Mechanisms Affecting Galaxies Nearby and Environmental Trends (MAGNET), which we introduce here.
}

   \maketitle
%

\section{Introduction}

Galaxy evolution is profoundly shaped by the intricate flow of gas within and out of galactic disks, which dictates the baryon cycle. Gas inflows are essential for sustaining star formation and enabling galaxies to grow in mass, while gas exhaustion and removal via various internal or external processes drives star formation quenching. A variety of physical mechanisms have been proposed to explain these gas outflows, ranging from internal feedback from energetic sources like supernovae and Active Galactic Nuclei \citep[AGN,][]{veilleux2005,erb2015, king2015} to external environmental effects \citep{boselli2006} such as ram pressure stripping \citep[RPS,][]{gunn1972}, tidal forces \citep{Barnes1992a}, mergers  \citep{toomre72}, harassment \citep{Moore1996}, and starvation  \citep{larson1980, balogh2000}. These mechanisms alter the star formation activity of galaxies by disturbing and/or removing their gas content, or induce short bursts of star formation \citep{Bekki2003, gavazzi2003, tomicic2018, vulcani2018_l, fujita2003, ostriker2011}, leading to a wide range of galaxy properties \citep[e.g.,][]{oemler1974, dressler1980, goto2003, Blanton2005, vulcani2023a}. While the occurrence of these processes is well established in some local environments, a key open question remains regarding their overall importance in regulating the entire galaxy population as a function of galaxy mass, redshift, and position within the large-scale cosmic structure.

From a theoretical perspective, the role of environment and internal processes in driving galaxy transformations has been explored extensively through both semi-analytic models \citep[SAMs; e.g.,][]{guo11, Henriques2015, Somerville2015, Xie2020, DeLucia2024}  and hydrodynamical simulations \citep[e.g., IllustrisTNG, EAGLE, SIMBA, Magneticum, FLAMINGO, COLIBRE;][respectively]{Pillepich2018, Schaye2015, Dave2019, Hirschmann2014, Schaye2023, Schaye2025}. These models include physical prescriptions for gas accretion, environmental quenching, and feedback processes, and offer predictions for galaxy scaling relations, star formation histories, and the spatial distribution of gas. However, despite their sophistication, many discrepancies remain between theoretical expectations and observations. 
For instance, simulations often overpredict the efficiency of environmental quenching at low stellar masses (e.g.,\citealt{Weinmann2006, font08}, but see \citealt{DeLucia2024}), leading to unrealistic suppression of star formation (the so-called "over-quenching problem"). In addition,  simulations tend to focus on clusters or field galaxies due to both historical interest in these environments and the relative scarcity of observational programs targeting galaxy groups, which limits  available constraints for simulations. Moreover, such simulations are computationally expensive, making it challenging to achieve uniform resolution across the full range of environments.

While various theoretical works have addressed the role of the environment and/or of different physical mechanisms, they do so typically without simultaneously connecting the detailed physical processes to the galaxies’ environmental histories or locations within large-scale structure, with only few exceptions \citep{bruggen08, DeLucia2012, Hirschmann2013, Xie2020, Galarraga2021, Herzog2023, Zakharova2024}. \cite{Hirschmann2013} found that isolated model galaxies have larger fractions of late-type, star-forming galaxies with respect to randomly selected samples of galaxies with the same mass distribution. Approximately 45\% of these isolated galaxies have experienced at least one merger event in the past. \cite{Hank2025} used the SIMBA \citep{Dave2019}  cosmological simulations to investigate the
relationship between the recent dynamical histories and HI spectral
and morphological asymmetries of $\sim$1100 spatially resolved galaxies
at low redshift.  They found that galaxies that had an interaction in the last 2 Gyr are, on average, more
asymmetric than galaxies in isolation. \cite{Zakharova2026} used the Galaxy Evolution and Assembly \citep[GAEA,][]{Hirschmann2016, DeLucia2024} semi-analytic model and the magnetohydrodynamic IllustrisTNG simulation \citep{Pillepich2018} to reconstruct the environmental histories of galaxies that today reside in filaments. They  found that filament galaxies at $z=0$ are a heterogeneous mix of populations with distinct environmental histories, and a clear dependency on the infall times into filaments. Their findings demonstrate that filaments regulate galaxy evolution in a mass-dependent way: group environments regulate the evolution of low-mass galaxies, while filaments favor the growth of massive galaxies. 

This paucity of theoretical studies trying to understand the role of environmental history in affecting galaxy evolution makes it difficult to fully interpret observations, especially in  environments where multiple mechanisms may act in concert or sequentially. 

In low-redshift galaxy clusters—the most massive gravitationally bound structures in the Universe—the relative importance and timescales of quenching processes are increasingly well constrained, although theoretical predictions and observational results do not always align. Among these processes, RPS, caused by the interaction between the galaxy’s interstellar medium (ISM) and the hot intracluster medium (ICM), has proven particularly effective at removing gas and suppressing star formation in dense environments \citep{Vulcani2022, Boselli2022, Moretti2022, Vulcani2020}. Observations have significantly advanced our understanding of how RPS unfolds, the conditions under which it is most efficient, and its role in driving transformation \citep[e.g.,][]{gunn1972, boselli2006, 
Jaffe2015, Marasco2024}.
Theoretical models have likewise highlighted its importance. 
SAMs have helped quantify the impact of RPS on the cold gas content of satellite galaxies \citep{Cora2018, Xie2020}, while hydrodynamical simulations have shown that stripping can be efficient even in group-scale halos, with its effectiveness strongly modulated by orbital parameters, halo mass, and gas density profiles \citep[e.g.,][]{Bahe2013, Steinhauser2016}. Recent SAM results further suggest that most cluster galaxies retain a significant fraction of their gas and continue forming stars after first pericentric passage, except for low-mass galaxies ($\log M_\ast/M_\odot <9.5$) within the most massive halos ($\log M_{halo}/M_\odot >15$) \citep{Xie2024}.

In addition to RPS, starvation  (also known as strangulation) is also recognized as a dominant quenching mechanism in clusters, particularly for satellite galaxies with intermediate stellar masses \citep[M$_\ast\sim 10^{10.5}$,][]{Stevens2017, Xie2020}. In this scenario, infalling galaxies lose their extended hot gas halos—through tidal interactions or ram pressure—cutting off the supply of fresh gas needed to sustain star formation \citep{larson1980, balogh2000}. Star formation then declines gradually as the remaining cold gas is consumed over several gigayears. Observational evidence for this slow quenching includes the elevated stellar metallicities of cluster satellites, which may reflect the lack of dilution from fresh gas inflows \citep{Peng2015}, but could also be influenced by metallicity gradients and stellar stripping processes \citep{Bahe2017, Tollet2017}. In either cases, observations are consistent with 
quenching timescales of 2–5 Gyr, which align with theoretical expectations \citep{wetzel13}.
Starvation arises naturally in hydrodynamical simulations as galaxies interact with the hot ICM and lose their gaseous halos. These simulations show that satellite galaxies can continue forming stars for 1–3 Gyr after infall 
\citep{Bahe2015, Donnari2021a, Donnari2021b}. They also reproduce the metallicity and gas content trends seen in observations \citep[e.g.][]{Bahe2016, Maier2022, Wang2023, Garcia2024}, reinforcing strangulation as a key pathway for gradual environmental quenching in clusters.

However, the majority of galaxies in the local Universe do not reside in massive clusters \citep{Cautun2014}, but rather in less extreme environments such as cosmic filaments and groups \citep{Eke2004, Tempel_2014_necklace}. Our understanding of the physical processes at play in these environments is considerably more limited. The interplay between tidal effects, starvation, preprocessing, and cosmic web stripping is expected to be particularly complex and subtle in these intermediate regimes \citep[e.g.,][]{Kawata2008, Benitez2013, Cortese2021, Oman2016, Kolokythas2022},  and a comprehensive characterization and census of these mechanisms is still lacking. 

To distinguish and characterize the various physical processes at play, spatially resolved information on the different gas phases, combined with precise knowledge of the host environment, is essential. Such data are key to understand galaxy transformations and to classifying galaxies according to their governing processes, since each mechanism leaves a distinct imprint on the spatial distribution of gas and stars \citep{Vulcani2021}. 

A wide range of surveys now provide spatially resolved data, relying either on deep imaging or on integral field unit (IFU) spectroscopy.
Imaging surveys offer the advantage of covering extremely large samples, often over contiguous areas of the sky, enabling statistical analyses across diverse environments and cosmic structures.  Imaging from surveys such as SDSS \citep{york00}, GAMA \citep{Driver2011}, and DESI Imaging Legacy Surveys \citep{Dey2019}  has been instrumental in probing galaxy morphology, stellar halos, and structural features with uniform coverage. 
Beyond global morphology, deep, high-quality imaging also provides spatially resolved information on the distribution of stellar populations, dust, and tidal features. For example, in the Virgo cluster, systematic imaging campaigns with the CFHT and other facilities have revealed low-surface-brightness structures, extended disks, and tidal debris, providing crucial insights into the environmental effects shaping galaxies \citep[e.g.,][]{Ferrarese2012, Boselli2016, Mihos2017}. These studies highlight how imaging, even without spectroscopy, can uncover the spatial footprints of processes such as RPS, harassment, or tidal interactions.
Looking forward, new facilities, such as the Legacy Survey of Space and Time (LSST), promise to extend this capability to unprecedented depths and scales. Also early results from Euclid have already demonstrated the survey’s ability to resolve faint stellar structures  across wide fields \citep[e.g.,][]{George2025, Mondelin2025, Urbano2025}, opening the door to statistical studies of environmental effects on galaxy outskirts and diffuse components.

IFU surveys  provide a far richer set of diagnostics on individual galaxies. Early efforts such as SAURON \citep{Bacon2001} were followed by large-scale programs including CALIFA \citep{Sanchez2012}, MaNGA \citep{Bundy2015}, SAMI \citep{Croom2012}, and the forthcoming Hector survey \citep{Bryant2016, Oh2025}. These projects have demonstrated the power of spatially resolved spectroscopy to constrain stellar populations, kinematics, star-formation patterns, and ionized gas properties. However, IFU surveys necessarily trade off sample size and spatial coverage against depth and detail: most current large surveys observe up to a few thousand galaxies in the local Universe, typically sampling only the inner 1–1.5 effective radii. This is because IFU mapping of extended regions remains challenging due to the limited field of view of individual instruments, and because the outer regions are fainter, requiring significantly longer integration times. As an alternative, programs such as PHANGS \citep{Leroy2016, Schinnerer2019} have employed mosaicking strategies, achieving exquisite detail on the resolved star formation and interstellar medium of tens of galaxies, albeit at the expense of large sample statistics.

Another fundamental limitation of IFU surveys is their reliance on target preselection, which shapes their scientific reach. Some surveys focus mainly on  field galaxies (e.g., MaNGA, CALIFA), while others prioritize dense environments (e.g., SAMI). The GAs Stripping Phenomena in galaxies (GASP) survey \citep{Poggianti2017, Poggianti2025}, using VLT/MUSE spectroscopy, 
has directly traced gas flows and identified their physical origin across a wide variety of environments, from galaxy clusters \citep{Jaffe2018, Poggianti2019, Radovich2019GASPGalaxies, Gullieuszik2020, Franchetto2020, Franchetto2021, Bellhouse2020, Tomicic2021, Tomicic2021b, Peluso2022, Sanchez2023} to groups and isolated systems \citep{Vulcani2017c, Vulcani2018_b, Vulcani2018_g, Vulcani2019_fil, Vulcani2021}. GASP’s spatially resolved data have been instrumental in disentangling diverse processes—including galaxy-galaxy interactions, mergers, RPS, cosmic-web stripping, gas accretion, enhancement, and starvation—particularly in non-cluster environments.
However, a crucial aspect of GASP’s design is that its primary targets were chosen among galaxies with optical signatures of gas stripping, while galaxies showing clear evidence of mergers or tidal interactions were purposely excluded. As a result, while GASP has yielded invaluable insights into gas removal phenomena, it cannot provide an unbiased statistical census of all physical processes acting across the galaxy population. 

It is clear that, both from the theoretical and observational perspectives, we have not yet achieved a comprehensive understanding of the incidence and relative importance of the different physical mechanisms shaping galaxies in the local Universe outside of clusters. Recent efforts are beginning to extend cluster studies toward their outskirts \citep[e.g., CHANCES, WEAVE][]{Haines2023, Cornwell2023}, yet our knowledge of intermediate-density environments, particularly beyond $\sim5 R_{200}$, remains very limited. In this paper, we take a step forward by introducing the Mechanisms Affecting Galaxies Nearby and Environmental Trends (MAGNET)\footnote{\url{https://sites.google.com/inaf.it/magnet/home}} multi-wavelength program, a new  survey designed to provide a transformative, holistic observational view of the processes affecting the gas content and star-formation activity of galaxies in the different environments.  At the same time, we present theoretical predictions for how galaxies are expected to evolve in diverse contexts, with particular emphasis on groups and filamentary structures, where the drivers of gas removal and quenching are subtler and less well understood than in dense cluster environments. The results presented here go beyond the specific case of  MAGNET   and provide important information to interpret a broader range of observations.

We use predictions from  GAEA \citep{Hirschmann2016, DeLucia2024}, a theoretical model that incorporates various environmental effects, including hot gas stripping, starvation, and RPS of cold gas \citep{Xie2020}. Because GAEA reproduces a broad range of observed galaxy properties across diverse environments, it provides an ideal tool for the investigation presented below. We examine whether the region targeted by MAGNET is representative in terms of environmental diversity and galaxy populations, and characterize the predicted global properties of galaxies in different environments. As MAGNET data are currently being acquired (see Sec.\ref{sec:magnet}), direct, systematic comparisons are deferred to future work. We further explore how galaxies' current environment and the physical processes that affected their evolution since $z\sim 2$ shape their positions in standard scaling relations, such as stellar mass versus star formation rate or gas content, and assess whether different evolutionary pathways imprint distinguishable observational signatures. 
The use of a semi-analytic model such as GAEA offers an optimal compromise between physical realism and computational efficiency, allowing us to model statistically representative galaxy samples and isolate the effects of different physical processes that would be more difficult to disentangle in hydrodynamical simulations. Unlike widely used hydrodynamical models such as IllustrisTNG \citep{Pillepich2018}, EAGLE \citep{Schaye2015}, 
and The Three Hundred project~\citep{Cui2018}, GAEA includes an explicit treatment of the partition of cold gas into its atomic and molecular components and is coupled to a large cosmological volume with relatively high resolution.

The paper is structured as follows. Section \ref{sec:magnet} introduces MAGNET, describing its goals, the reference field, the included environments, and the status of the project. Section \ref{sec:gaea} introduces the adopted SAM and describes how we selected the sub-volumes to be analyzed (Sect. \ref{sec:boxes}), how we defined the environments (Sect. \ref{sec:env}) and distinguished the different mechanisms that affected galaxies (Sect. \ref{sec:process}). Section \ref{sec:results} presents our results, including direct comparisons with the observed data (Sect. \ref{sec:comp_magnet}), the relative fraction of galaxies in different environments and experiencing different mechanisms (Sect.\ref{sec:frac}) and their properties (Sect. \ref{sec:prop}); and their evolution since $z\sim2$  (Sect. \ref{sec:hist}). In Section \ref{sec:discussion} we discuss the results and the limitations of our analysis.  We summarize our work in Section \ref{sec:summary}. Readers mainly interested in the theoretical approach and results can skip Sec. \ref{sec:magnet} and Sec. \ref{sec:comp_magnet}.

\begin{figure*}
    \centering
    \includegraphics[trim = 0 20 0 0, width=0.9\linewidth]{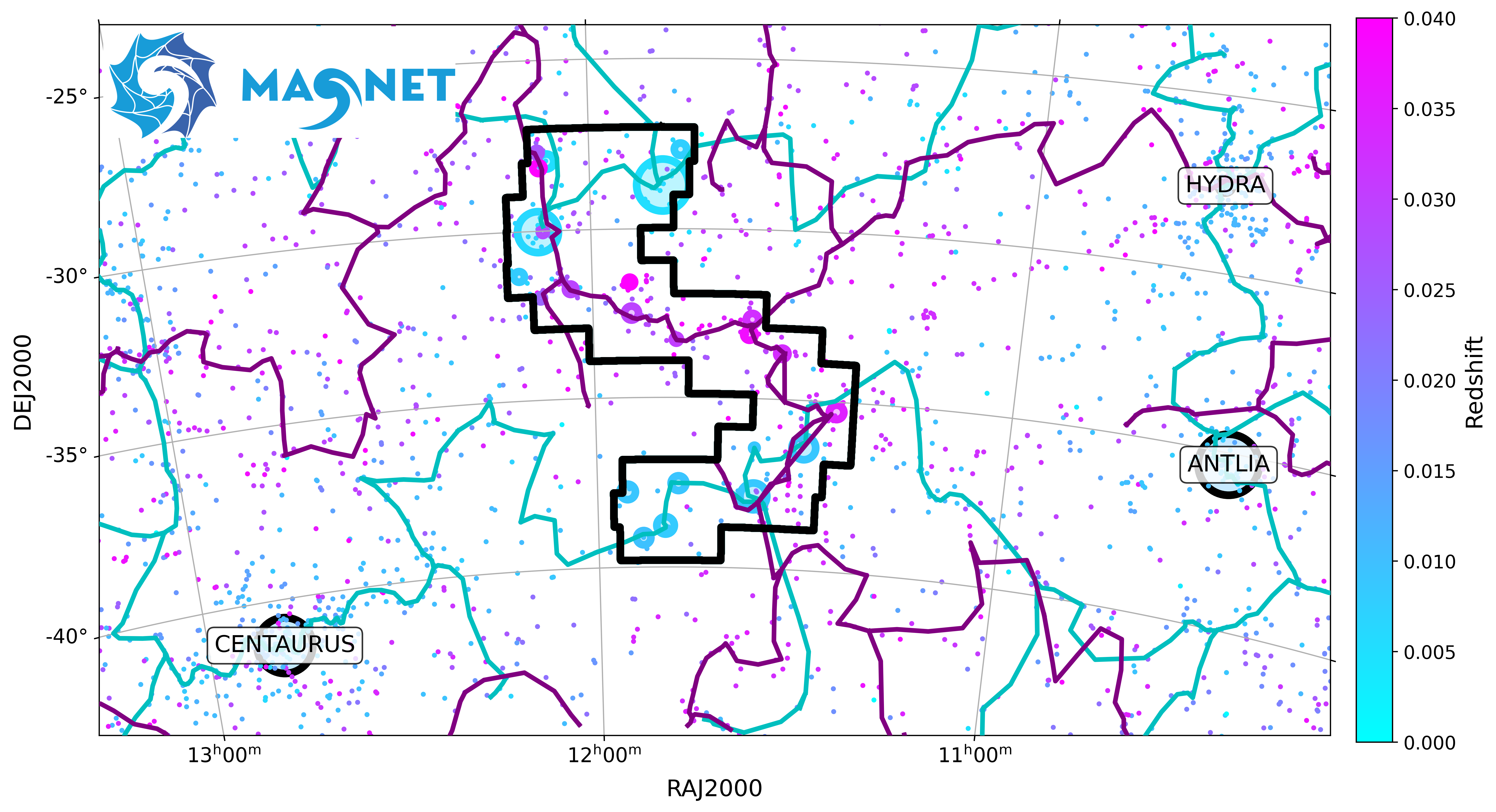}
    \caption{Overview of the MAGNET field. Small points show existing spectroscopically confirmed galaxies at $z<0.04$. Larger circles show the location of the groups falling in the MAGNET footprint, shown in black. The size of the circles is proportional to the group virial radii. Both circles and points are color coded by redshift. Cyan (purple) solid lines indicate the filament network identified by DisPerSE using galaxies in the range $0<z<0.02$ ($0.02<z<0.04$). The only structures shown outside the MAGNET footprint are Centaurus, Hydra and Antlia, three clusters that are located more than 10$^\circ$ away from our field.}
    \label{fig:MAGNET}
\end{figure*}

\section{A new multi-wavelength program: Mechanisms Affecting Galaxies Nearby and Environmental Trends}\label{sec:magnet}

The overarching goal of the MAGNET project is to obtain a comprehensive, spatially resolved view of galaxies across diverse environments—particularly groups and filaments—by investigating how the multi-phase composition and distribution of the ISM evolve under different types of interactions. Comparing atomic, molecular, and ionized gas on matched spatial scales will reveal the interplay between gas phases and star formation. At the time of writing, MAGNET is still in its initial phases, and, as described below, the first observations are currently underway.

\subsection{The observed reference field}\label{sec:snake}
The MAGNET survey  targets a carefully selected, contiguous sky region with an “S”-shaped footprint, named SNAKE, covering approximately 65 square degrees (Fig.~\ref{fig:MAGNET}). This footprint is deliberately designed to be large enough to robustly capture environmental processes and is comparable in scale to individual GAMA fields \citep{Driver2011} and to the 4MOST WAVES survey \citep{Driver2019}.

The SNAKE field was identified through a meticulous selection process aimed at encompassing a diverse range of environments. The selection procedure followed several criteria. First, we focused on the southern sky, where most current state-of-the-art instrumentation has coverage. 
We use a homogenized compilation of major spectroscopic redshift catalogs including 2MRS v2.4 \citep{Huchra2012}, 6dFGS DR3 \citep{Jones2009}, 2dFGS \citep{colless01}, GAMA DR4 \citep{Baldry2018, Driver2022}, and SDSS DR15 \citep{Abazajian2009, Aguado2019}. Duplicate observations have been identified through friends-of-friends position matching on the spectral target coordinates, using a tiered set of priorities to resolve which survey(s) measurements to prefer when multiple redshifts exist. These priorities have been set based on clipped root mean square scatter in repeat measurements within and across surveys, with preference given to the more accurate surveys. Spectroscopic redshifts are in heliocentric reference system.

Additionally, multiple catalogs of galaxy groups \citep{Tully2015, Kourkchi2017, Tempel2016, Tempel2018, Lim2017, Lambert2020} were used. The analysis was limited to galaxies with  $z<0.04$ to ensure high physical resolution, which is crucial for mapping the spatially resolved distribution of different ISM phases with current instrumentation. Rectangular regions containing at least ten galaxy groups, each with a minimum of five members, were extracted.

A key objective was to avoid massive galaxy clusters. To this end, we excluded any region containing a cluster with an X-ray luminosity greater than $3\times10^{43}$ erg s$^{-1}$ (from eROSITA DR1; \citealt{Kluge2024}) or a halo mass above $10^{14}$ M$_\odot$ (from OmegaWINGS; \citealt{Fasano2006, Gullieuszik2015} and CHANCES \citealt{Haines2023}) within a 10 Mpc radius from its center.

The final choice maximized environmental diversity, prioritizing regions that include galaxy associations and groups of varying richness, as well as filaments of the cosmic web, to evaluate environmental effects associated with large-scale structures. The S-shaped footprint was carved to exclude areas with a sparse galaxy population across the redshift range of interest, thereby maximizing survey efficiency. The final survey volume is expected to be 32,000 Mpc$^3$.

The SNAKE field has well-defined characteristics. Its southern arm, oriented east–west at declinations below –35°, incorporates a known structure called the Antlia Wall. The northern arm, at declinations above –30°, contains a filament of galaxies and groups at approximately $z=0.01$, aligned in the direction connecting the Centaurus and Hydra clusters. The diagonal section of SNAKE includes numerous galaxies and groups at $0.02<z<0.04$, which may be part of a larger filamentary structure (see Sect.~\ref{sec:magnet_env_fil}). The nearest known galaxy clusters (Centaurus, Antlia, and Hydra) are all located more than 10 degrees away, corresponding to over 10 cluster virial radii, reinforcing the field’s focus on less extreme environments.

Within the SNAKE field, approximately 450 galaxies currently have a spectroscopic redshift in the range 0<z<0.04 and a stellar mass \citep[taken from][]{Biteau2021} above $10^9$ M$_\odot$. Our ongoing VST campaign (Sect.~\ref{sec:magnet_status}) is designed to reach much deeper than existing observations, thereby increasing the number statistics. Details on group sizes and properties are given below.

\subsection{Environmental definitions}\label{sec:magnet_env}

\subsubsection{Groups}
For the time being, we rely on group identifications from existing catalogs. However, these definitions are provisional: they will be revised in the future as new data—both proprietary and from upcoming public surveys—become available. 

\begin{figure}
    \centering
    \includegraphics[trim=0 145 0 120, clip, width=0.9\linewidth]{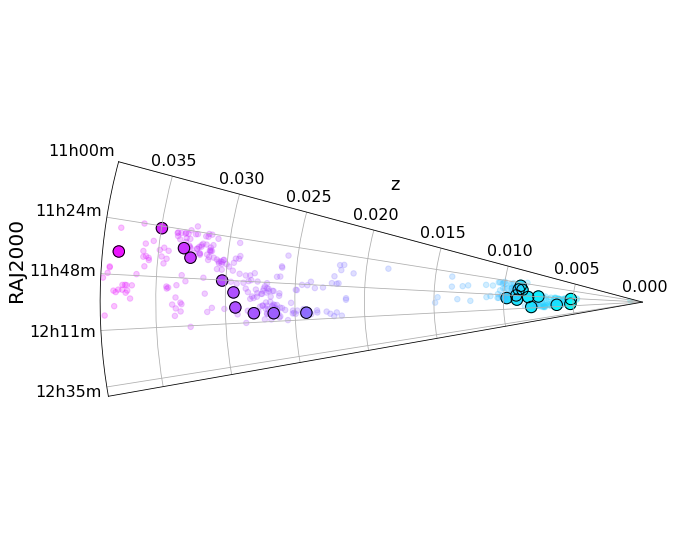}
    \caption{Polar plot showing the redshift and right ascension distribution of the galaxies (shaded small circles) and groups (larger points) within the MAGNET footprint.}
    \label{fig:polar}
\end{figure}

We inspected several galaxy group catalogs that cover the local Universe: \cite{Tully2015, Kourkchi2017, Tempel2016, Tempel2018, Lim2017, Lambert2020}. Our aim is to identify and characterize systems over a wide range of environments, explicitly distinguishing galaxy pairs from groups. Galaxy groups span a wide range of scales, and we aim to take advantage of grouping methods down to the level of galaxy pairs. 

We follow the approach of \cite{Tully2023Cosmicflows-4}, prioritizing physically motivated group catalogs that make use of distance information when available. 
For galaxies with systemic velocities below 3,500 km s$^{-1}$ ($z\leq 0.011$), we use the group catalog of \cite{Kourkchi2017}. This catalog assembles all known galaxies with $v < 3,500$ km s$^{-1}$ into groups using scaling relations that relate halo mass, velocity dispersion, and projected radius. Groups are roughly bounded by the second turnaround radius, and the availability of Cosmicflows-3 distance data \citep{Tully2019} significantly improves group identification, especially in resolving projected pairs and small systems. At higher redshifts, where the 2MASS Redshift Survey is more complete, we adopt the group catalog of \cite{Tully2015}, which includes $\sim$43,000 galaxies brighter than Ks = 11.75 and $|b| > 5^\circ$, covering systemic velocities out to $\sim$15,000 km s$^{-1}$. Group definitions here are based on the same physical framework as in the nearby catalog, maintaining consistency across the transition. In regions of overlap ($v \leq 3,500$ km s$^{-1}$), we resolve discrepancies by favoring the specifications of \cite{Kourkchi2017}. 

We additionally include two groups identified in the catalog by \cite{Tempel2016}, which are missed by the above catalogs but clearly seen in the galaxy spatial distributions. This SDSS-based catalog applies a friends-of-friends algorithm with a redshift-dependent linking length, optimized to recover virialized systems.  
When new spectroscopic coverage is available, we will explore why these were missed by \cite{Tully2015}.  

We consider a total of 24 groups, shown as empty circles in Fig.~\ref{fig:MAGNET} and whose properties will be discussed in Sect. \ref{sec:distr}. A posteriori, we also verified that only two groups have X-ray emission, appearing in the AXES and eROSITA catalogs \citep{Khalil2024, Kluge2024}. These have $L_X = 1.09$–$1.53\times 10^{43}$ erg s$^{-1}$, and are not the most massive systems in our field ($\sigma_v= 170$–300 km s$^{-1}$). A more detailed investigation of the group X-ray properties is underway (see Sect. \ref{sec:magnet_status}).  

Figure \ref{fig:polar} shows the group distribution in space and redshift, revealing a clear double-peaked structure with a relative paucity of data around $z\sim0.02$.
Nonetheless,  we do not anticipate any evolutionary differences between the two redshift ranges.

\subsubsection{Filaments}\label{sec:magnet_env_fil}

We identify filaments using the Discrete Persistent Structures Extractor \citep[DisPerSE;][]{Sousbie2011a, Sousbie2011TheIllustrations}. This tool reconstructs the cosmic web by estimating a density field from the galaxy distribution and extracting the filament spines through topological persistence. 
We use all galaxies with currently available spectroscopic redshifts (see Sect.\ref{sec:snake}) and stellar mass above $10^{9} M_{\odot}$ \citep[from][adopting a Chabrier (2003) IMF]{Biteau2021}. \cite{Zakharova2023} have shown that including lower-mass galaxies does not improve filament determination.  
We run DisPerSE separately for 2D galaxy positions in the RA–DEC plane, using redshift slices $0<z<0.02$ and $0.02<z<0.04$, to detect two sets of filament structures. 
In all runs, we adopt a 3$\sigma$ persistence level, which is the typical value used in most studies employing this algorithm. The distance between galaxies and filaments is defined as the projected angular distance (in degrees) to the nearest filament, converted to comoving distances. We consider galaxies within 1 Mpc/h of the filament spines to be filament members.  
We also extract filaments both considering only the SNAKE region and the full extended area (shown in Fig. \ref{fig:MAGNET}) to test the robustness of the extracted web. Filament identification inside the SNAKE does not depend on the size of the surrounding area; here, we adopt the structures obtained from the full available area. Figure \ref{fig:MAGNET} shows the resulting network: DisPerSE confirms the existence of previously known filaments as well as structures  visible by eye.  

As with the group determination, we will revise our filament structures once new spectroscopic redshifts become available. Additionally, we plan to characterize other environmental features—including walls and voids—provided that spectroscopic completeness is sufficient.

\subsection{Driving Questions and Objectives}

As discussed above, MAGNET is designed to probe galaxies across a wide range of environments, from isolated systems to groups, while avoiding massive clusters. The survey will combine spatially resolved H$\alpha$ narrow-band imaging, H\,\textsc{i} interferometric mapping, and ancillary multi-wavelength datasets to address several key questions in galaxy evolution. 

{\it Which physical processes dominate galaxy transformation in different environments?}
A primary goal of MAGNET is to assess the relative importance of hydrodynamical and gravitational interactions in shaping galaxies. By analyzing galaxy morphologies and spatially resolved gas and stellar kinematics and identifying nearby companions, we will distinguish between RPS and tidal effects. Galaxies exhibiting disturbed gas kinematics but regular stellar rotation are likely dominated by hydrodynamical processes, whereas gravitational interactions are expected to most likely perturb both stellar and gaseous components.

{\it How do internal mechanisms and the environment work together to quench star formation?}
MAGNET will investigate the interplay between external environmental processes and internal secular mechanisms—such as AGN and stellar feedback—in driving gas removal and star formation quenching. Observations of both central and outer galaxy regions will allow us to characterize gas ionization mechanisms (including AGN-driven outflows) and to assess how environmental processes affect the availability and physical state of the ISM, particularly in the outskirts where galaxies interact with the circumgalactic medium.

{\it Over what timescales does star formation shut down in different environments?}
The survey will constrain the timescales of star formation quenching associated with different physical processes. By linking gas removal signatures (e.g., truncated gas disks, one-sided tails) with spatially resolved star formation rate maps and stellar population properties, we will reconstruct the quenching histories of galaxies undergoing distinct environmental interactions.

{\it How does the environment impact the multiphase ISM?}
MAGNET will provide a spatially resolved view of the multiphase ISM—comparing the distributions of atomic, molecular, and ionized gas—to trace the impact of the environment on gas physics and the star formation cycle. Variations among gas phases on small scales will reveal how different processes suppress or enhance star formation and help identify galaxies at various stages of their evolutionary paths.

{\it How does feedback from supermassive black holes interact with environmental processes?}
Recent theoretical and numerical work has highlighted the role of self-regulated Black Hole Weather or Chaotic Cold Accretion (CCA) cycles in governing the exchange of gas between galaxies, their halos, and central black holes \citep[e.g.,][]{Gaspari2020}.
In this framework, turbulence and cooling instabilities in the circumgalactic medium lead to multiphase condensation and episodic black hole fueling, while AGN feedback reheats and stirs the gas, maintaining a quasi-equilibrium state \citep[e.g.,][]{Wittor2020}.
Environmental mechanisms such as RPS, tidal interactions, mergers, and starvation can modulate this cycle by altering gas supply, turbulence, and entropy structure—either triggering brief condensation episodes or suppressing cooling altogether \citep{Gaspari2012, Gaspari2013, Gaspari2015, Gaspari2017, Gaspari2018}.
MAGNET’s combined H$\alpha$ and HI mapping will allow us to identify galaxies caught in these transient phases—e.g., systems showing turbulent, multiphase gas in group environments—and to test how external perturbations regulate the balance between cooling, feedback, and star formation across different mass scales \citep[e.g.,][]{Maccagni2021, Olivares2022}.

By combining a statistically representative and environmentally diverse sample with these multi-wavelength observations, MAGNET will enable robust comparisons with theoretical models and simulations, providing critical observational constraints on the physical drivers of galaxy transformation in the low-redshift Universe.

\subsection{Current status of the project} \label{sec:magnet_status}
The  survey is still in its early stages, but several efforts are already underway to establish the foundations of the project. From the outset, MAGNET was designed with a multiwavelength strategy, and its sky footprint has been optimized to enable follow-up across the necessary wavebands. As a first step, we have assembled a comprehensive set of ancillary data, including spectroscopic redshifts (see Sect.~\ref{sec:snake}), and broad-band imaging from GALEX, DeCALS \citep{Dey2019}, and NEOWISE, covering wavebands ranging from the UV to the mid-IR.
We have followed the approach of \cite{Marasco2025}  to derive photometric measurements and, for galaxies with spectroscopic redshifts, their main properties via SED modelling techniques.
A full description of these data products will be presented in a forthcoming paper.

Looking ahead, MAGNET will strongly benefit from its location within the footprints of major ongoing surveys. The LSST will deliver deeper, higher-resolution imaging, enabling a more precise structural parameterization of galaxies and the characterization of low surface brightness features which are very sensitive to the environment, while the 4MOST surveys \citep{deJong2019}—in particular 4HS \citep{Taylor2023}—will, once completed (expected by 2030), provide spectroscopic redshifts with high completeness. Together, these resources will greatly enhance the accuracy of our environmental characterizations and expand the survey’s statistical reach.

To complement these large-scale survey datasets, we are conducting targeted observations within the SNAKE region. Dedicated VST imaging (PI Gullieuszik, projectID: 114.27UC.001) is being carried out to produce a uniform mosaic in four narrow bands, yielding spatially resolved maps of H$\alpha$ emission for all galaxies up to $z=0.04$. At present, 6 out of 65 fields have been acquired and reduced, with completion expected by 2028. These observations will provide a crucial view of the ionized gas and ongoing star formation activity across the region.

In parallel, we are building a multiwavelength view of the intragroup and intergalactic medium in the SNAKE. Public eROSITA \citep{Bulbul2024, Kluge2024} data are being analyzed to measure the temperature and density of the hot gas, while complementary XMM observations have been requested. At the same time, we plan 
to obtain spatially resolved H\,\textsc{i} and radio continuum maps at sub-kpc scales, which will provide an unprecedented view of the cold gas reservoirs in these galaxies. 

The long-term vision of MAGNET is to extend this effort by securing spatially resolved observations of both ionized and molecular gas for carefully selected subsets of galaxies. While current instrumentation does not allow us to cover the entire SNAKE region, targeted representative samples will enable a comprehensive multiphase view of the gas and, combined with the ancillary datasets, will establish MAGNET as a unique resource for studying the interplay between galaxies and their environments across all cosmic structures.

To summarize, the key strength of MAGNET, compared with both existing surveys such as MaNGA \citep{Bundy2015}, CALIFA \citep{Sanchez2012}, SAMI \citep{Bryant2015}, PHANGS \citep{Schinnerer2019}, VESTIGE \citep{Boselli2018}, MAUVE \citep{Watts2024}, MAGPI \citep{Foster2021}, and CAVITY \citep{Perez2024}, and upcoming ones such as 4MOST \citep{deJong2019}, WEAVE \citep{Jin2024}, and EMPOWER (PI: Popesso), lies in its ability to achieve sub-kiloparsec resolution over a contiguous region of the sky without the need to preselect targets. This enables us to obtain data for all galaxies above a low stellar mass threshold, with the exact limit depending on the instrument used. Such an unbiased, volume-limited approach is essential for a comprehensive characterization of the incidence and interplay of different quenching and transformation mechanisms.
In contrast to surveys like MAUVE, VESTIGE, and GASP, which focus specifically on rich cluster environments, MAGNET explicitly avoids massive clusters. This allows us to target a less-explored regime, probing how galaxy evolution unfolds in low- to intermediate-density environments.
Importantly, even with the observations already secured, the available data are sufficient to deliver new, impactful insights into environmental processes, independent of future expansions.

\section{Simulated data}\label{sec:gaea}
As mentioned in the Introduction, in this paper we use simulations to extract a set of environments comparable to those found in the MAGNET region, and to characterize the predicted global properties of galaxies across environments. In this section we hence introduce the model used.

\subsection{The semi-analytic model}
We rely on predictions from the SAM GAEA \citep{DeLucia2014, Hirschmann2016, DeLucia2024}, applied to dark matter halo merger trees derived from the Millennium Simulation \citep{Springel2005}. GAEA represents an advanced version of the original framework \citep{deluciablaiz07}, incorporating several significant enhancements. Specifically, we adopt the most recent implementation of the model described by \citet{DeLucia2024}, which features: (i) a non-instantaneous recycling scheme for gas, metals, and energy that enables the tracking of individual elemental abundances \citep{DeLucia2014}; (ii) an updated stellar feedback prescription that reproduces the observed evolution of the galaxy stellar mass function up to  $z \sim 3$, along with other  scaling relations \citep{Hirschmann2016}; (iii) an explicit treatment of cold gas partitioning into atomic and molecular phases, and a detailed implementation of RPS affecting both hot and cold gas reservoirs in satellite galaxies, as well as the tidal stripping of the hot gas reservoir \citep{Xie2017, Xie2020}; and (iv) a revised AGN feedback model that includes a physically motivated description of cold gas accretion onto supermassive black holes and quasar-mode winds \citep{Fontanot2020}. 

\citet{DeLucia2024} and \citet{Xie2024} demonstrated that this latest version  offers improved agreement with observed distributions of specific star formation rates in the local Universe and provides a reliable match to passive galaxy fractions up to $z \sim 3$. GAEA also reproduces the observed gas fractions of satellite galaxies reasonably well \citep{Xie2020} and is broadly consistent with the radial dependence of H\,\textsc{i} fractions in group halos \citep{Chen2024}. Despite its simplified treatment of ram pressure, the method adopted in GAEA shows good agreement with predictions from the IllustrisTNG hydrodynamical simulation \citep{Xie2025}. The model also correctly captures both the distribution of galaxy populations in the large-scale structure and the interplay between the main physical processes regulating their baryonic content, both for central and satellite galaxies \citep{Fontanot2025}. Given this consistency with both observations and simulations, GAEA represents an ideal framework for this analysis.

We employ model outputs based on  the Millennium Simulation, which traces $10^{10}$ dark matter particles within a comoving box of 500 Mpc $h^{-1}$ on a side, from $z = 127$ to $z = 0$, and has a gravitational softening of 5 $h^{-1}$ kpc. The simulation adopts a WMAP1 cosmology with $\Omega_{\mathrm{m}} = 0.25$, $\Omega_{\mathrm{b}} = 0.045$, $\Omega_{\Lambda} = 0.75$, $h = 0.73$, and $\sigma_8 = 0.9$, and achieves a dark matter particle mass resolution of $8.6 \times 10^{8}$ M$_\odot$ $h^{-1}$, enabling robust tracking of galaxies down to stellar masses of $\sim 10^{8.5}$ M$_\odot$.
As a consequence of its N-body foundation, the model naturally captures assembly bias—the tendency for halos that assemble earlier to be more strongly clustered than their later-forming counterparts of similar mass \citep{gao05}. This effect can leave observables imprints on galaxy properties \citep{Croton2007, Wang2013}. It is worth noting that, apart from this assembly bias, the current version of GAEA does not include any explicit treatment of either interactions between galaxies and the cosmic web (e.g., filaments) or tidal interactions between galaxies.

\begin{figure*}
    \centering
    \includegraphics[trim=0 105 0 120, clip, width=0.9\linewidth]{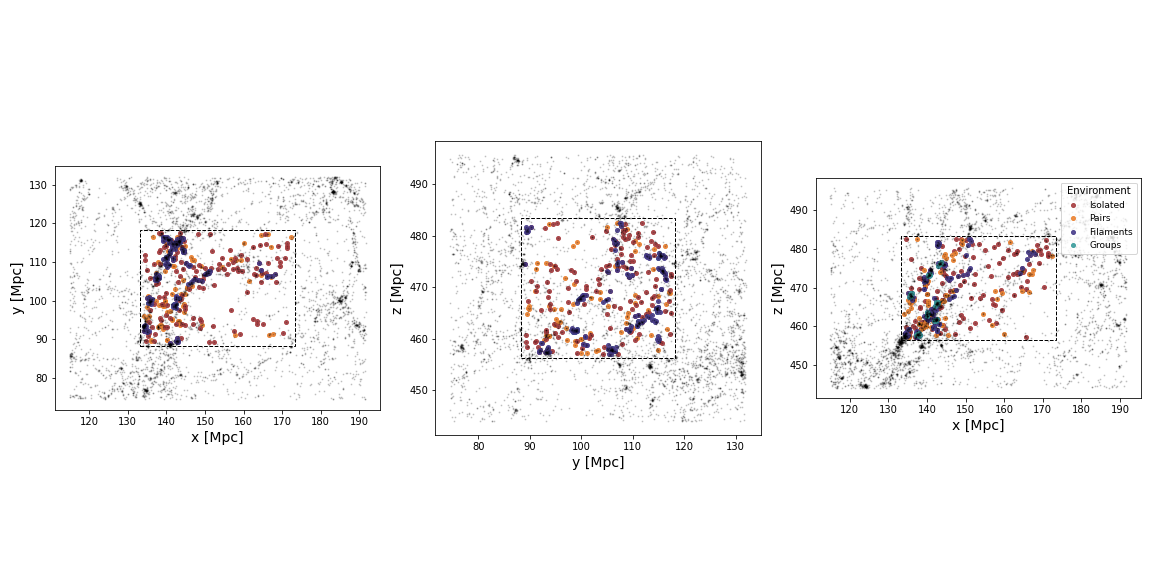}
    \caption{Example of one of the extracted sub volumes (within the dashed rectangle) and the surrounding region. Each panel shows a different projection. Within the sub-box, galaxies belonging to different environments are highlighted: isolated (rusty red), pairs (orange), filaments (steel blue), groups (forest green). This is based on the 2D environmental definition, i.e. mimicking the observational approach, but no significant changes are seen when adopted the other definitions. 
    }
    \label{fig:cube_example}
\end{figure*}

\subsection{Halos and galaxy extraction}\label{sec:boxes}
As our aim is to replicate the environments probed by MAGNET, we select sub-regions from  GAEA  based on the survey’s environmental constraints. We note, however, that the MAGNET field is designed to sample a broad range of typical cosmic environments, making the analysis presented here broadly applicable beyond the specific survey.

We systematically exclude massive groups (halo mass $> 10^{14.5}\,{\rm M_\odot}$) while ensuring a controlled presence of medium ($10^{13.5} < M/{\rm M_\odot} < 10^{14.5}$) and small ($10^{12.8} < M/{\rm M_\odot} < 10^{13.5}$) groups within each selected sub-box, as described below.

To mimic the MAGNET volume (see Sect.~\ref{sec:magnet}), we focus on the snapshot corresponding to $z = 0.02$ and extract a combination of sub-boxes of varying shapes but fixed total volume, equivalent to the MAGNET survey volume ($3.2 \times 10^4\,{\rm Mpc^3}$).\footnote{We do not reproduce the actual S-shaped geometry of MAGNET, having verified that this does not affect our results.} The side lengths of the sub-boxes range from 13 to 64~Mpc$/h$.
We proceed as follows:
\begin{itemize}
\item \underline{Identify groups:} We select the central galaxy (type~0) of each halo and classify halos into three categories:
    \begin{itemize}
    \item Massive groups: $\log({\rm M_{vir}/M_\odot}) > 14.5$;
    \item Medium groups: $13.5 < \log({\rm M_{vir}/M_\odot}) \leq 14.5$;
    \item Small groups: $12.8 < \log({\rm M_{vir}/M_\odot}) \leq 13.5$. The lower limit ensures that halos containing only one galaxy are not classified as groups.
    \end{itemize}

\item \underline{Select valid sub-boxes:} We randomly sample sub-box centers and apply the following conditions:
    \begin{itemize}
    \item No massive groups within the sub-box or within a 10~Mpc buffer outside its boundaries.
    \item No medium groups within the sub-box; up to three medium groups may lie within 10–15~Mpc of its edges, mimicking the presence of nearby structures such as Centaurus, Antlia, and Hydra, which have similar halo masses.
    \item At least five small groups must be contained within the sub-box.
    \end{itemize}
    We extract at least ten valid sub-boxes for each selected size.

\item \underline{Extract galaxies:} 
    From each valid sub-box, we extract all 
    galaxies, along with galaxies in a surrounding region twice as large, which will be used for filament identification (see Sect.~\ref{sec:env}). We also flag sub-boxes that have medium groups within 15~Mpc of their edges to assess whether their presence affects our results.
\end{itemize}

For  comparison, if we randomly extract the same number of boxes with identical geometry, but without applying any selection on the valid sub-boxes, we find that 62$\pm6$\% of the sub boxes contain medium groups,  93$\pm2$\% include small groups, and 52$\pm$6\% contain at least 5 of them. This confirms that our extraction—and therefore the MAGNET field—is representative of low-density environments.

For each model galaxy and its progenitors, we consider the following quantities: 3D positions and velocities; host halo mass ($M_{200}$, defined as the mass enclosed within a region of mean density $200\,\rho_{\rm crit}$, where $\rho_{\rm crit}$ is the critical density of the Universe); stellar mass; cold gas and H$_2$ masses; hot gas and ejected masses; galaxy type (central: type 0 or satellite: type 1); star formation rate; stellar disk size; and broad-band magnitudes. Details of how these quantities are computed are provided in \citet{Xie2017, Xie2020, DeLucia2024}. All model predictions assume a \citet{Chabrier2003} initial mass function.
In our analysis,  we  assume that helium, dust and ionized gas account for 26 per cent of the cold gas at all redshifts. The remaining gas is partitioned in H\textsubscript{I} and H$_2$, as derived in \cite{Xie2017}.
We consider only galaxies with stellar masses $> 10^{8.5}\,{\rm M_\odot}$, which approximately corresponds to the completeness limit of the VST survey. On average, each sub-box contains $\sim$650 galaxies above this mass threshold, of which $\sim$60\% are centrals, 15\% are satellites associated with a distinct dark matter subhalo, and 25\% are orphan galaxies (defined as type=2 in the model). The latter are excluded from our main analysis, though their impact is discussed in Sect.~\ref{sec:type}.

\subsection{Environmental definitions}\label{sec:env}
Before selecting the sub-boxes for analysis, we perform a comprehensive environmental classification of galaxies. We adopt three complementary environmental definitions, based on both intrinsic model outputs and projected observational quantities.

According to the ``Halo'' definition—directly derived from model outputs—we classify as isolated those type~0 galaxies that have no gravitationally bound satellites above the adopted stellar mass limit. Similarly, we define pair galaxies as central galaxies hosting a single companion, while groups correspond to haloes containing at least three members (including the central).

The second and third definitions are designed to mimic observational classifications, using both three-dimensional and projected (2D) quantities. Following the AMIGA sample selection criteria \citep{Verdes2005}, we define isolated galaxies as those with no companions within 0.5~Mpc having (i) a relative velocity $< 500\,{\rm km\,s^{-1}}$ and (ii) a stellar mass exceeding 10\% of the target galaxy’s mass. The same criteria identify pair galaxies as systems with exactly one such companion. Group members are defined as type~0 or 1 galaxies residing in haloes with $\log(M_{\rm vir}/{\rm M_\odot}) > 12.8$. For the 2D case, we further require that group members lie within one projected $R_{200}$ of their central galaxy.

To trace the large-scale structure, similarly to what we did in observations, we identify filaments using the DisPerSE algorithm, applied to the 3D spatial distribution of galaxies with $M_\ast > 10^9\,{\rm M_\odot}$.\footnote{\citet{Laigle2018} showed that two- and three-dimensional filament reconstructions yield broadly consistent results, justifying the use of 3D filaments in this context.} We follow the approach of \citet{Zakharova2023}, adopting a persistence threshold of $5\sigma$.\footnote{A lower threshold of $3\sigma$ is used for the observational data, owing to their lower sampling density. The higher resolution of the simulation allows for a stricter $5\sigma$ threshold; adopting $3\sigma$ would introduce spurious filament detections.} We apply five smoothing iterations to suppress small-scale noise, averaging over five adjacent segments, and discard filaments shorter than 3~Mpc$/h$. Galaxies located within 1~Mpc$/h$ of a filament spine are considered filament members.

We examine the environmental composition of all extracted sub-boxes and retain those containing at least four identified groups under all definitions. This selection yields a final sample of 15 sub-boxes, which form the basis of the analysis presented in the following sections. Most selected sub-boxes lack massive haloes in their surroundings; only four contain haloes with masses comparable to those of Antlia, Centaurus, or Hydra. In these cases, visual inspection reveals that the sub-boxes are connected to these haloes by prominent filaments, resembling the configuration observed in MAGNET. 
Figure~\ref{fig:cube_example} illustrates one of the selected sub-boxes, highlighting the various identified environments.

\subsection{Physical processes affecting galaxies}\label{sec:process}
We follow model galaxies up to $z\sim2$, identifying the physical processes that affected them over the last 9 Gyr to assess their role in driving galaxy evolution.
We identify the following evolutionary paths:

\begin{itemize}
\item \underline{Main branch progenitor}: we reconstruct the main branch (or main progenitor line) for a target galaxy by recursively following the firstProgenitorId field. This quantity indicates the most massive progenitor of the galaxy in the previous redshift bin. We stop tracing at $z=2$ or when  no further progenitor exists.

\item \underline{Merger detection}: we scan all progenitors of a target galaxy to detect merger events using the descendantId field given by the model. When multiple progenitors share the same descendant, a merger is counted. We consider all mergers with a mass ratio down to 1:10, and we will discuss in Sect.~\ref{sec:discussion} the impact of this choice. We allow for multiple mergers occurring simultaneously. For galaxies undergoing mergers, we also compute the net total mass that has been accreted onto the main galaxy.

\item \underline{Tidal (close) interaction detection}: 
We define the tidal radius $R_{\text{tidal}}$ as the maximum distance within which material in the target galaxy remains gravitationally bound, given the tidal perturbation exerted by a companion galaxy at a separation $r$. 
Following \citet{Vollmer2005, Merluzzi2016}, we compare the tidal acceleration induced by the companion ($a_{\text{tid}}$) to the self-gravitational acceleration of the target galaxy ($a_{\text{gal}}$). 
The two quantities are expressed as

\begin{equation}
a_{\text{tid}} = G M_{\text{comp}} \left( \frac{1}{(r - R)^2} - \frac{1}{r^2} \right), \qquad
a_{\text{gal}} = -\frac{G M_{\text{gal}}}{R^2},
\end{equation}

where $r$ is the separation between the centers of the companion and target galaxies, 
$M_{\text{gal}}$ and $M_{\text{comp}}$ are their total masses, $R$ is the distance from the center of the target galaxy at which the balance is evaluated, and $G$ is the gravitational constant. 
In this formulation, $a_{\text{gal}}$ represents the gravitational acceleration exerted by the target galaxy on a test particle located at a distance $R$ from its center, 
while $a_{\text{tid}}$ corresponds to the differential acceleration (tidal force) exerted by the companion on the same particle.
The tidal radius $R_{\text{tidal}}$ is obtained by solving

\[
\mu R^2 \left[ \frac{1}{r^2} - \frac{1}{(r -  R)^2} \right] = a_{\text{tid}}/a_{\text{gal}},
\]

where $\mu = M_{\text{comp}}/M_{\text{gal}}$. 
We adopt $a_{\text{tid}}/a_{\text{gal}} = 0.15$, following \citet{Vulcani2021, Watson2025}, and solve for $R = R_{\text{tidal}}$ using Brent's method within the interval $[10^{-3}, 0.99r]$. 

For each target galaxy, all possible companions are scanned. 
If the derived $R_{\text{tidal}}$ exceeds the stellar disk size, defined as StellarDiskRadius/3 $\times$ 1.68,\footnote{In the model, \texttt{StellarDiskRadius} corresponds to three times the disk scale length; for an exponential profile, $R_e = 1.68$ times the scale length.} 
the tidal effect is deemed dynamically negligible and excluded from further analysis. 
If instead $R_{\text{tidal}}$ is smaller than the stellar disk radius, the system is flagged as a potential tidal interaction. 

We note that this method is based on an analytic  proxy rather than fully dynamical stripping. The current version of GAEA does not include an explicit treatment of tidal interactions between galaxies. 
To avoid double-counting pre-merger configurations, interactions between galaxies belonging to the same merger tree (i.e., progenitor–descendant pairs) are excluded.

\item \underline{Ram pressure stripping}: we classify a galaxy as undergoing RPS if its RPradius—defined by \citet{Xie2020} as the radius beyond which ram pressure is effective—drops below half the stellar disk radius at any point between $z=2$ and $z=0$. Specifically, for each type = 1 galaxy, we trace its main progenitor branch back to $z=2$ and check whether this condition is ever met.\footnote{A galaxy is classified as RPS-affected if the criterion is satisfied at least once since $z=2$. In some cases, the RPradius decreases temporarily and then increases again.}
In the GAEA model, cold-gas stripping can occur only if the radius of the hot-gas halo becomes smaller than the RPradius—that is, after substantial removal of the hot gas. This ensures that our criterion selects strong cases of RPS.
As a result, galaxies identified as RPS-affected in this work have also experienced significant starvation, although they are not included in the starvation sample.
Since we rely on a semi-analytic approach, it is not possible to trace the spatial distribution of the stripped gas and therefore to identify the counterparts of the observed galaxies exhibiting long ram-pressure–stripped tails \citep[e.g.,][]{Poggianti2025}.

\item \underline{Starvation}: to select galaxies undergoing starvation, for each type = 1 galaxy that was a type = 0 at $z=2$, we compare the total mass (HotGas + EjectedMass + ColdGas + StellarMass) they had the last time they were a type = 0 galaxy—i.e., at the redshift before entering their current halo and becoming satellites—with the amount of mass they have lost since then due to stripping or tidal effects on the hot gas (Mhot\_rps+Mhot\_tidal). If more than 40\% of the mass is lost, we consider the galaxy to have undergone starvation.\footnote{Slightly changing the threshold does not qualitatively alter the results, even though absolute numbers change.}

\end{itemize}

Finally, we group galaxies that do not fall into any of the previously defined categories. {We stress that this class does not imply the absence of physical processes affecting galaxy evolution, but rather that  these galaxies do not meet the criteria adopted to be classified as undergoing the specific, strong environmental mechanisms considered here. Their evolution is therefore expected to be dominated by internal processes or by environmental effects that are weaker, more transient, or below the thresholds of our classification scheme.} From now on, we refer to this category as “None”.

\begin{figure*}
    \centering
    \includegraphics[trim=0 15 0 0, clip, width=0.9\linewidth]{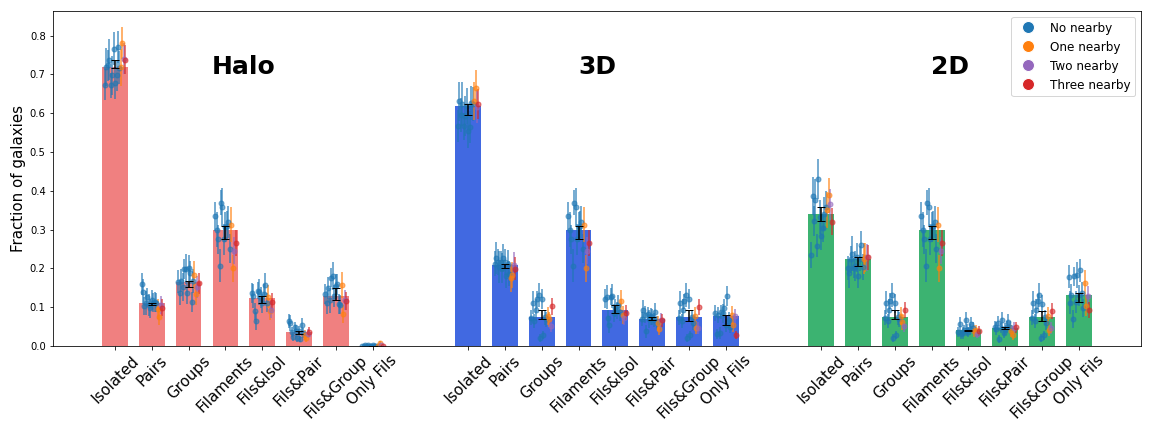}
    \caption{Frequency of galaxies in the different environments, based on the three adopted definitions of environment. Reported fractions are the median value of the 15 realizations, with the black errorbars showing the binomial errors.  Small  circles represent the results from the individual extracted sub-boxes and colors refer to sub-boxes with different numbers of neighboring groups, as defined in the labels.}
    \label{fig:envs_stats}
\end{figure*}

\begin{figure*}
    \centering
    \includegraphics[trim=0 0 0 0, clip, width=0.4\linewidth]{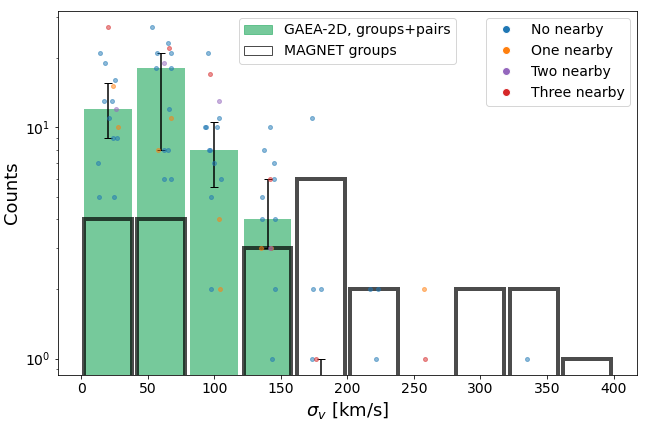}
   \includegraphics[width=0.4\linewidth]{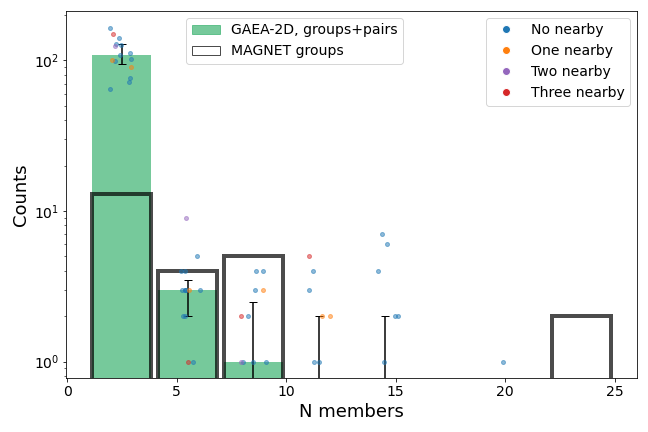}
    \caption{Comparison between the velocity dispersion distribution (left) and the number of members (right) of the MAGNET field (empty histogram) and the pairs and groups identified in GAEA (green) when using the 2D environment definition. For GAEA, median values are reported.  Small  circles represent the results from the individual extracted sub-boxes and colors refer to sub-boxes with different numbers of neighboring groups, as defined in the labels. }
    \label{fig:cfr_obs}
\end{figure*}

\section{Results}\label{sec:results}

\subsection{Comparing model and observations}\label{sec:comp_magnet}In Sect. \ref{sec:env}, we identified galaxies in different environments according to three definitions: halo-based, 2D, and 3D.
Figure \ref{fig:envs_stats} shows the fraction of galaxies with $\log (M_\ast/M_\odot) > 8.5$ belonging to each environment, also distinguishing those that simultaneously reside in both filaments and any of the other environments, as well as galaxies located only in filaments. The reported fractions correspond to the median values across the 15 individual realizations; for completeness, we also show as small circles the fractions derived from each individual extraction. In general, the spread among different realizations is at most of order 10\%.
We also analyzed sub-boxes characterized by different numbers of groups in the surrounding regions, but no clear trends emerge, suggesting that the large-scale environment outside the selected regions does not play a dominant role.

Considering the halo-based definition, we note that by construction the fractions of isolated galaxies, pairs, and groups necessarily sum to unity, whereas other environmental categories may overlap. About 75\% of the galaxies are found in isolation, followed by galaxies in filaments, groups, and pairs. The latter constitute only $\sim$10\% of the entire sample. Clearly, these fractions depend on the adopted mass cut; in the following, we will also consider different mass regimes.
\begin{figure*}
    \centering
    \includegraphics[trim=0 20 0 0, width=0.9\linewidth]{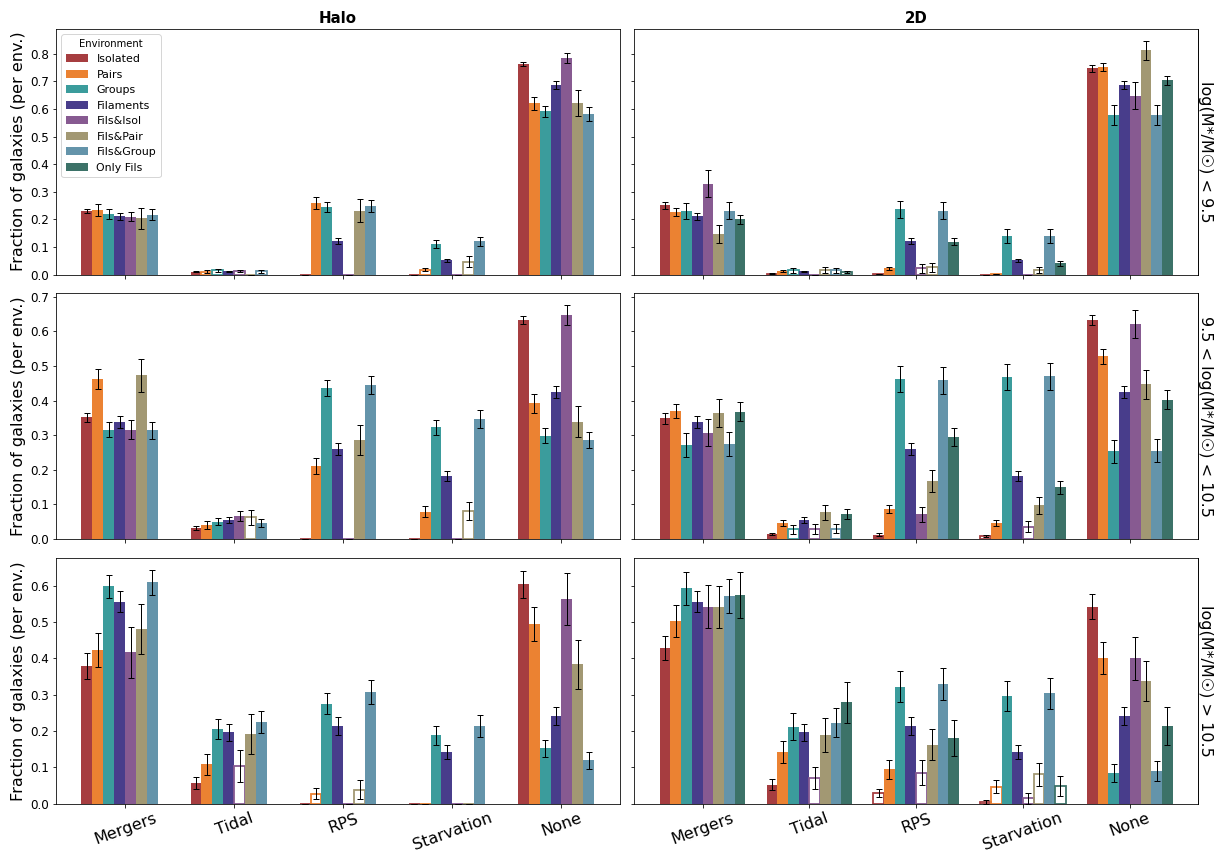}
     \caption{Fractions of galaxies undergoing different processes in different environments, in three different mass bins. Fractions are normalized per environment. Errorbars report the binomial errors. Empty bars indicate cases where there are less than 10 galaxies in that environment and are subject to that process. Left: Halo-based environment definition, Right: 2D-based environment definition.}
    \label{fig:distr-env-proc}
\end{figure*}
Overall, most galaxies found in groups are also located in filaments, in agreement with previous theoretical and observational studies \citep[e.g.,][]{Tempel_2014_necklace, Poudel2017, Castignani2022, Zakharova2024}. About 15\% of isolated galaxies are also in filaments, indicating that filaments extend over scales larger than individual halos.  

Focusing on the filament population, roughly half of the galaxies also belong to groups, about half are isolated, and fewer than 5\% are in pairs. We note that the combined fractions of Filaments \& Groups, Filaments \& Pairs, and Filaments \& Isolated correspond to the total filament fraction reported in Fig.~\ref{fig:envs_stats}. The fraction of galaxies located only in filaments (i.e., not in groups) is almost zero. According to the halo-based definition, galaxies should not exist solely in filaments; these few cases correspond to filament satellites at the edges of the extractions whose central galaxy lies outside the sub-box. Since there are fewer than 20 such galaxies, from now on we will ignore this population.  

 Using the 2D definition, the fractions of isolated galaxies, pairs, and groups no longer sum to unity, as given their definition (see Sect.\ref{sec:env}), these classifications are not mutually exclusive.
Approximately 30\% of galaxies remain unassigned to any of the three categories. Overall, the fractions of isolated and group galaxies decrease by nearly a factor of two, while the fraction of pairs doubles with respect to the halo based definition, highlighting how galaxies classified as isolated in the 2D framework are often misassigned. {It is indeed important to note a key conceptual difference between the halo-based and the 2D definitions. In the halo-based framework, isolated galaxies are, by construction, centrals and cannot experience processes such as RPS or starvation. In contrast, in the 2D definition, galaxies classified as “isolated” can in fact be satellites projected outside their host halo, and therefore may still undergo environmental effects.} This emphasizes the intrinsic limitation of observational definitions, which are susceptible to projection effects and contamination. In this definition, the fraction of galaxies exclusively in filaments increases to $\sim$20\%\footnote{We note that the same filament structure is used across all three environmental definitions, so the filament fractions remain identical.}. Discrepancies between the halo-based definition and the observational-like definition are partially alleviated when using 3D distances, although they do not disappear entirely.  
Since the 3D definition produces results intermediate between the halo and 2D definitions, we will not present them in detail here, having verified that they consistently fall between the other two sets of results.  

As noted, the 2D definition is intended to be directly comparable to the MAGNET field. In Fig.~\ref{fig:cfr_obs}, we compare the velocity dispersion and number of group members in MAGNET groups to those identified in GAEA. We note that in MAGNET, pairs and groups are identified using the same method; therefore, in GAEA, we consider both pairs and groups together.  

To mimic observational measurements of velocity dispersions in GAEA, we consider the central galaxy of each group or pair and extract a sub-box containing all galaxies above our mass limit within three projected virial radii and with redshifts\footnote{Redshifts are derived from peculiar velocities as described in Appendix B of \cite{Zakharova2024}.} within $\pm 0.002$. We then compute the projected velocity dispersions using a 3$\sigma$ clipping procedure.  
Overall, the two samples span a similar range in velocity dispersion, although very few groups in GAEA have $\sigma_v > 200$ km/s, compared to seven in MAGNET. These MAGNET groups, however, contain few members ($<8$), suggesting that their velocity dispersions may be overestimated. We will revisit these measurements as more data become available.  
Regarding group membership, GAEA identifies many more pairs than observed and relatively few groups with more than ten members. Only one (two) system in GAEA (MAGNET) contains more than 20 galaxies. In MAGNET, both systems have reported velocity dispersions of $\sim 180$ km/s \citep{Kourkchi2017}. These differences may result from variations in the methodologies used to identify galaxy groups and pairs. Nevertheless, the overall numbers are broadly comparable, providing confidence in the predictions presented in the following sections.

\subsection{The distribution of galaxies across environments and physical processes}\label{sec:frac}

In Sect.~\ref{sec:process}, we have identified galaxies undergoing different physical processes based on their current and past properties, offering a theoretical context for interpreting observations. Comprehensive comparative analyses are planned for a subsequent phase of the MAGNET project.

Figure~\ref{fig:distr-env-proc} shows the fraction of galaxies subject to different environmental processes across three stellar mass bins and for both the halo-based and 2D environment definitions. For clarity, we do not present results for individual sub-boxes, but instead show median values obtained by combining galaxies from all 15 sub-boxes.  
Fractions strongly depend on stellar mass and, to a lesser extent, on the environment classification—particularly for tidal interactions and mergers. The fraction of galaxies that have experienced at least one merger event\footnote{Here we do not distinguish between galaxies that had only one or multiple merger events in the considered time frame. A discussion is deferred to Sect.~\ref{sec:mergers}.} ranges from about 20\% for galaxies with $\log(M_\ast/M_\odot) < 9.5$ to 40--60\% for galaxies with $\log(M_\ast/M_\odot) > 10.5$, regardless of the  environment definition. In the highest mass bin, mergers are more common in group and filament galaxies and in groups in filaments, and less so in isolated galaxies or pairs.  

Tidal interaction cases also become more frequent with increasing stellar mass: less than 5\% of low-mass galaxies are affected, rising to approximately 20\% among the most massive galaxies. Also the incidence of starvation depends on stellar mass, but its dependence is not monotonic. It goes from 10--15\% at low masses up to 30\% (40-45\%) among group and filament galaxies in the halo (2D) definition  at intermediate masses and then decreases again to 20--30\% in the highest mass bin.

The trend of RPS depends sensitively on the environment definition (2D vs. halo-based) and on the type of environment considered. Among pairs in the halo-based definition, RPS is most prevalent at low masses, affecting up to 30\% of galaxies, and decreases to less than 5\% in the highest mass bin. In the 2D definition, it is almost absent among low-mass galaxies, while its incidence increases with stellar mass. In groups, for both definitions, RPS affects about 30\% of galaxies in the lowest and highest mass bins and up to 50\% at intermediate masses. In filaments, the fractions are highest at intermediate masses.  

A significant fraction of galaxies do not appear to have been affected by any of the considered physical mechanisms, though this fraction varies with environment. This is especially true for isolated galaxies (both within and outside filaments), where up to 80\% of low-mass galaxies and around 60\% of high-mass galaxies show no signs of environmental processing (halo-based definition). In the 2D classification, a similarly large fraction of filament galaxies and galaxy pairs also appear unaffected, though this fraction tends to decrease with increasing stellar mass.

\begin{figure*}
    \centering
    \includegraphics[width=0.9\linewidth]{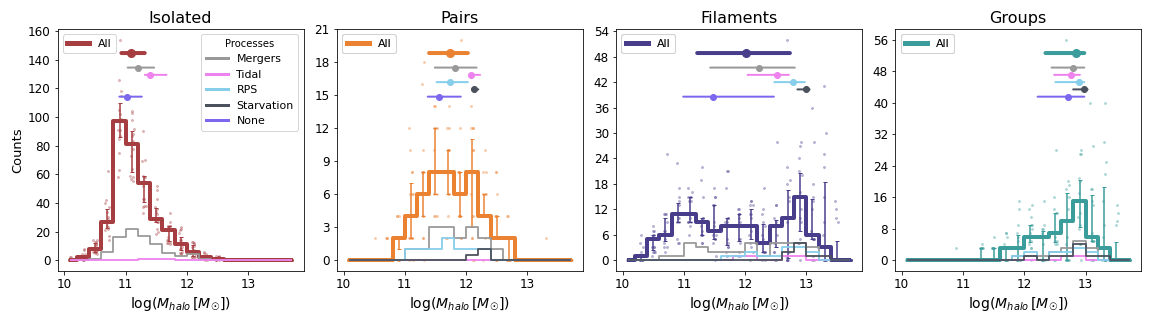}
    \caption{Halo mass distribution for galaxies in different environments (based on the Halo definition) and affected by different mechanisms.  Small circles show the distributions for the individual realizations. When the physical processes are taken into account, neither single realizations nor errobars are shown, for the sake of clarity. The distribution of galaxies not undergoing physical mechanisms is not reported, but can be inferred by the difference between the total and the sum of the other mechanisms.  Circles and horizontal lines show median values along with the 25 and 75 percentiles of the distributions. Median values for the None categories are also given. Thicker lines with errorbars shows the median distribution in each environment, regardless of the mechanisms. }
    \label{fig:halo_distr_halo}
\end{figure*}

Since the definitions of the different physical processes are independent, galaxies can simultaneously meet the criteria for more than one process. This explains why, in a given environment, the sum of the fractions discussed above can exceed unity. The occurrence of multiple processes acting on the same galaxy is a well-documented and expected phenomenon \citep[e.g.,][]{Fritz2017, Serra2024, Watson2025, Finn2025}.  
To explore this further, we quantify how many galaxies experience multiple processes simultaneously, 
for each stellar mass bin and environment. Overall, at all masses and in all environments, more than 60\% of the galaxies have experienced only one mechanism since $z\sim 2$. The fraction is the highest in isolated galaxies (almost 100\%) and the lowest in groups. The fraction of galaxies that underwent a merger event and at least one other mechanism increases with stellar mass and ranges from 20 to $>50\%$. The secondary mechanism is more often RPS. As for tidal interactions, at all masses only $\sim20\%$ of the galaxies also underwent another mechanism and again RPS is the most common. The fraction of objects feeling both RPS and starvation  increases with stellar mass and can be as high as 50\% at the highest masses. More details can be found in Appendix \ref{app:multiplicity}. 

{We emphasize that throughout this work galaxies are classified according to their environment at $z \sim 0$, while the physical processes considered here may have occurred at earlier times and potentially in different environments along the galaxy evolutionary history. As a consequence, all results should be interpreted as describing the association between present-day environments and the past occurrence of specific physical mechanisms, rather than implying a direct causal link between the two. This limitation is intrinsic to any classification based on instantaneous environmental definitions (see Sect.~\ref{sec:caveat}). 

Moreover, some trends may partly reflect the built-in halo mass dependence of the GAEA model, while processes such as tidal interactions or filament-specific effects are not explicitly modeled and therefore may contribute only indirectly via assembly bias. Similarly, RPS is identified only in its most advanced stages, which may affect the inferred fractions. Despite these caveats, robust trends—such as the dependence of mergers, tidal interactions, and starvation on stellar mass—remain meaningful within the context of the model’s strengths.}

\subsection{Properties of galaxies in different environments and subject to different processes }\label{sec:prop}

\begin{figure*}
    \centering
    \includegraphics[width=0.9\linewidth]{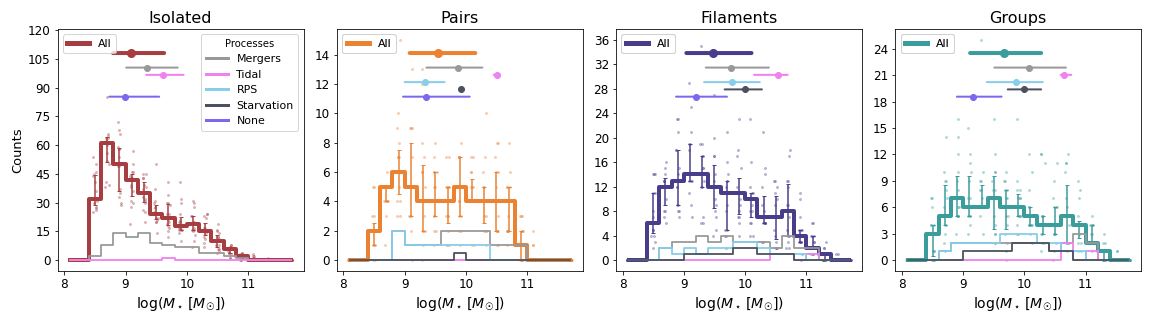}
    \caption{Stellar mass distribution for galaxies in different environments  (based on the Halo definition) and affected by different mechanisms. Panels, colors, lines and symbols are as in Fig.\ref{fig:halo_distr_halo}.}
    \label{fig:mass_distr_halo}
\end{figure*}

\begin{figure*}
    \centering
    \includegraphics[trim = 0 50 0 0, width=0.9\linewidth]{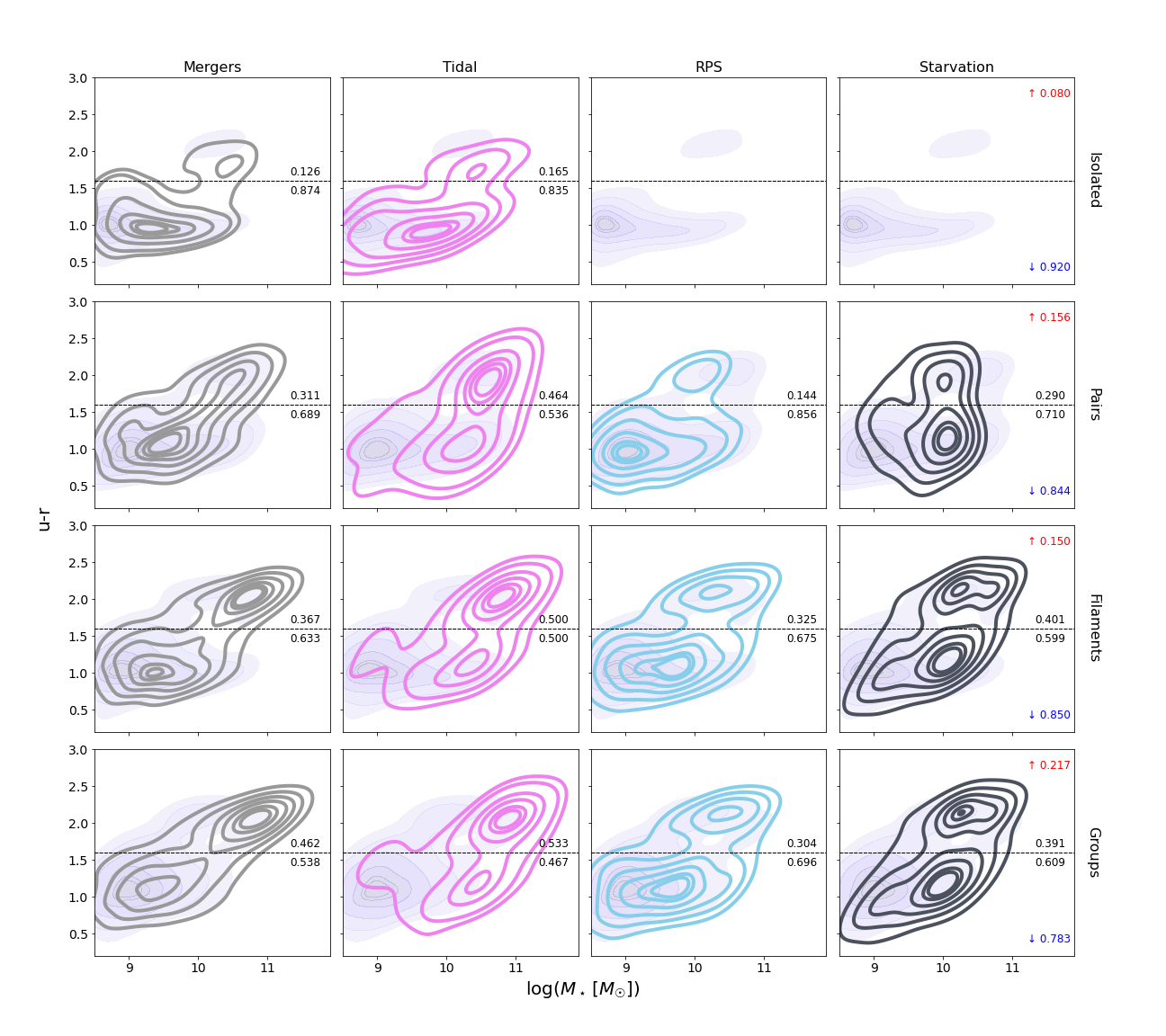}
    \caption{$u-r$ vs. stellar mass diagram for galaxies experiencing different physical mechanisms (columns) and residing in different halo-defined environments (rows). Solid lines represent the 10th, 25th, 50th (median), 75th, 90th, and 95th percentiles of the distributions. In each panel, the curves correspond to the mechanism indicated at the top and the environment indicated on the left. As a reference, the background shaded distribution shows galaxies in the same environment not affected by any mechanism. These red curves are identical across all columns within a given row. A horizontal line at $u - r = 1.6$ is shown as a tentative division between blue and red galaxies. The numbers above and below this line indicate the fraction of galaxies in the red and blue regions, respectively. In the rightmost column, the red and blue fractions refer to the population of galaxies not undergoing any process. }
    \label{fig:coloromass_halo}
\end{figure*}

In the previous sections, we examined the occurrence of different environments and physical processes. We now turn to the analysis of galaxy properties within the various environmental and process categories. For clarity, the main text focuses on results obtained using the halo-based environment definition. The corresponding figures based on the 2D definition are provided in Appendix~\ref{app:2d}.

\subsubsection{Halo and stellar mass distributions} \label{sec:distr}
Figure~\ref{fig:halo_distr_halo} shows the distribution of halo masses for galaxies residing in different environments and affected by different physical processes. Focusing first on the environmental classification (represented by thicker lines), the distributions behave as expected: isolated galaxies are skewed toward lower halo masses, with a median value of 11.08 and an interquartile range (IQR) of 10.90--11.34. The distribution shifts to higher values for galaxy pairs (median: 11.74, IQR: 11.36--12.06), and further for groups (median: 12.83, IQR: 11.31--13.01). Notably, the distributions of isolated and group galaxies show minimal overlap.

Filament galaxies exhibit a broad, bimodal distribution that spans the full range of halo masses observed for isolated, pair, and group galaxies (median: 12.02, IQR: 11.19--12.77). This behavior reflects the intrinsic heterogeneity of filament environments. Indeed, when we break down the filament category into finer sub-classes—simultaneously in filaments and isolated galaxies, filaments and pair galaxies, and both filament and group members (plots not shown)—we recover the respective halo mass distributions of isolated, pair, and group galaxies. This further supports the idea that filaments are not distinct environments per se, but rather encompass a mixture of the other environmental categories \citep[e.g.,][]{Zakharova2023, Zakharova2024}.

In general, the median values in Fig.~\ref{fig:halo_distr_halo} indicate that galaxies affected by any physical mechanism tend to reside in more massive halos compared to galaxies in the same environment that do not experience such processes. The only notable exception is found within the group category, where, though—as shown in Fig.~\ref{fig:distr-env-proc}—less than 20\% of galaxies are unaffected by environmental mechanisms.

Galaxies affected by RPS are found in lower-mass halos when in pairs than when in groups or filaments. In pairs, their halo masses are similar to those of galaxies that have undergone mergers, while in filaments they tend to occupy somewhat higher-mass halos. This shift toward larger halo masses is not seen in groups.  Although based on limited statistics, starved galaxies are systematically located in the most massive halos. Among galaxies that have experienced at least one merger event, we find that median halo masses increase progressively from isolated galaxies to pairs and then to groups, with filament galaxies consistently showing intermediate values.

The fact that distributions strongly depend on environment and less on the physical mechanism experienced suggests that the physical mechanisms are tightly related to the environment in which galaxies live today. However, it also reflects the intrinsic nature of the environmental classification itself (see Sect.~\ref{sec:env}). 

Figure~\ref{fig:mass_distr_halo} shows the distribution of stellar masses for galaxies in different environments and undergoing different physical mechanisms. Focusing first on the environmental classification, isolated galaxies are skewed toward lower stellar masses (median: 9.09, IQR: 8.76--9.66). The median stellar mass increases when moving to pairs (median: 9.54, IQR: 9.05--9.91) and groups (median: 9.67, IQR: 9.08--10.21), with filament galaxies showing intermediate values (median: 9.47, IQR: 9.01--10.14).

When considering the various physical processes, isolated and filament galaxies that have not experienced any environmental mechanism tend to have lower stellar masses, as indicated by their systematically lower median values. In contrast, galaxies that experienced tidal interactions exhibit higher median stellar masses, suggesting a history of more significant mass growth. Starved galaxies, when sample sizes allow reliable statistics, also tend to be more massive than those affected by mergers or RPS. In all environments, mergers have a rather flat stellar mass distribution, suggesting that galaxies of all masses have similar chances of undergoing this type of process.
Galaxies having undergone RPS are systematically the least massive across all environments where this mechanism is present. This trend likely reflects the fact that galaxies with lower stellar masses are more vulnerable to ram-pressure effects, rather than RPS directly suppressing or reducing stellar mass over time. In particular, galaxies in pairs affected by RPS show the lowest stellar masses among all galaxies experiencing any mechanism (median: 9.33, IQR: 8.97--9.69), and their distribution is skewed toward low values.

Comparing galaxies in isolation, pairs, and groups with their counterparts located also  within filaments (plots not shown), no significant differences emerge. This indicates that the filament environment, in itself, does not play a major role in determining stellar mass.

\begin{figure*}
    \centering
    \includegraphics[width=0.9\linewidth]{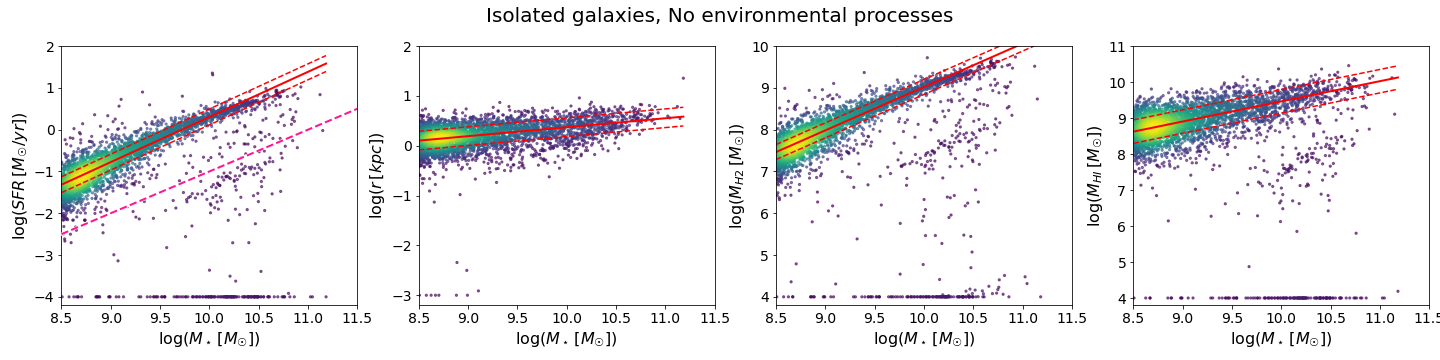}
    \caption{Scaling relations that will be investigated as a function of environment and process. From left to right: SFR-mass, size-mass, H2-mass, HI-mass. Only isolated galaxies with no process are considered and the best fit of the relation is shown. This line will be used as reference in the following analysis. In the SFR-mass plot, the separation used to identify star forming and passive galaxies ($\log(SSFR[yr^{-1}])= -11$) is also shown.}
    \label{fig:ref_relations}
\end{figure*}

{We note that some of the trends discussed above— particularly those for RPS and starvation—are influenced by the modeling choices in GAEA, such as the identification of RPS only in advanced stages and the mass-dependent quenching prescriptions. Furthermore, the present-day environmental classification captures a snapshot, while galaxies may have experienced these processes at earlier times in different halos, potentially broadening the distributions. 
}

\subsubsection{Color - stellar mass diagrams}

Figure~\ref{fig:coloromass_halo} presents the $z=0.05$ $u-r$ color versus stellar mass diagrams, shown as density plots. Galaxies are separated by environment and by the physical mechanism they experience: each row corresponds to a different environment, and each column to a different mechanism. For reference, in each panel, the distribution of galaxies in the same environment but not affected by any mechanism (i.e., the None category) is shown as shaded purple background contours. The corresponding red and blue fractions for these ``unaffected'' galaxies are reported in the upper right (red) and lower right (blue) corners of each plot. 
We adopt a horizontal line at $u - r = 1.6$ as a rough separation between red and blue galaxy populations. The red and blue fractions for each specific population (i.e., galaxies affected by the mechanism under consideration) are shown in black, above and below the dividing line, respectively.

The vast majority of galaxies in isolation are characterized by blue colors. Among isolated galaxies not affected by any physical mechanism, the fraction of red galaxies is nearly zero, even at the high-mass end. As expected from their definitions, no ram-pressure stripped or starved galaxies are found in the isolated category.  Merger or tidal interaction cases show a slightly higher probability of being red, although in all cases the fraction of red galaxies remains below 20\%.

Regardless of the specific mechanism considered, the fraction of red galaxies increases progressively from isolated galaxies to pairs, and then to groups, with filaments showing intermediate values between pairs and groups. 
Within each environment, the highest blue fractions are typically associated with galaxies classified as  RPS. Their color--mass distribution largely overlaps with that of galaxies not experiencing any mechanism. This contrasts with the other mechanisms, where the blue cloud is systematically shifted toward higher stellar masses, suggesting a more evolved or transformed population. As far as the other mechanisms are concerned, the red and blue fractions are very similar, suggesting that they are all efficient in populating the red sequence. 

\begin{figure*}
    \centering
    \includegraphics[trim=0 20 0 0, width=0.9\linewidth]{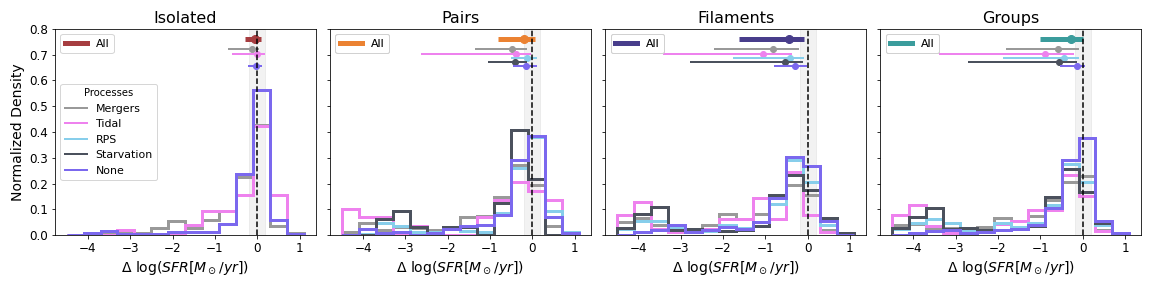}
    \caption{Normalized distribution of the difference between each galaxy SFR and the expected value given its mass and the best fit for isolated galaxies with no process, shown in Fig.\ref{fig:ref_relations}. Each plot shows a different environment (based on the halo definition), each color shows a different process, as indicated in the legend. The vertical dashed line and the shaded area shows the the location of the best fit and the scatter.  Circles and horizontal lines show median values along with the 25th and 75th percentiles of the distributions. Thicker lines with errorbars shows the median distribution in each environment, regardless of the mechanisms, following the color scheme of Fig.\ref{fig:distr-env-proc}.}
    \label{fig:delta_sfr}
\end{figure*}

To conclude, both environment and physical mechanisms play a significant role in shaping the color distribution of galaxies. The environment sets the stage, with red fractions increasing from isolated systems to pairs and groups. At the same time, specific physical processes—particularly starvation and tidal interactions—further enhance the likelihood of galaxies transitioning to the red sequence. {We remind the reader, however, that galaxies are classified by their present-day environment, while the physical mechanisms responsible for color transformation may have acted earlier and potentially in different environments. Therefore, the trends reported here should be interpreted as statistical associations rather than direct causal links. 
Results might be driven by some inherent limitations 
of the model and should be kept in mind when interpreting these trends (see Sec.\ref{sec:caveat}).}

\subsubsection{Deviations from scaling relations}\label{sec:deviations}
We now aim to investigate whether galaxies residing in different environments and subject to different physical mechanisms occupy distinct regions within standard galaxy scaling relations. Specifically, we focus on the following key relations: the star formation rate (SFR)–stellar mass, the size–mass, the H\textsubscript{I}–mass, and the H\textsubscript{2}–mass relations. 

To quantify potential differences arising from various environments or mechanisms, the choice of the reference sample used for comparison is crucial. We chose isolated galaxies that have not undergone any physical processes, regardless of their environment, because their evolution is expected to be unaffected by external influences. Figure \ref{fig:ref_relations} presents the scaling relations derived for this population. A linear fit is applied to these data points to determine the best-fit relations, which are shown as solid lines, with the associated scatter represented by dashed lines. Subsequently, we analyze galaxies in different environments and undergoing distinct processes separately. For each subgroup and for each galaxy property, we compute the distribution of deviations from the reference relation. Specifically, for each galaxy, we calculate the difference between its actual y-axis value and the value predicted by the best-fit relation at its stellar mass. These deviations are then used to construct the corresponding distributions.

\begin{figure*}
    \centering
    \includegraphics[trim=0 20 0 0,width=0.9\linewidth]{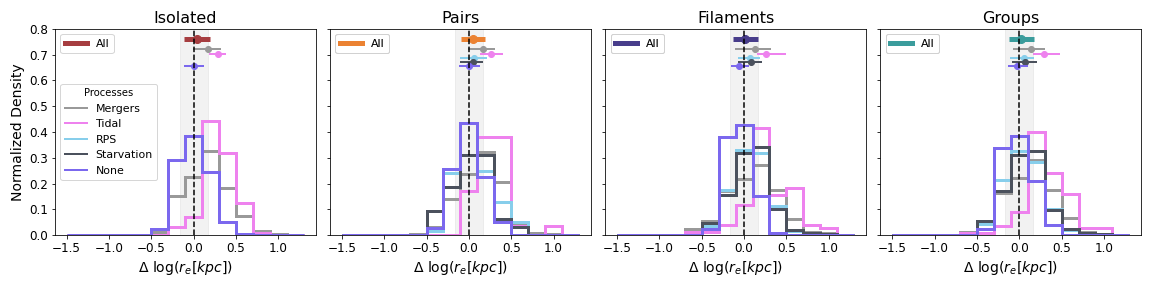}
    \caption{Normalized distribution of the difference between each galaxy size and the expected value given its mass and the best fit  for isolated galaxies with no process, shown in Fig.\ref{fig:ref_relations}. Panels, colors, lines and symbols are as in Fig. \ref{fig:delta_sfr}.}
    \label{fig:delta_size}
    \end{figure*}

\begin{figure*}
    \centering
    \includegraphics[trim=0 20 0 0,width=0.9\linewidth]{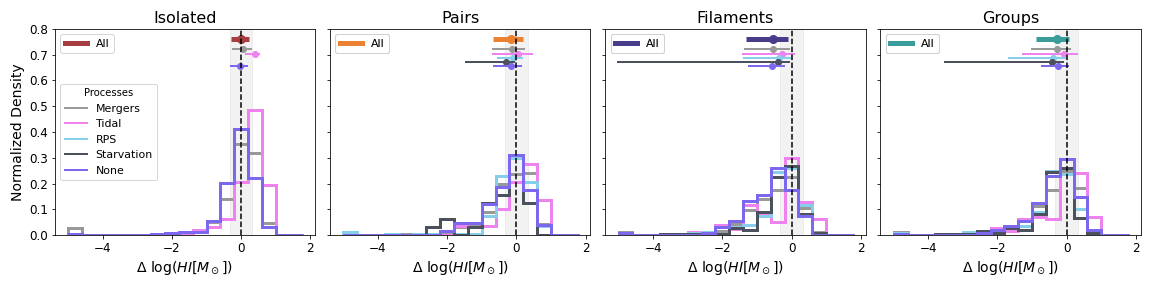}
    \caption{Normalized distribution of the difference between each galaxy HI mass and the expected value given its mass and the best fit for isolated galaxies with no process, shown in Fig.\ref{fig:ref_relations}. Panels, colors, lines and symbols are as in Fig. \ref{fig:delta_sfr}.}
    \label{fig:delta_Hi}
\end{figure*}

Figure \ref{fig:delta_sfr} shows the normalized distributions of residuals from the SFR--mass relation of isolated galaxies with no process. When focusing on isolated galaxies, no significant trends are observed across the different physical mechanisms. Median values and overall distributions are largely indistinguishable, and pairwise Kolmogorov--Smirnov (KS) tests fail to detect any statistically significant differences among the samples. This confirms that, in isolation, the impact of the considered mechanisms on star formation is either negligible or not distinguishable within the available statistics. 
In contrast, clear trends emerge in denser environments. A tail of galaxies with suppressed SFR is evident across all mechanisms and is present in all environments. As a result, the median SFR residuals are systematically lower than those of the reference (unaffected) sample, regardless of the mechanism considered. The KS test confirms that all subsamples significantly differ from the reference population ($p$-value$<0.05$), supporting the idea that environmental mechanisms contribute to star formation suppression. 
The effect is particularly strong for filament and group galaxies having undergone either starvation or tidal interactions, where nearly all galaxies show $\Delta \log( \mathrm{SFR} \ [M_\odot/\mathrm{yr}]) < 0.5$, and the distribution is markedly skewed toward low values. 

\begin{figure*}
    \centering
    \includegraphics[trim=0 20 0 0,width=0.9\linewidth]{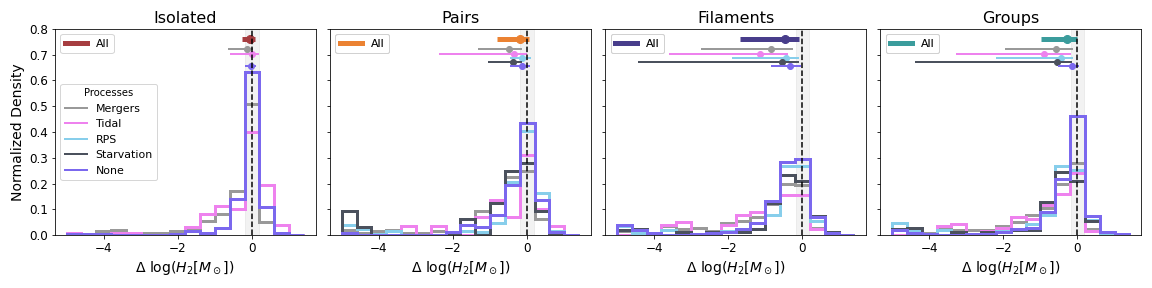}
    \caption{Normalized distribution of the difference between each galaxy H$_2$ mass and the expected value given its mass and the best fit  for isolated galaxies with no process, shown in Fig.\ref{fig:ref_relations}. Panels, colors, lines and symbols are as in Fig. \ref{fig:delta_sfr}.}
    \label{fig:delta_h2}
\end{figure*}

Shifting the focus to the size--mass relation (Figure \ref{fig:delta_size}), no significant deviations from the best-fit relation are observed when galaxies are separated only by environment. However, when considering physical mechanisms, clear differences emerge.
In all environments, galaxies that experienced a tidal interaction show a distribution shifted toward larger sizes, resulting in systematically higher median values compared to the unaffected population. This trend is statistically supported by KS tests, which confirm that these distributions are significantly different from the reference sample ($p$-value$<0.05$). Given that tidal interactions between galaxies are not implemented in the model, this result might indicate that interactions between halos could ultimately influence the galaxies, as angular momentum is inherited from halos.
Galaxies that underwent mergers also appear to have slightly larger sizes on average, but this difference is not statistically significant according to the KS test. For the remaining mechanisms, no meaningful differences are detected.

Figure \ref{fig:delta_Hi} shows the normalized distributions of residuals from the HI--mass relation. An increasing fraction of galaxies in groups and filaments have lower gas content, with median values falling below the best-fit relation. As a consequence, for all physical mechanisms considered within these environments, the corresponding median HI masses are also systematically lower than the reference relation. Despite the large scatter, galaxies that underwent starvation and ram-pressure stripping in groups and filaments show a systematic deficit in HI mass compared to the other categories. This is confirmed by KS tests, which detect statistically significant differences both between these two populations and with respect to all other subsamples ($p$-value$<0.05$).
In contrast, for galaxies in isolated and pair environments, the median HI mass residuals for all mechanisms are consistent, within the scatter, with the best-fit relation. 

Finally, Figure \ref{fig:delta_h2} focuses on the distribution of H\textsubscript{2} mass residuals. Galaxies in isolation exhibit median values and overall distributions that are consistent with the best-fit H\textsubscript{2}--mass relation. In contrast, galaxies located in pairs, groups, and filaments show a systematic deficit in H\textsubscript{2} content, with median residuals falling below the reference relation. 
As seen for HI, this trend holds across all physical mechanisms within these denser environments, with the median H\textsubscript{2} masses of affected galaxies systematically lower than those of the unaffected population. The effect is particularly pronounced in groups and filaments, while only marginally detectable in pairs. 

To conclude, while a careful inspection of the various scaling relations reveals some trends, the overall differences between galaxies affected by different physical mechanisms remain relatively subtle {when considering global, integrated quantities alone. This subtlety does not necessarily imply that the underlying physical processes have a weak impact on galaxy evolution. Rather, it largely reflects the fact that many environmental mechanisms act in a spatially localized and time-dependent manner, so that their signatures are diluted when averaged over entire galaxies.
In addition, galaxies affected by the same physical mechanism are often observed at different evolutionary stages or may have experienced the process in a different environment than their present-day one, further increasing the scatter in global scaling relations and reducing the contrast between different populations. As a result, integrated properties may remain within the intrinsic dispersion of the relations, even in the presence of strong ongoing environmental processing. This interpretation is consistent with the findings of \cite{Vulcani2021}, who showed that, based on observational data, scaling relations and integrated quantities alone are insufficient to robustly identify galaxies affected by specific physical processes, while spatially resolved diagnostics reveal clear and diverse signatures of environmental transformation.}

\subsubsection{Quiescent fractions}\label{sec:quie_frac}

\begin{figure*}
    \centering
\includegraphics[trim=0 20 0 0,width=0.9\linewidth]{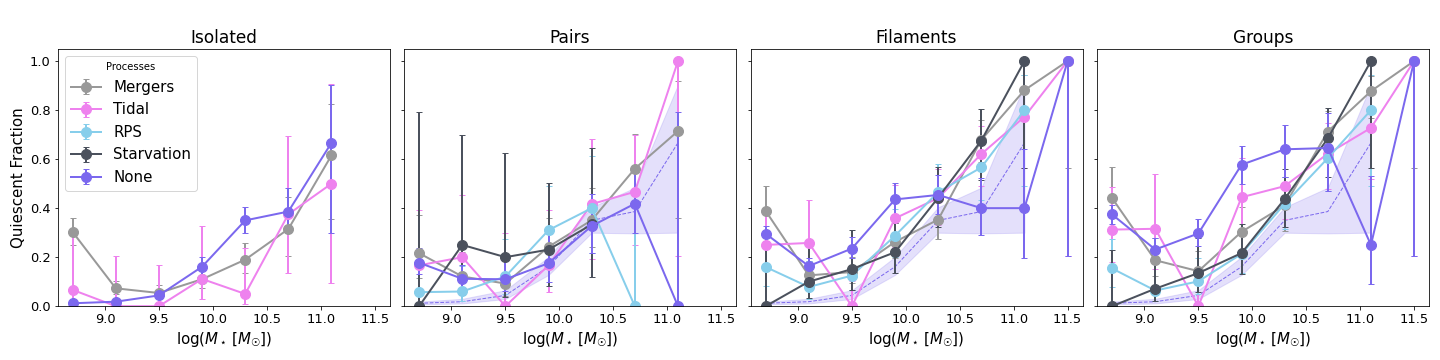}
    \caption{Quenched fraction as a function of stellar mass for galaxies in different categories and environments, as described in the legend. In the second, third, and fourth panels, the quenched fractions of the isolated galaxies not undergoing any process are also reported, as shaded regions, for reference. 
    }
    \label{fig:quiefrac}
\end{figure*}

\begin{figure*}
    \centering
    \includegraphics[trim=0 20 0 40,width=0.9\linewidth]{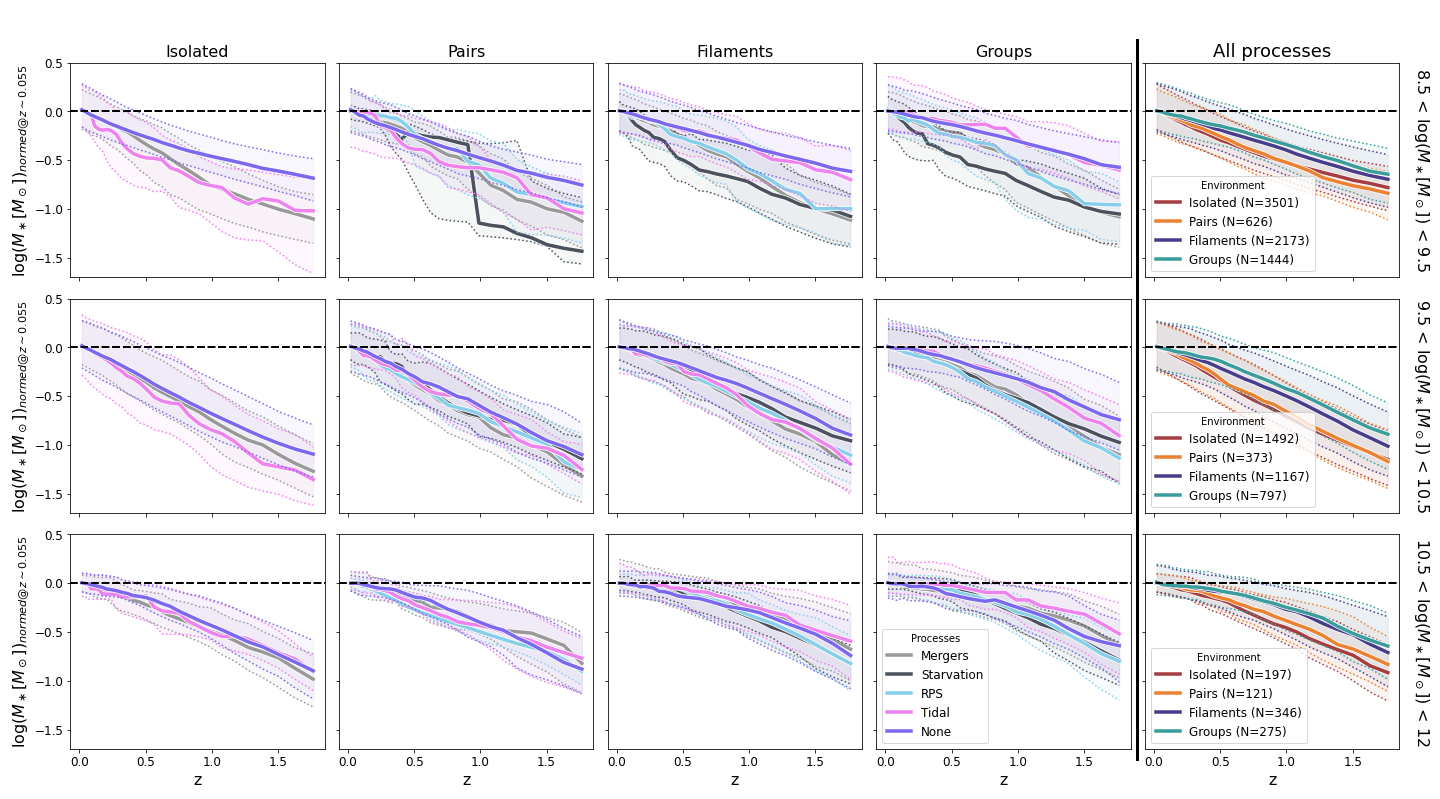}
     \caption{Stellar mass growth as a function of time for  galaxies in different environments (from the first to the fourth columns), undergoing different mechanisms (as indicated in the legend in the panel in the bottom row, fourth column) and in three different stellar mass bins. Lines are normalized to the z=0.05 galaxy mass. Solid colored lines represent the median values, considering all galaxies in all sub-boxes. Shaded regions show the 25th and 75th percentiles of the distributions. In the last column, a direct comparison of the mass growth for galaxies in different environments, independent on the process is also reported. Numbers in parenthesis are the number of galaxies in each environment, in each stellar mass bin. }
    \label{fig:mass_growth}  
\end{figure*}

An additional key aspect to explore is whether physical mechanisms influence the quenched fractions of galaxies. Numerous studies in the literature \citep[e.g.,][just to cite a few]{Peng2010, Wetzel2012, darvish16, Guglielmo2019, Donnari2021a, Donnari2021b} have examined quiescent fractions as a function of environment, consistently identifying clear trends. Here, we extend this analysis by incorporating information on the underlying physical mechanisms to inspect if they drive the  quenching.

The first panel of Figure \ref{fig:ref_relations} displays the commonly adopted threshold to separate star-forming from quiescent galaxies, defined as $\log(\mathrm{SSFR[\mathrm{yr}^{-1}])} = -11$. Using this criterion, Figure \ref{fig:quiefrac} presents the quiescent fractions as a function of stellar mass, separated both by {present-day} environment and by physical mechanism.
As well established in the literature \citep[e.g.,][]{Lin2014, darvish16, Gu2021}, the fraction of quiescent galaxies increases with stellar mass and shows a strong dependence on environment: at fixed mass, galaxies in isolation have a lower quiescent fraction than those in pairs or groups.

When examining one environment at a time, some subtle yet systematic differences emerge when comparing galaxies affected by different physical mechanisms. For example, among isolated galaxies, those affected by tidal interactions exhibit lower quiescent fractions across the entire stellar mass range, suggesting a systematically reduced efficiency of quenching for this category.
In filaments and groups—with a more pronounced effect in filaments—starved galaxies display the strongest dependence on mass: the quiescent fractions are consistent with the other mechanisms at low masses but dominate the overall quenched population at high masses. Interestingly, in all environments, a non-negligible fraction ($\sim$40\%) of galaxies not undergoing any identifiable physical mechanism are quenched, suggesting they have been either quenched by internal processes or by the presence of alternative or unresolved quenching channels. Alternatively, they were already quenched by $z \sim 2$.

When fixing the mechanism and varying the environment, clearer trends emerge. For most processes, the quiescent fraction is lowest in isolation and progressively increases in pairs, filaments, and groups at fixed stellar mass. These trends are most evident for mergers, where the environmental dependence is strongest, and least clear for RPS, where the curves for pairs, filaments, and groups largely overlap. This could indicate a more  less environment-sensitive or complex behavior for RPS overall: in addition to quenching, galaxies in filaments could also be influenced by cosmic web enhancement \citep{Vulcani2019_fil}, a mechanism in which galaxies moving through or along filaments experience facilitated gas cooling, leading to extended star formation in the densest regions of their CGM. It is interesting to note that at low stellar masses, quenched fractions in ram-pressure stripped galaxies are always lower than for galaxies in the same environment but not labeled with any mechanism. They are, however, above the corresponding line of the Isolated-None galaxies, still showing an enhancement in quenched fractions. In pairs, there are no galaxies with $\log(M_\ast/M_\odot)>10.5$ undergoing RPS that are quenched. 
As far as mergers and tidal interactions are concerned, an excess of quenched galaxies is observed in the lowest mass bin. The fraction then decreases before rising again with increasing stellar mass. 

\begin{figure*}
    \centering
    \includegraphics[trim=0 20 0 40,width=0.9\linewidth]{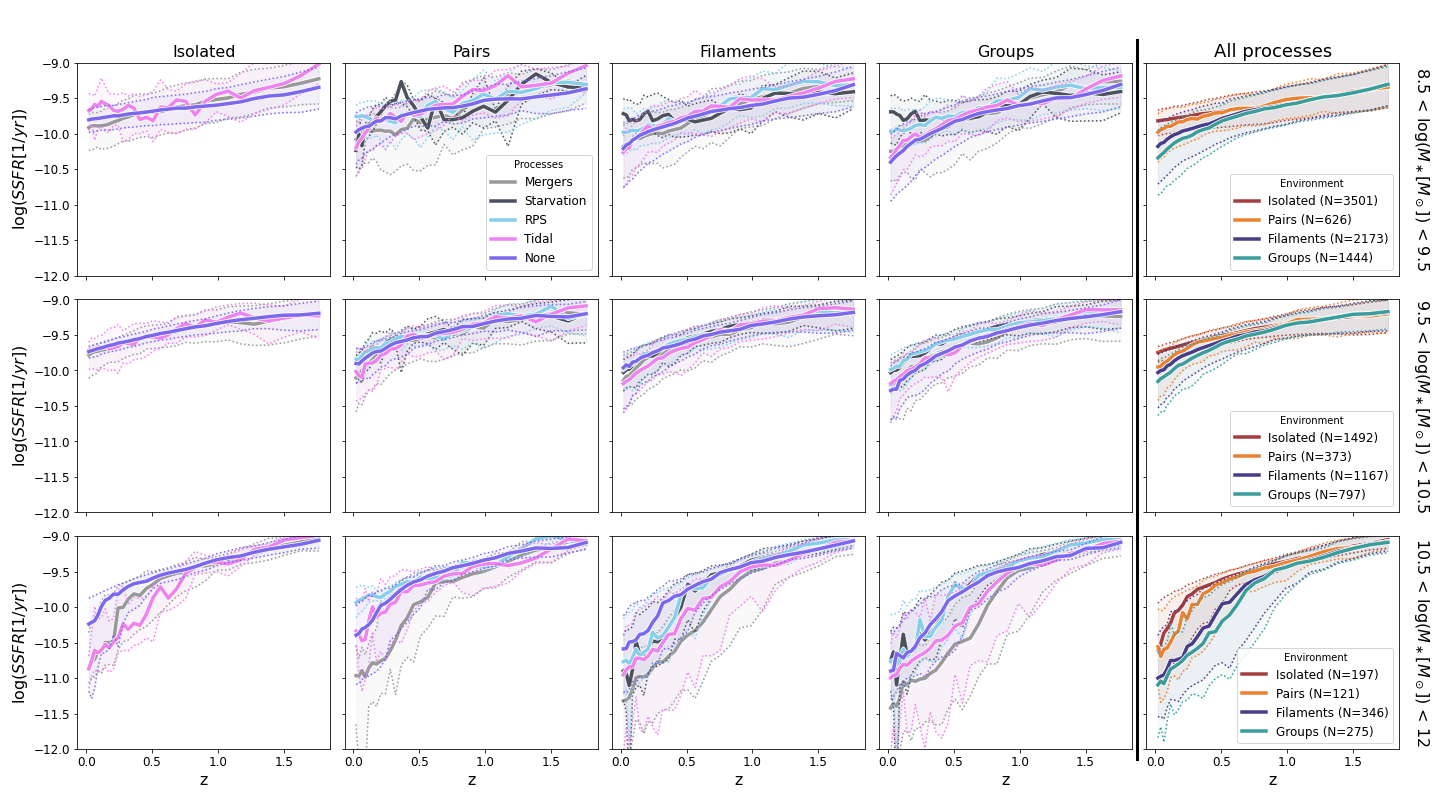}
      \caption{Evolution of SSFR a function of time for  galaxies in different environments, undergoing different mechanisms and in three different stellar mass bins. Columns, panels and symbols are as in Fig. \ref{fig:mass_growth}. }
    \label{fig:sfr_growth}  
\end{figure*}

\subsection{Mass assembly and star formation rate variation}\label{sec:hist}
In the previous sections, we have explored the role of {present-day} environment and physical mechanisms in shaping the present-day properties of galaxies. We now shift our focus to understanding how these galaxies evolved by tracing the growth of stellar mass, HI and H$_2$ content, as well as the evolution of their SFR from $z=2$ to $z=0$. Also in this case, the corresponding figures based on the 2D definition are provided in Appendix~\ref{app:2d}. 

For this analysis, we follow the main progenitor branch of each galaxy. In addition to differentiating galaxies by environment and physical mechanism, we further separate them into three stellar mass bins to account for the known mass dependence of galaxy evolution.

Figure \ref{fig:mass_growth} shows the stellar mass assembly. To characterize the populations as a whole, and avoid being driven by single objects, we consider only the median values and the 25th and 75th percentiles of the distributions. To better investigate the growth and minimize the effect of differences in the mass distributions, we normalize the curves to the galaxy stellar mass at $z \sim 0.05$.  
At all masses, the mass growth depends on both environment and physical process: focusing on one mechanism at a time, the median mass growth becomes progressively flatter as we move from isolated galaxies to pairs, filaments, and groups. This trend is also visible in the last column of Fig. \ref{fig:mass_growth}, where all processes are considered together (the number of galaxies in the different environments is reported in parentheses).  
Focusing on isolated galaxies, at low masses, galaxies that have undergone mergers or tidal interactions exhibit steeper mass growth than unaffected galaxies, while differences disappear at higher masses. In pairs, no significant differences are observed across all mass bins.  
In groups and filaments, at low masses, galaxies affected by starvation, mergers, or RPS show stronger mass growth in the last 2 Gyr than undisturbed galaxies in the same environment, suggesting that these environments favor mass growth. In contrast, galaxies labeled as undergoing tidal interactions are indistinguishable from the undisturbed population. We note, however, that the statistics for starved galaxies in this mass bin are very low ($\leq 5$ galaxies), making these results less reliable.  
At intermediate and high masses, no clear differences are seen, except for group galaxies at intermediate masses, where ram-pressure stripped, merged, and starved galaxies show steeper growth compared to galaxies undergoing tidal interactions or remaining undisturbed.

\begin{figure*}
    \centering
    \includegraphics[trim=0 20 0 40,width=0.9\linewidth]{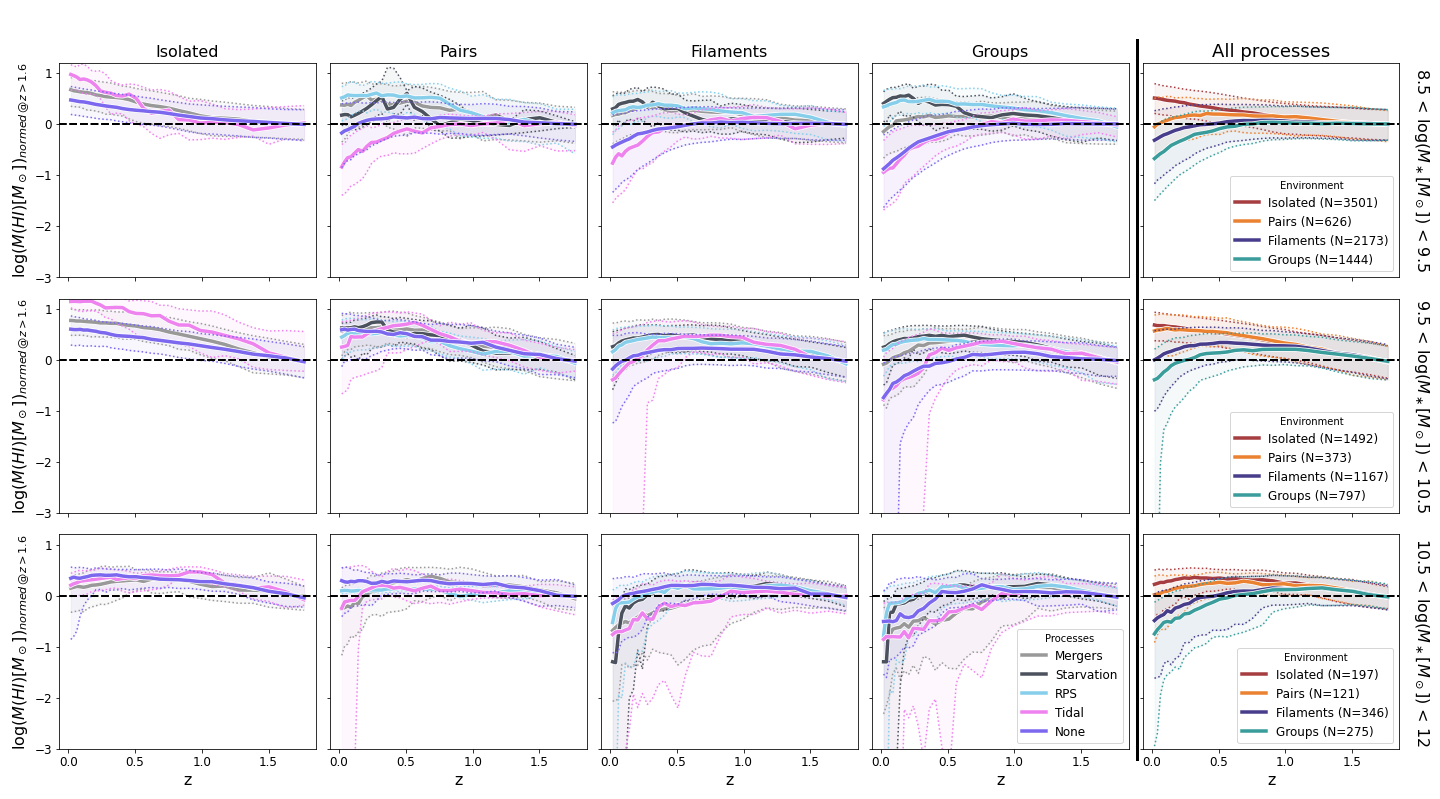}
    \caption{Evolution of normalized HI mass content as a function of time for  galaxies in different environments, undergoing different mechanisms and in three different stellar mass bins. For each mass bin and environment, we compute the mean median HI of the None (unperturbed) population at $z>1.6$ (a different z range is used for Pairs-RPS, and Pair-Tidal, as explained in the text) and shift all curves so that this baseline is zero. Each process curve is additionally aligned to this baseline using its own mean in the same redshift interval. In the left columns, we use the isolated galaxies as baseline. The dashed line at y=0 shows the case of no evolution. Columns, panels and symbols are as in Fig. \ref{fig:mass_growth}.}    \label{fig:HI_growth}  
\end{figure*}

\begin{figure*}
    \centering
     \includegraphics[trim=0 20 0 40,width=0.9\linewidth]{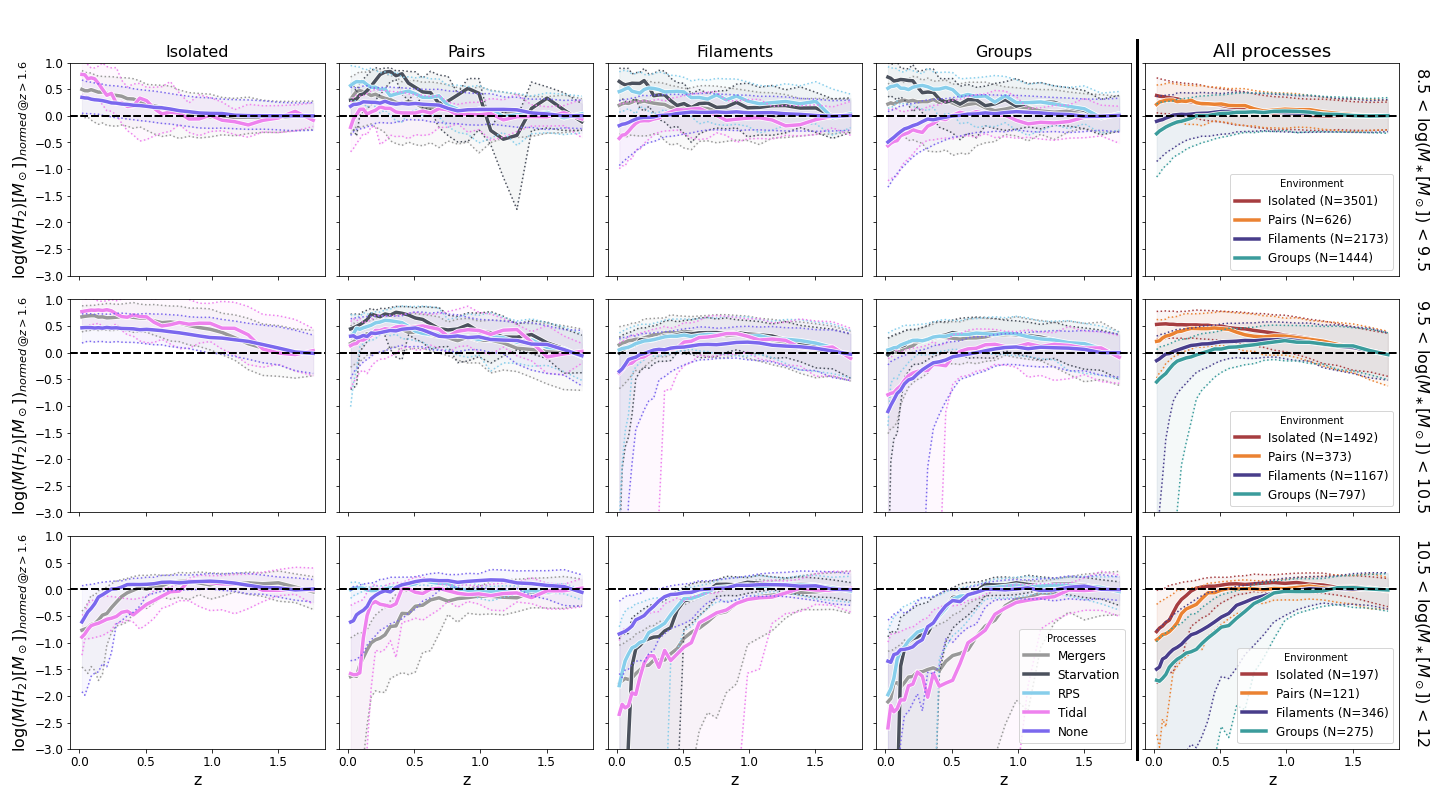}
    \caption{Evolution of normalized H$_2$ mass content as a function of time for  galaxies in different environments, undergoing different mechanisms and in three different stellar mass bins.  Columns, panels and symbols are as in Fig. \ref{fig:HI_growth}.}
    \label{fig:H2_growth}
\end{figure*}

Figure \ref{fig:sfr_growth} shows the evolution of the SFR over time. To account for the SFR–mass relation, we consider the specific star formation rate (SSFR) instead of the absolute SFR. Comparing galaxies across different environments, in all mass bins the SSFR declines from high to low $z$.  
In agreement with previous studies \citep[e.g.,][]{Whitaker2012, Tomczak2016, Davidzon2018}, the SSFR decline is environment dependent, with isolated systems exhibiting a more gradual decrease compared to galaxies in filaments and groups. The decline is also steeper for more massive galaxies in all environments, consistent with the downsizing scenario \citep[e.g.,][]{cowie96, fontanot09, Pacifici2016}.  
Considering the different physical processes, significant differences emerge only at the highest stellar masses, where galaxies affected by any mechanism other than RPS exhibit a steeper decline in SSFR compared to those unaffected since $z\sim0.6$. In particular, galaxies experiencing mergers show a pronounced drop in star formation across all environments. A comparable decrease is also evident in groups and filaments for galaxies undergoing starvation or tidal interactions.  For ram-pressure stripped galaxies, the decline closely follows that of galaxies in the same environment that are unaffected, with the exception of pairs, where the decline is slightly less steep.  
Due to small number statistics, results for galaxies undergoing starvation are noisy and should be interpreted with caution.

We now focus on the evolution of HI and H$_2$ gas content. Although we have so far binned galaxies into three stellar mass ranges, this may not fully capture the variations in HI and H$_2$ content. Well-established relations exist between these gas components and stellar mass, but they exhibit significant scatter, as also shown in Fig.~\ref{fig:ref_relations}. To minimize systematic offsets among different evolutionary tracks, we normalize all curves relative to a common baseline within each environment and mass bin.  
First, we select the unaffected galaxies and compute their median HI (or H$_2$) content at $z>1.6$. We then apply a rigid vertical shift to all curves—including both the baseline and those corresponding to other physical mechanisms—so that this reference value is set to zero. When all processes are considered together, isolated galaxies are used as the baseline. For each perturbed category, we then compute its median value over the same redshift interval ($z>1.6$) and shift the entire curve to align with the baseline. By construction, all curves have zero average in the high-$z$ regime. 

The leftmost panels in Figure \ref{fig:HI_growth} show that the HI content of isolated galaxies remains roughly constant over time at all masses, if not  increasing. For galaxies with $\log(M_\ast/M_\odot)<10.5$, environmental differences emerge at $z<1$, with pairs, filament, and group galaxies showing an overall decline, except for pairs at $9.5 <\log(M_\ast/M_\odot)<10.5$ which show the same trend as for isolated systems. Similar trends are observed at higher masses, although intermediate-mass pair galaxies are largely indistinguishable from isolated ones. At $z=0$, the relative difference in HI mass between isolated and group galaxies is approximately 1 dex.  
The other panels in Fig. \ref{fig:HI_growth} show that in all environments except for isolated galaxies, and where statistics are sufficient, galaxies not undergoing any mechanism still exhibit a decline in HI content from $z=2$ to $z=0$, indicating natural gas consumption over time. Overall, galaxies undergoing RPS show the smallest decline, contrary to the expectation that HI should be the first component removed. This result may be driven by the normalization procedure, which sets all galaxies to the same gas content at $z=2$, potentially masking early gas loss in galaxies already affected by ram pressure.  

Figure \ref{fig:H2_growth} presents the redshift evolution of H$_2$ content. Unlike HI, isolated galaxies show a strong stellar mass dependence: at low masses, H$_2$ content is consistent with almost being flat with time, while moving to higher masses it systematically decreases as time goes by, regardless of the physical mechanism. This decline highlights the enhanced susceptibility of massive galaxies to molecular gas loss or consumption over cosmic time.  
The decrease is also environment dependent: the H$_2$ content is the lowest in groups and filaments. Mergers and tidal interactions are the processes that seem to have the largest effect in decreasing the H$_2$ content, especially in groups and filaments, but also in pairs and is isolation. Also galaxies affected by RPS show a decrease in $H_2$ content at the highest masses, while no significant trends are visible in the lowest mass bins.

{We note that these evolutionary trends are derived from the GAEA model and are therefore subject to its intrinsic assumptions and limitations. In particular, the halo mass dependence of some processes, the simplified treatment of RPS, and the absence of explicit cosmic web or tidal interactions may affect the detailed timing and amplitude of mass, gas, and SFR evolution.}

\section{Discussion}\label{sec:discussion}
This section provides a discussion of the results, as well as the assumptions and caveats underlying our analysis, offering further insights and explanations for the findings presented above. Specifically, Section \ref{sec:future} briefly summarizes the main results and their implications, with a focus on the consequences for the MAGNET project. Section \ref{sec:mergers} investigates the nature and role of mergers in more detail, Section \ref{sec:type} examines the properties and impact of central and satellite galaxies, and Section \ref{sec:threshold} addresses the influence of the $z=2$ threshold adopted for studying physical processes. Finally, Section \ref{sec:caveat} discusses the main caveats and limitations of the adopted model.

\subsection{Key results and implications in the context of the MAGNET survey}\label{sec:future}

Our analysis, designed to reproduce the MAGNET survey volume and both observational-like and intrinsic (halo-based) environmental classifications, offers insights into the frequency and impact of different physical processes acting on galaxies outside of clusters since $z=2$.

Comparing GAEA observational-like and MAGNET classifications (Fig. \ref{fig:cfr_obs}) reveals systematic differences: mimicking 2D observational criteria in the model increases the fraction of pairs while decreasing isolated and group fractions. GAEA also underproduces high-dispersion ($\sigma > 200$ km/s) groups and rich systems ($N_\mathrm{members} > 8$), which are more common in MAGNET. These discrepancies likely reflect a combination of survey incompleteness and methodological differences in group identification.

The prevalence of each process is primarily mass-dependent, with environmental effects becoming significant in groups—where the fraction of undisturbed systems drops sharply, confirming them as key sites of environmental processing (Fig. \ref{fig:distr-env-proc}). Mergers dominate at high stellar masses, consistent with hierarchical growth \citep[e.g.,][]{White1978, Fakhouri2010}, while low-mass galaxies are most susceptible to RPS, as their shallow potential wells cannot retain loosely bound gas in intragroup or filamentary environments \citep{gunn1972, boselli2006}.

Starvation shows a clear mass dependence and acts gradually but effectively, especially in groups where hot halo gas suppresses accretion and quenches star formation over long timescales \citep{larson1980, vandenbosch08a}. Its measured fraction depends strongly on the adopted threshold (see Sect.~\ref{sec:process}); lowering this value increases the count but includes galaxies marginally affected. The overall rarity of tidal interactions and starvation underscores the need for large, statistically complete samples such as MAGNET to capture the full diversity of environmental effects.

Galaxy colors (Fig. \ref{fig:coloromass_halo}) reveal a clear environmental trend: systems transition toward higher stellar mass and redder colors from isolated, undisturbed galaxies to group members affected by starvation or tidal forces. In contrast, most galaxies undergoing RPS remain blue, indicating that in non-cluster environments, ram-pressure stripping by itself is generally insufficient to produce lasting quenching, or that subsequent gas reaccretion can replenish the fuel for star formation.

Gas content (Figs. \ref{fig:HI_growth}, \ref{fig:H2_growth}) further clarifies these behaviors. Both components show a clear decline with cosmic epoch which is environment and mass dependent.  At the highest masses, mergers and tidal interactions 
drive the steepest molecular gas declines over time,  
likely through dynamical heating and morphological transformation that inhibit gas retention and compression \citep{Moore1996, Park2009}. This naturally explains why tidally disturbed galaxies are larger (Fig. \ref{fig:delta_size}) and less efficient at star formation (Fig. \ref{fig:delta_sfr}) \citep{Mihos1996, Ellison2018}.

The prevalence of quenching (Fig. \ref{fig:quiefrac}) increases steadily from isolated galaxies to pairs, filaments, and groups, supporting a pre-processing scenario in intermediate-density environments \citep{Zabludoff1998, Fujita2004, Bianconi2018}. MAGNET’s emphasis on groups and filaments will thus be crucial for probing galaxy transformation prior to cluster infall.

{Several of the trends discussed above are particularly robust, as they rely on key strengths of the GAEA model. 
In particular, the ability to simultaneously track stellar mass assembly, star formation, and the partitioning of cold gas into HI and H$_2$ allows us to consistently interpret color, SFR, and gas-content trends across environments and physical mechanisms. 
Results based on relative differences between populations within the same environment—such as the systematic shifts in color, SSFR, and molecular gas content among galaxies affected by different processes—are therefore less sensitive to the absolute calibration of the model and remain stable under different environmental definitions.

At the same time, some aspects of our results are more directly influenced by model assumptions and should be interpreted with caution. 
The explicit dependence of several mechanisms on halo mass, together with the absence of a fully resolved treatment of cosmic web effects and tidal interactions, may affect the detailed incidence and timing of specific processes, particularly in filaments and low-mass groups. 
Similarly, the simplified implementation of RPS may limit our ability to capture mild or early-stage gas removal, potentially contributing to the relatively weak impact of RPS inferred from integrated galaxy properties.

Importantly, these limitations do not undermine the qualitative trends identified in this work, but rather constrain the level at which physical interpretation can be pushed. For this reason, throughout the paper, we have  emphasized statistical associations rather than direct causal links between environment, physical mechanisms, and galaxy evolution, and we consistently compared populations within homogeneous environmental classes to minimize biases introduced by model assumptions.
}

Overall, these results provide a coherent framework for interpreting upcoming MAGNET data. Simulations track physical mechanisms directly, while observations must infer them through morphology, kinematics, and star formation history. Bridging this gap—particularly through multiphase ISM observations—will be key to connecting processes across mass and environment. Despite residual differences in environmental definitions and group properties, the main evolutionary trends are robust and highlight the need for survey strategies that minimize projection effects and classification ambiguities.

\begin{figure*}
    \centering
    \includegraphics[trim=0 20 0 0,width=0.4\linewidth]{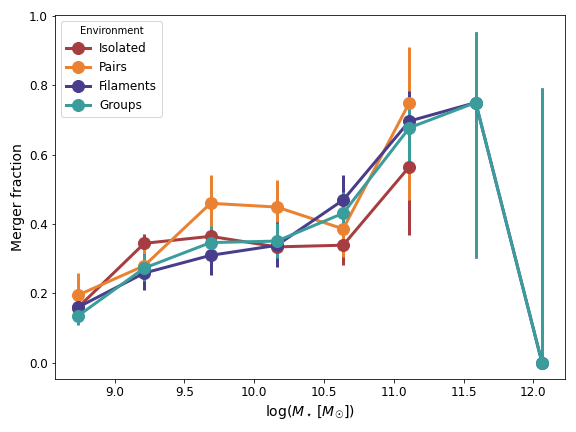}
    \includegraphics[trim=0 20 0 0,width=0.4\linewidth]{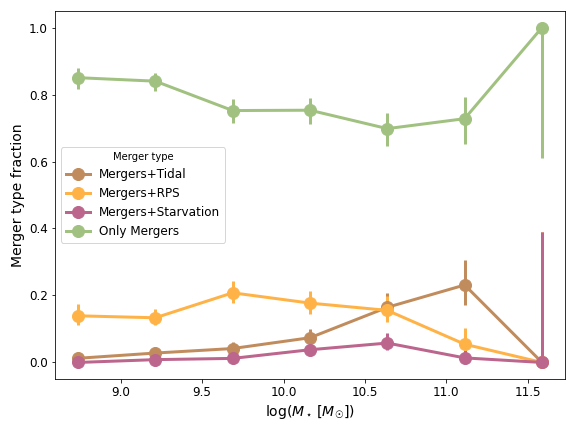}
    \caption{Merger fraction as a function of stellar mass. Left: different environments are considered. Right: different merger categories are considered, as described in the label.}
    \label{fig:merger_frac}
\end{figure*}

\begin{figure*}
    \centering
    \includegraphics[trim=0 20 0 0,width=0.9\linewidth]{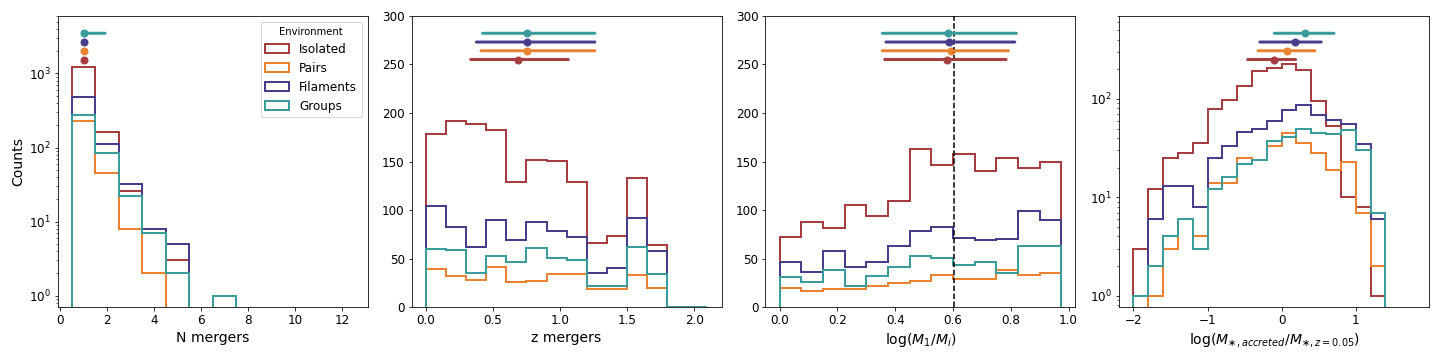}
    \caption{Distributions characterizing the nature of mergers,  in the different environments. From left to right: number of merger events, redshift at which mergers happened, mass ratio between the galaxy involved, mass accreted by mergers in galaxies. In each panel, circles and horizontal lines show median values along with the 25th and 75th percentiles of the distributions. In the third panel, vertical dashed lines indicate the threshold commonly adopted to separate minor and major mergers (1:4). }
    \label{fig:merger_distr}
\end{figure*}

\subsection{The nature of the mergers}\label{sec:mergers}

In the previous sections, we found that mergers constitute a significant fraction of the evolutionary processes under consideration. Galaxies that have experienced at least one merger tend to reside in more massive halos\footnote{We remind the reader that this is the halo mass in which the galaxy reside today, not the halo mass of the environment in which the merger happened.} and span a wide range of stellar masses. In group environments, approximately half of the galaxies identified as mergers are red (Fig.\ref{fig:coloromass_halo}). Interestingly, a significant fraction of isolated galaxies experience mergers, indicating that one cannot always assume these galaxies have evolved solely under the influence of internal processes. 
However, mergers appear to have a limited impact on the quenched fraction and do not significantly affect global scaling relations. It is important to emphasize that the merger category is highly heterogeneous. It encompasses galaxies that have undergone varying numbers of merger events, with different mass ratios, gas fractions, and corresponding amounts of accreted mass. Consequently, the role mergers have played in shaping galaxy evolution can differ substantially across individual systems. 
In this section, we explore this category in greater detail, analyzing how various aspects of mergers—such as frequency, timing, mass ratios, and total accreted mass—vary with stellar mass and environment.

We begin by examining the incidence of mergers as a function of stellar mass across different environments. The left panel of Fig.~\ref{fig:merger_frac} shows the fraction of galaxies that have undergone at least one merger, relative to the total number of galaxies, as a function of stellar mass and environment. Tthe merger fraction increases with stellar mass, ranging from $\sim 0.2$ at low masses to $\sim 0.8$ at the high-mass end. No clear systematic trends with environment are observed. 
The right panel of the same figure shows the relative contribution of different merger sub-categories among galaxies classified as mergers. The Only merger category—representing galaxies that experienced mergers but were not affected by any other environmental mechanisms—dominates across the full stellar mass range, with fractions between 0.8 and 1. Moreover, 96\% of these galaxies are type 0 galaxies. The fraction of galaxies classified as merger + tidal interactions increases with stellar mass, while the merger + RPS component decreases. The merger + starvation category remains nearly flat and consistently contributes less than 10\%. 

To better understand how mergers contribute to galaxy assembly, we now turn to the detailed properties of merger events themselves—specifically their frequency, timing, and physical characteristics—across different environments. Figure \ref{fig:merger_distr} shows the distribution of merger characteristics across environments. 
First, we examine the number of merger events per galaxy. While the majority of galaxies experienced only one merger since $z=2$, a non-negligible tail extends to higher merger counts. In particular, some galaxies experienced up to five mergers, and one group galaxy underwent as many as seven. In general, galaxies with a larger number of mergers are group central galaxies. 
The number of mergers clearly depends on environment: the distribution is narrowest for galaxies in pairs and broadest for those in groups. A KS test confirms, with high statistical significance ($p$-value $<0.05$), that the distributions for isolated and filaments, isolated and groups, and pairs and groups are drawn from distinct parent populations, while no statistically significant differences are retrieved in the other comparisons.

Next, we study the redshift distribution of merger events across different environments. Overall, the distributions appear relatively flat, with the exception of isolated systems, which exhibit a noticeable skew toward lower redshifts. While median redshift values do not differ significantly across environments, a KS test supports this trend: the redshift distribution for isolated galaxies is statistically distinct from that of galaxies in other environments. No significant differences are found among the remaining environments. This result suggests that mergers in isolated galaxies tend to occur more recently, whereas no strong redshift dependence is observed for mergers in pairs or groups.

We now turn to the characterization of mass ratios in merger events. These are defined as the stellar mass of the target galaxy just before the merger divided by that of its merging companion, with the environmental classification based on the target galaxy. As previously mentioned, our analysis includes only mergers with stellar mass ratios up to 1:10, and we adopt a mass completeness limit of 
$\log (M_\ast/M_\odot) > 8.5$ throughout. Although the GAEA model becomes statistically unreliable below this threshold due to resolution limitations, we still retain mergers involving lower-mass companions, while acknowledging the resulting incompleteness. For example, a galaxy with 
$\log (M_\ast/M_\odot) = 8.5$ may merge with a companion of 
$\log (M_\ast/M_\odot) = 7.5$, and such events are included in our sample.
In all environments, the target galaxy is typically more massive than its merging companion, with very few events approaching a 1:1 mass ratio. Figure~\ref{fig:merger_distr} shows the distribution of mass ratios, with vertical dashed lines indicating the 1:4 threshold commonly used to distinguish between major and minor mergers. Approximately half of the mergers fall below the 1:4 threshold, while the rest lie between 1:4 and 1:10. A KS test confirms that the mass ratio distributions do not differ significantly across environments.

Finally, we characterize the total amount of stellar mass accreted through mergers. For each merging system, we compute the relative cumulative accreted mass by summing the stellar masses of all galaxies that merged with any of its progenitors since $z=2$ (see Sec.~\ref{sec:process} for details), divided by the stellar mass the galaxy has today. We note that we do not consider any mass loss. The rightmost panel of Fig.~\ref{fig:merger_distr} shows the resulting distributions, which span 
about three orders of magnitude, from 0.01 to 10. 
These distributions show a clear dependence on environment: isolated galaxies tend to accrete the least stellar mass, followed in order by galaxies in pairs, filaments, and groups. A KS test confirms that the distributions are statistically distinct across all environments.

To summarize, the incidence of mergers increases steadily with stellar mass and shows only mild environmental dependence. The strongest environmental effect is that in pairs and among isolated galaxies there are no mergers with $\log (M_\ast/M_\odot) > 11$, while in groups and filaments mergers can reach $\log (M_\ast/M_\odot) \sim 11.5$. The majority of merger-affected galaxies have not experienced other environmental mechanisms. However, the merger category is highly diverse, encompassing a range of mass ratios, frequencies, and accreted stellar mass. While most galaxies experience only one merger, group environments host more frequent and earlier mergers, along with higher cumulative accreted mass.

\begin{figure*}
    \centering
    \includegraphics[trim = 0 30 0 0 , width=0.9\linewidth]{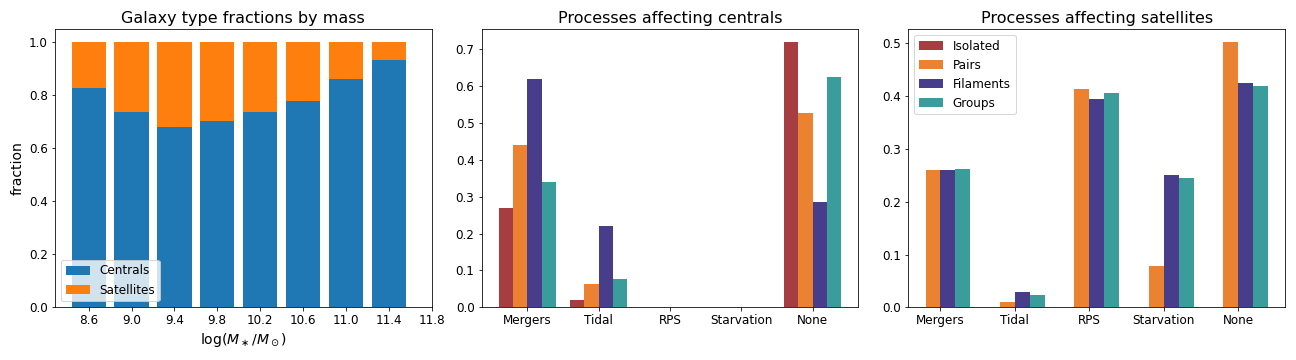}
     \caption{Fraction of galaxies of different types, for the halo definition. 
     Left: fraction of galaxies of each type (centrals in blue, satellites in orange) as a function of stellar mass. Center: fraction of central galaxies as a function of the physical process. Right: fraction of satellite galaxies as a function of the physical process.}
    \label{fig:cen_sat}
\end{figure*}

\subsection{Central and satellite galaxies}\label{sec:type}

Model galaxies in the GAEA simulation are classified into three categories: centrals, satellites, and orphans. Centrals and satellites are associated with resolved dark matter subhaloes—centrals reside in the main halo (i.e., the bound component) of a friend-of-friend  group, while satellites are linked to subhaloes within larger halos. Orphan galaxies, instead, are those whose original subhalo has been stripped below the resolution limit of the simulation.
In this work, we focus on central and satellite galaxies, which are analyzed together throughout our study. Orphan galaxies are excluded, except for the calculation of group velocity dispersions used for comparison with observations (see Sect.~\ref{sec:comp_magnet}). As mentioned in Sect.~\ref{sec:env}, each sub-box contains about 150 orphan galaxies (25\% of the total), with 60\% found in groups, $<30\%$ among isolated systems, and $\sim10\%$ in pairs. We tested the impact of including orphan galaxies in the computation of the fractions of galaxies undergoing different mechanisms shown in Fig.\ref{fig:distr-env-proc}  (Appendix~\ref{app:orphans}) and found that their inclusion produces only moderate variations—mainly in group environments and at low stellar masses—while the overall trends and high-mass results remain robust.

Here we explicitly separate central and satellite galaxies, and investigate whether the trends identified in the previous sections vary between the two galaxy types.
As expected by definition, 100\% of isolated galaxies are central, while in pairs centrals represent 50\% and in groups the 33\%. In filaments, satellites dominate, accounting for approximately 80\% of galaxies. These fractions exhibit a mild dependence on stellar mass, as illustrated in the left panel of Fig.~\ref{fig:cen_sat}. Central galaxies dominate at all stellar masses, although their relative abundance decreases with increasing mass: from about 80\% at low masses, to $\sim$60\% at intermediate masses, and then rising again to nearly 100\% for $\log(M_\ast/M_\odot) > 11.3$.
The central and left panels of Fig.~\ref{fig:cen_sat} show the relative contribution of different physical mechanisms across environments, separately for central and satellite galaxies. Among central galaxies, more than 70\% of isolated galaxies have not experienced any physical interaction, while the remaining have undergone mergers. In pairs, over 40\% of central galaxies experienced at least one merger event, with fewer than 10\% showing signs of tidal interactions; the remainder appear unaffected. In filaments, the majority of centrals ($>60\%$) underwent mergers, followed by a smaller fraction unaffected ($\sim30\%$) and a minor contribution from tidal interaction instances.  While filaments are generally found to represent an intermediate environment between groups and pairs/isolated systems, an interesting exception emerges here: central galaxies in filaments experience a significantly higher merger fraction compared to those outside filaments. In groups, by contrast, central galaxies are mostly unaffected.
It is important to note that, by definition, central galaxies cannot be influenced by RPS or starvation.
Focusing on satellite galaxies, we find that RPS is by far the most prevalent physical mechanism across all environments where satellites are present. It affects up to 40\% of satellites in pairs, groups, and filaments. Approximately 25\% of satellite galaxies have undergone mergers, $<30\%$ are affected by starvation, while less than 5\% show evidence of tidal interactions. About 40\% remain unaffected by any identified process. Overall, there is no significant variation in the distribution of physical mechanisms across different environments: being a satellite appears to have a similar effect regardless of host environment.

These findings highlight the fundamental role of galaxy type—central versus satellite—in shaping the evolutionary pathways of galaxies. While centrals are predominantly unaffected by environmental mechanisms beyond mergers, satellites are uniformly and strongly influenced by ram-pressure stripping, independent of their specific environment. This dichotomy partly reflects the construction of SAMs such as GAEA, where centrals and satellites are treated distinctly. However, comparison studies with state-of-the-art hydrodynamical simulations reveal qualitatively similar evolutionary trends \citep[e.g.,][]{Gabor2015, Engler2021, Marasco2016}, suggesting that these results are not merely artifacts of the semi-analytic approach but capture a more general physical picture.

\begin{figure*}
    \centering
    \includegraphics[trim=0 20 0 0,width=0.85\linewidth]{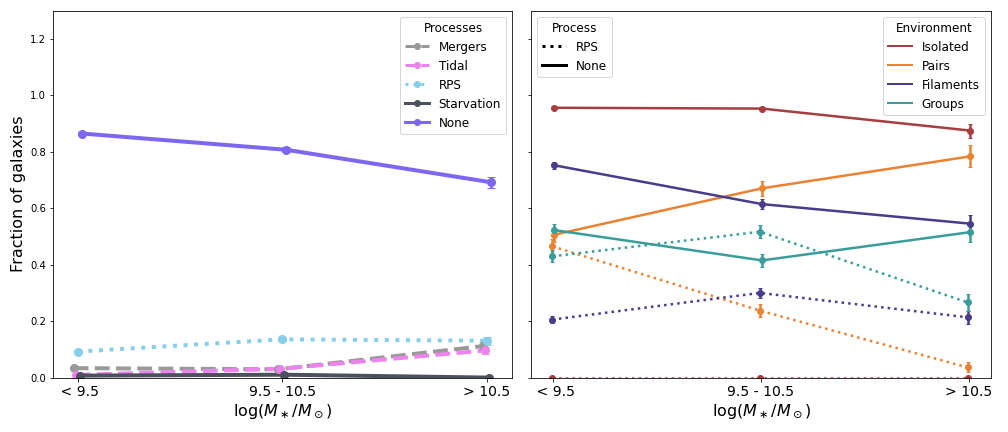}
    \caption{Fraction of galaxies as a function of stellar mass, separated by physical processes and environments, when the extraction of the physical  is limited to 1.5 Gyr prior to observations. The left panel shows the incidence of the different processes, regardless of environment.  The right panel shows the incidence of RPS and None in the different environments. These are the two processes showing the stronges environmental dependence (see text for details). Binomial error bars indicate the statistical uncertainties on each fraction. A small horizontal offset is applied to the points for visualization purposes.}
    \label{fig:frac_envs_1.5}
\end{figure*}

\subsection{Focusing on the recent past: tracing physical processes in the last 1.5 Gyr}\label{sec:threshold}

In this analysis, we trace the environmental histories of galaxies back to a redshift of 
$z=2$, under the assumption that evolutionary events occurring at earlier epochs are less likely to leave clear, observable imprints on the present-day properties of galaxies. This choice is motivated by the goal of capturing the most recent and observationally relevant processes, particularly those linked to quenching and morphological transformation in the local Universe. By limiting the analysis to $z<2$, we focus on transformation channels that are more likely to leave discernible signatures in low-redshift galaxies, in line with the scientific objectives of the MAGNET survey. While some galaxies may have undergone significant events—such as major mergers or early quenching—at higher redshifts ($z>2$), the impact of such processes may have been diluted or overwritten by subsequent evolution, making their present-day effects difficult to isolate. Extending the analysis to earlier epochs could potentially increase the completeness of the transformation census, but at the cost of interpretability and direct observational relevance.

In the following, we further narrow our temporal focus  on galaxies that experienced key environmental processes in the last 1.5 Gyr, in order to isolate the most recent evolutionary imprints and better assess the timescales and efficiency of quenching mechanisms in the current epoch. We therefore re-run all analyses stopping at $z=0.15$, which corresponds to approximately 1.5 Gyr prior to the epoch of observation.

Figure~\ref{fig:frac_envs_1.5} shows the fraction of galaxies as a function of stellar mass, separated by physical processes (left) and environment (right). To avoid overcrowding the figure, we show the environmental dependence only for the None and RPS categories—cases where different environments exhibit notable deviations from the global trend. 
Within this short timescale, and as expected, the fraction of galaxies not affected by any physical process is the highest, reaching nearly 90\% in the lowest mass bin. This fraction decreases steadily with increasing stellar mass but remains above 60\% in the highest bin. A similar decline is observed among isolated and filament galaxies. The trend is instead rather flat in groups, where the fraction of None is always about $\sim$50\%. the trend is even reversed in pairs,  where the None fraction increases with stellar mass.
The most prominent mechanism is RPS, whose total fraction shows a mild but steady increase with stellar mass, reaching $\sim$15\%. This is largely due to the fact that no ram-pressure stripped galaxies are found in isolated environments. The right panel of Fig.\ref{fig:frac_envs_1.5} shows that, among the remaining environments, RPS fractions vary: in pairs, they decrease from $\sim$50\% in the lowest mass bin to below 10\% in the highest; in groups and filaments, they follow an inverted-V trend—rising from the lowest to the intermediate mass bin and then declining at higher masses. The fraction of ram-pressure stripped galaxies reaches a peak of $\sim$60\% in groups and $\sim$30\% in filaments.
Moving to the other mechanisms, the fraction of mergers remains low but increases with stellar mass, reaching up to $\sim$10\% in the highest mass bin. The fraction of starved galaxies is nearly zero at all masses. The fraction of tidally affected galaxies closely follows that of mergers, with a mild but steady increase as stellar mass grows. 

Compared to the cumulative analysis up to $z=2$ (Fig.~\ref{fig:distr-env-proc}), this short-timescale investigation reveals that environmental processing is not only mass-dependent but also time-dependent. Most mechanisms—such as mergers, tidal interactions, and starvation—leave imprints only over long timescales, showing minimal immediate impact. In contrast, RPS remains effective even over short timescales, particularly for low-mass galaxies in dense environments. Despite these temporal differences, the physical properties of galaxies undergoing a given process remain largely consistent: an inspection of colors and residuals in SFR, size, HI, and H$_2$ content (relative to best-fit scaling relations) shows no significant variation across environments (plots not shown).

\subsection{Intrinsic limitations of  GAEA  and of our analysis} \label{sec:caveat}
Understanding galaxy evolution through theoretical frameworks is essential, yet inherently constrained by model assumptions and simplifications. While GAEA represents an advanced generation of SAMs, incorporating improved treatments of hot and cold gas stripping and a revised AGN feedback scheme, several caveats must be acknowledged when interpreting its predictions, particularly in the context of environmentally-driven processes.
\begin{itemize}

\item Built-in halo mass dependence: The trends we have observed may partly reflect the intrinsic nature of SAMs. By construction, many physical processes in these models are parameterized as explicit functions of halo mass (e.g., cooling efficiency, feedback strength). As a result, correlations between galaxy properties and halo mass are not purely emergent predictions, but are at least in part built into the framework. This means that some of the tight connections we observe between environment, galaxy evolution, and halo mass could be driven by these underlying prescriptions, rather than arising entirely from first-principle dynamics.

\item Lack of explicit interactions with the cosmic web and galaxy-galaxy tidal interactions: The current implementation of GAEA does not explicitly model interactions between galaxies and the cosmic web, nor does it include direct tidal interactions between galaxies. This means that environmental trends observed filaments are likely driven solely by assembly bias.

\item Identification of RPS:  
In GAEA, RPS is only identified in its most advanced stages, i.e., when the ram pressure is sufficient to remove gas from within the stellar disk. This modeling choice limits the model's ability to reproduce galaxies with extended ionized gas tails which represent earlier or ongoing stages of ram pressure stripping. Consequently, direct comparisons between model predictions and spatially resolved observations of RPS events remain difficult.  

\end{itemize}

Beyond the intrinsic limitations of the model, it is important to also consider the assumptions and boundaries of our analysis:

\begin{itemize}
\item  
Throughout this work, galaxies have been classified according to their environment at z$\sim$0. However, the physical process responsible for a given transformation may have occurred 
{at earlier times and potentially in different environments along the galaxy’s evolutionary path. As a result, the present-day environmental classification does not encode the full environmental history of galaxies, preventing us from establishing direct causal connections between current environment and past physical mechanisms.}

\item  
While we have identified dominant physical processes associated with each galaxy, we cannot exclude the possibility that a different, unaccounted-for mechanism may have been primarily responsible for its transformation. In some cases, a galaxy could have been quenched by internal mechanisms (e.g., mass quenching, AGN feedback) before any environmental influence took place. Similarly, galaxies that underwent a particular process may have already been affected by earlier environmental mechanisms. Disentangling the full sequence of evolutionary events for each galaxy would require a dedicated, case-by-case analysis—an effort beyond the scope of this work, but one that could be addressed in future studies focused on individual processes.

\item  
It is important to recognize that galaxies classified as None may, in fact, be experiencing milder forms of environmental processes that fall below our selection thresholds. This is particularly relevant for RPS and starvation, which in this work are limited to strong, well-identified cases. As a result, the trends observed in the None population—such as elevated quenched or red fractions—could still be shaped by low-level environmental influence. These subtle effects may not be sufficient to meet the criteria for classification, yet can still contribute to the galaxies' evolutionary paths. Future work may benefit from exploring a more continuous treatment of process intensity to capture this broader spectrum of environmental impact.

\item  
An important aspect not captured in our current classification is the duration of each physical process. For example, a galaxy may undergo intense RPS during a brief passage on a highly eccentric orbit, while another may experience weaker but prolonged stripping over several gigayears. The cumulative impact on the galaxy's evolution—such as quenching efficiency or gas removal—could be greater in the latter case, despite the lower instantaneous strength of the interaction. Incorporating a time-integrated measure of environmental influence could provide a more nuanced understanding of these processes and their long-term consequences.

\item {We do not explicitly consider AGN as a separate physical mechanism in this work. Observational and simulation studies indicate that AGN feedback is a key driver of quenching in massive galaxies (\( \log M_\ast/M_\odot \gtrsim 11 \)), even outside cluster environments \citep[e.g.,][]{Hirschmann2014, Terrazas2016, Cheung2016SuppressingWinds, Weinberger2017, Fontanot2020}. AGN activity can suppress star formation both directly, by heating or expelling gas \citep{Fabian2012, DiMatteo2005}, and indirectly, by interacting with other processes: for instance, gas removal via RPS or starvation can facilitate AGN-driven quenching by reducing the cold gas reservoir \citep{Croton2006, Peng2015, Terrazas2020, Ricarte2020}, while mergers can trigger AGN activity and modify galaxy structure, enhancing the efficiency of star formation suppression \citep{Hopkins2006, Hopkins2008} Taken together, these effects suggest that the interplay between AGN and environmental processes may be important in shaping the evolution of the most massive galaxies in our sample.
In GAEA, AGN feedback is included and naturally affects the evolution of massive galaxies \citep[e.g.,][]{delucia06, Hirschmann2014}, but for simplicity we have not explicitly considered it in our classification of physical mechanisms. This choice is also motivated by the fact that AGN-driven quenching tends to act in tandem with other processes (e.g., RPS can induce AGN activity, \citealt{Poggianti2017Ram-pressureHoles, Peluso2022}), making it difficult to isolate as the dominant mechanism in most cases. If AGN feedback were the dominant quenching mechanism in these galaxies, our analysis may slightly overestimate the impact of the environmental mechanisms considered here, particularly mergers, which are the most frequent process among massive galaxies. Similarly, some of the trends observed in the ``None'' category for massive galaxies—such as elevated red fractions or suppressed star formation—could be partially driven by AGN activity. Future work could systematically explore the interplay between AGN activity and environmental processes in shaping galaxy evolution.
}

\end{itemize}

In summary, both the GAEA model and our methodological choices impose constraints on the interpretation of results. While these caveats do not undermine the main conclusions of this work, they highlight areas where further investigation will be critical to fully disentangle the interplay between internal and external mechanisms in galaxy evolution.

\section{Summary}\label{sec:summary}
The Mechanisms Affecting Nearby Galaxies and Environmental Trends (MAGNET) survey is designed to provide a comprehensive, spatially resolved view of galaxies in diverse low-$z$ environments,  while deliberately avoiding massive clusters. MAGNET's multiwavelength approach will address key open questions, including the interplay between hydrodynamical and gravitational interactions, the efficiency of gas removal and quenching in different environments, and the timescales associated with star formation suppression. This paper provides a theoretical foundation that is not only valuable in its own right, but is also crucial for interpreting the observational forthcoming results. 

Our analysis leverages predictions from the GAEA semi-analytic model \citep{Hirschmann2016, DeLucia2024}, carefully mimicking the MAGNET survey volume by selecting sub-regions consistent with its environmental constraints—excluding massive clusters but including small and medium groups. Galaxies were classified both by intrinsic halo properties ("halo definition") and observation-like projected criteria ("2D" and "3D" definitions), enabling reconstruction of their environmental histories and the identification of mergers, tidal interactions, RPS, and starvation since redshift $z=2$. It is worth highlighting that the MAGNET field was selected to encompass a diverse set of representative cosmic environments, ensuring that the results of this analysis are widely applicable and extend beyond the scope of the MAGNET survey.

As a first step, we compared GAEA predictions with MAGNET  observations, focusing on environmental definitions and group properties. We find that observationally motivated (2D) definitions can significantly reclassify galaxies relative to intrinsic halo-based assignments, reducing the fraction of isolated and group galaxies while increasing that of pairs. Despite these classification shifts and some quantitative mismatches (e.g., the absence of low-dispersion groups in observations and a higher fraction of pairs in the model), the overall distribution of environments and group types is qualitatively comparable, supporting the use of GAEA predictions for MAGNET interpretation.

Our results show that more than half of the galaxies remain unaffected by the considered environmental mechanisms since $z=2$. Among affected systems, mergers are the most common process overall—especially at high stellar mass—consistent with hierarchical growth, while RPS 
has a high incidence in galaxies in halos with $\log M_{\rm halo}>12$, in all environments, and is as common as mergers in pairs. Starvation and tidal interactions are less frequent but leave measurable imprints on galaxy properties. 

The incidence of each mechanism is strongly dependent on both mass and environment, yet their signatures on global properties (e.g., SFRs, gas fractions, sizes) are often subtle, highlighting the importance of spatially resolved data to detect them. Environment exerts a stronger influence than individual mechanisms in shaping galaxy colors and the quenched fractions, which increase steadily from isolated galaxies to pairs, filaments, and groups. Ram pressure stripped galaxies tend to be gas-poorer but remain blue, while starvation emerges as one of the most effective processes in driving quenching in groups. Tidal interactions and mergers display mass-dependent trends, with tidal galaxies appearing systematically larger, and mergers driving more rapid stellar mass growth.

The evolutionary pathways of galaxies further reflect these differences: especially at low masses, starvation- and merger-driven systems exhibit a steeper mass growth  star formation suppression and lower HI content, while undisturbed galaxies evolve more gradually. In the mass growth, mechanism-driven trends disappear as moving towards more massive systems, indicating that most of the assembly already occurred. Nonetheless, we observe a strong dependence of the SSFR evolution especially at the highest masses, where galaxies in groups and filaments stopped forming stars earlier on, especially if affected by gravitational mechanisms. The monotonic increase in quenching fraction across environments supports the idea that galaxy transformation is not confined to clusters, but is already largely established in group and filament regimes, consistent with pre-processing scenarios.  The HI content is mostly affected at the highest masses  in groups and filaments. while H$_2$ is affected also in the least dense environments and at intermediate masses.  

An analysis of the past 1.5 Gyr highlights contrasting timescales: RPS emerges as efficient mechanism even on short timescales, while mergers, starvation, and tidal interactions show a lower incidence in the same time frame. 

This theoretical investigation provides critical context for interpreting MAGNET’s forthcoming multi-wavelength observations, clarifying the complex interplay of internal and external processes shaping galaxies in low-redshift environments. It underscores the importance of combining short- and long-timescale diagnostics, and highlights the need for spatially resolved data to distinguish the often subtle signatures of different mechanisms outside of cluster regimes. 
{While GAEA provides valuable predictions over cosmological volumes, its limited ability to capture the spatial and dynamical complexity of environmental processes (e.g., tidal interactions, AGN feedback, early-stage ram-pressure stripping) motivates complementary approaches.  In particular, hydrodynamical simulations are essential to perform direct, spatially resolved comparisons with MAGNET \citep[e.g.,][]{Hank2025}, making localized signatures of gas stripping, structural disturbances, or star formation suppression clearly discernible and directly connected to the underlying physical drivers \citep{Vulcani2021}.

At the same time, insights from hydrodynamical simulations can inform future developments of GAEA and other SAMs by motivating a more explicit treatment of cosmic web interactions, tidal effects, and mild, time-extended environmental processes. Incorporating these processes in a physically motivated and continuous manner will help bridge the gap between idealized environmental classifications and the complex, gradual nature of galaxy transformation revealed by spatially resolved observations, thereby strengthening the interpretation of environmental processing across cosmic time.}

\begin{acknowledgements}
We thank Ned Taylor and the 4HS team for sharing their internal spectroscopic catalog.  BV and AEL acknowledge support from the INAF GO grant 2023
“Identifying ram pressure induced unwinding arms in cluster spirals” (P.I. Vul-
cani). SB acknowledges the support from the Physics Foundation through the Messel Research Fellowship. KK acknowledges funding support from the South Africa Radio Astronomy Observatory (SARAO) and the National Research Foundation (NRF) (grant UID: 97930). MG acknowledges support from the ERC Consolidator Grant \textit{BlackHoleWeather} (101086804). YLJ acknowledges support from the Agencia Nacional de Investigaci\'on y Desarrollo (ANID) through Basal project FB210003, FONDECYT Regular projects 1241426 and 1230441, and  Millennium  Science Initiative Program NCN2024\_112.

\end{acknowledgements}
\bibliography{references.bib}{}
\bibliographystyle{aa}

\begin{appendix}
\begin{figure*}

\section{Simultaneous occurrence of different mechanisms}
\label{app:multiplicity}
As discussed in Sec.~\ref{sec:frac}, galaxies can be affected by multiple physical mechanisms since $z\sim2$. Here, we quantify the fraction of galaxies experiencing simultaneous processes for each stellar mass bin and environment. Figure~\ref{fig:multipl} presents, for each mass bin and classification scheme, the fraction of galaxies affected by exactly one mechanism versus two or more mechanisms simultaneously. Environments are color-coded consistently with the main analysis to maintain visual continuity. Across all masses and environments, galaxies undergoing only one process typically dominate ($>50\%$), reaching nearly 100\% in isolated systems. Multiple processes are most frequent in groups, where, particularly at intermediate masses and under the 2D classification, their incidence can exceed that of single-process galaxies. The environment dependence of the incidence of the different mechanisms is clear. 

\begin{center}
       \includegraphics[width=0.9\linewidth]{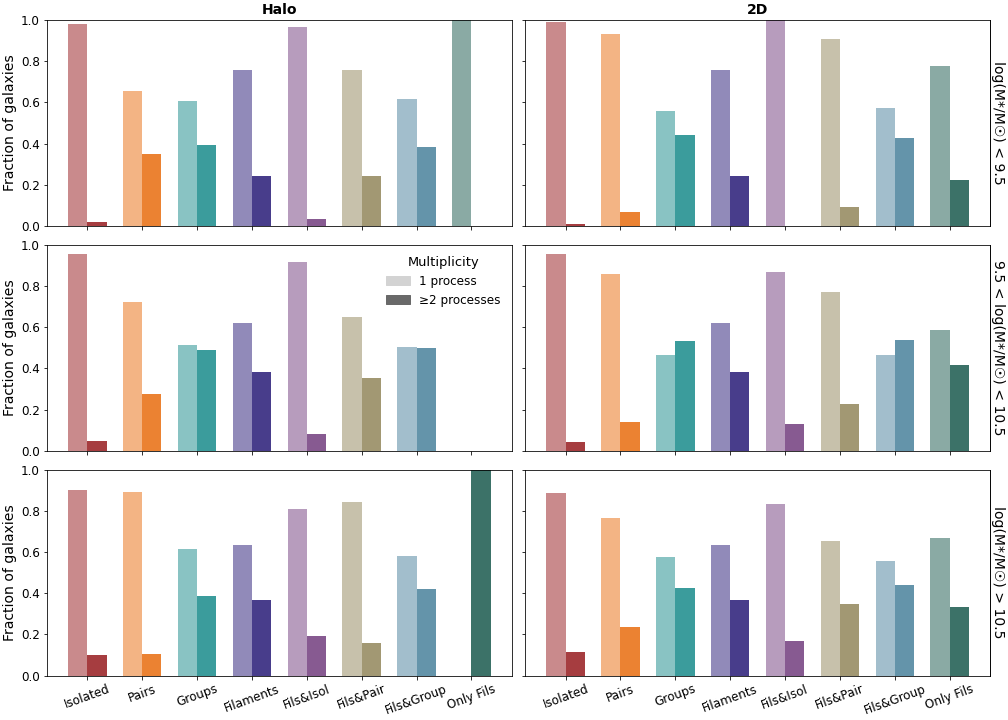}
        \caption{    Fraction of galaxies experiencing exactly one (light gray) or multiple (dark gray)  processes, split by environment and stellar mass. 
    Panels show results for three stellar mass bins (rows) and two environmental classifications: halo-based (left) and 2D (right).}
    \label{fig:multipl}
\end{center} 

Next, to further quantify the simultaneous action of multiple processes on galaxies, we compute conditional co-occurrence matrices per stellar mass bin. 
For each pair of mechanisms we measured the fraction of galaxies experiencing the second process given that the first one occurs. 
These fractions are computed separately for different environments (Isolated, Pairs, Filaments, Groups). 
The resulting matrices are shown in Fig.~\ref{fig:cooccurrence_env}. 
In each panel, the color of a cell encodes the overall conditional fraction (averaged over all environments), while the four corners display the conditional fractions in the individual environments. 
The central value in each cell provides a representative overall fraction. 
This visualization highlights which combinations of mechanisms are most frequent and how their interplay depends on environment and stellar mass.

At all stellar masses, the majority of galaxies undergoing mergers show no evidence of additional environmental mechanisms since $z = 2$. This is particularly true for isolated systems at low and intermediate masses, where the fraction approaches unity. At low masses, galaxy pairs are the most likely to experience multiple processes, with roughly 75\% also affected by RPS; this fraction decreases with increasing stellar mass. Conversely, the fraction of galaxies experiencing both mergers and tidal interactions rises with stellar mass, exceeding 20\% in groups and filaments in the highest mass bin. Similarly, the fraction of galaxies simultaneously affected by RPS and starvation increases with stellar mass, reaching ~50\% in groups and filaments in the highest mass bin. No significant differences are observed between the Halo and 2D classifications.

Focusing on tidal interactions, we find that at low and intermediate stellar masses, over 60\% of galaxies affected by tidal forces are influenced solely by this mechanism, while the remainder have also undergone a merger.
\end{figure*}

\begin{figure*}
\begin{center}
    \includegraphics[width=0.9\textwidth]{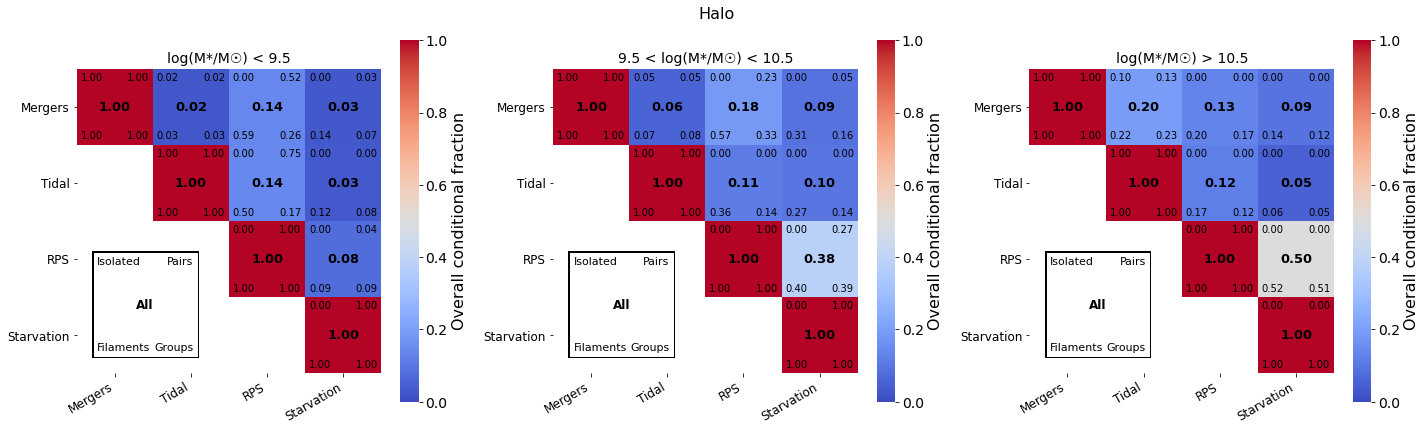}
    \includegraphics[width=0.9\textwidth]{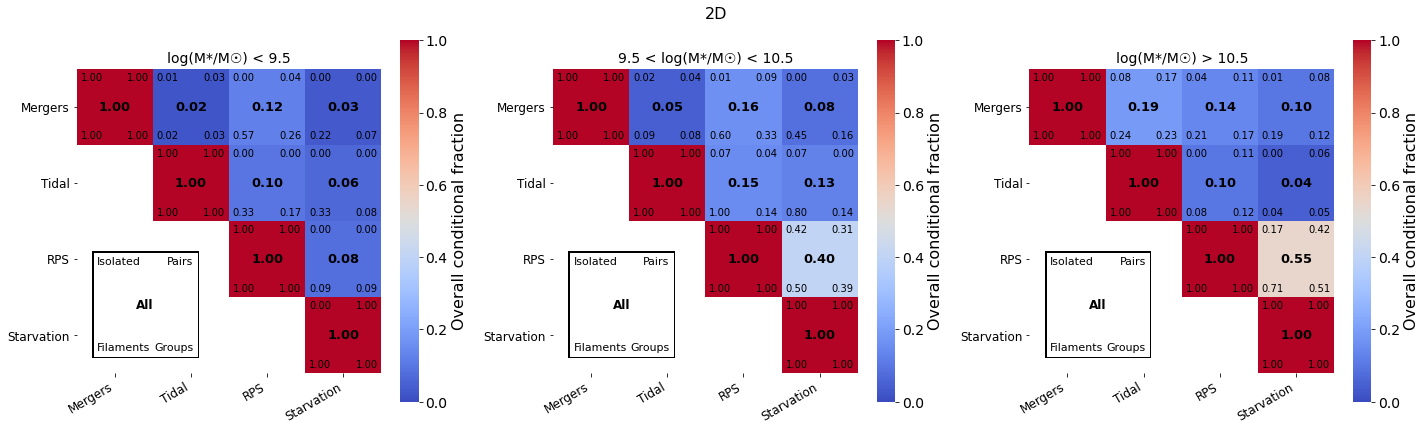}
    \caption{Conditional co-occurrence of physical mechanisms per stellar mass bin, split by environment. 
    Upper panels correspond to Halo-based classification, lower panels to 2D observational-like classification. 
    Each cell shows the fraction of galaxies experiencing the second process given the first one: the cell color represents the overall fraction across all environments, the four corners indicate the fractions in Isolated (top-left), Pairs (top-right), Filaments (bottom-left), and Groups (bottom-right). 
    Only the upper triangle is displayed, as co-occurrence is symmetric.}
    \label{fig:cooccurrence_env}

\end{center}    

\section{Properties of galaxies in different environments and subject to different processes using the 2D environment definition}
\label{app:2d}


In this Appendix, we present a subset of the results discussed in the main text, but using the 2D (projected) definition of environment. The relevant differences and interpretations are discussed where appropriate.

It is important to highlight a key conceptual difference between the halo-based and the 2D definitions. In the halo-based framework, isolated galaxies are by construction central objects, and therefore cannot be affected by mechanisms such as ram pressure stripping or starvation. However, in the 2D (observationally motivated) classification, galaxies identified as ``isolated'' can in fact be satellites projected outside the boundaries of their host halo, and thus may still be subject to environmental processes. This discrepancy reveals an intrinsic limitation of observational definitions of environment: they are more susceptible to projection effects and contamination. This should be carefully considered when interpreting observational results based on projected environmental metrics.

Figure~\ref{fig:distr_2D} shows the halo mass and stellar mass distributions, discussed in Sect.~\ref{sec:distr}. As expected, the 2D environment definition is less tightly coupled to halo mass, and as a consequence the influence of the physical mechanisms becomes more apparent when investigating the halo mass distributions of galaxies in different environments and undergoing different mechanisms (top panels in Fig.~\ref{fig:distr_2D}). In both isolated and pair categories, galaxies undergoing mergers, tidal interactions, ram-pressure stripping, or starvation exhibit systematically higher halo masses than those unaffected by any mechanism. In contrast, within the group category—where halo mass is effectively fixed by definition in the 2D classification—there is insufficient dynamic range to detect such differences. Differences in the stellar mass (bottom panels in Fig.~\ref{fig:distr_2D}) distribution are less evident when comparing the halo-based and 2D environment definitions.

Figure~\ref{fig:coloromass_2D} presents the color–mass diagram for galaxies undergoing different mechanisms and in different environments. The main difference between these plots and those presented in the main text is the presence of ram pressure stripped galaxies among isolated and starved galaxies in pairs. The former have a higher incidence and are shifted toward more massive galaxies with respect to all the other ram pressure stripped galaxies in the other environments; the latter are all red.

Figures~\ref{fig:delta_2d} show the normalized distribution of the deviations from the scaling relations for the SFR, size, HI, and H$_2$ mass, respectively, discussed in Sect.~\ref{sec:deviations}. We do not show the corresponding scaling relations, as they are almost indistinguishable from those shown in Fig.~\ref{fig:ref_relations}. Overall, all the results discussed in the main text hold.

Figure~\ref{fig:quiefrac_2d} shows the quiescent fractions as a function of environment and process discussed in Sect.~\ref{sec:quie_frac}. Trends are much noisier than those based on the halo definition, but still in place. This again suggests that the observational approach is significantly contaminated.

Figures~\ref{fig:mass_SFR_growth_2D} and \ref{fig:HI_H2_growth_2D} present the evolution of the stellar mass, SSFR, HI, and H$_2$ mass content over time. Overall, the results discussed in the main text remain valid under this alternative classification, with all the main trends being qualitatively and quantitatively preserved.
\end{figure*}

\begin{figure*}
    \centering
    \includegraphics[trim = 0 12 0 8, clip, width=0.9\linewidth]{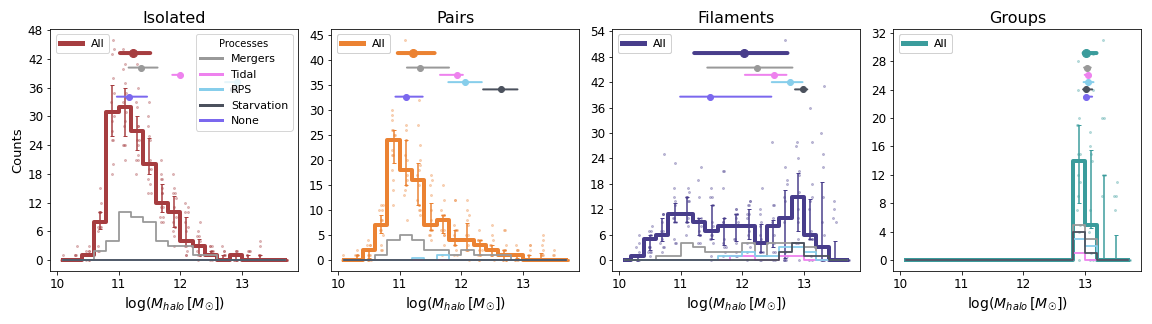}
        \includegraphics[trim = 0 12 0 29, clip, width=0.9\linewidth]{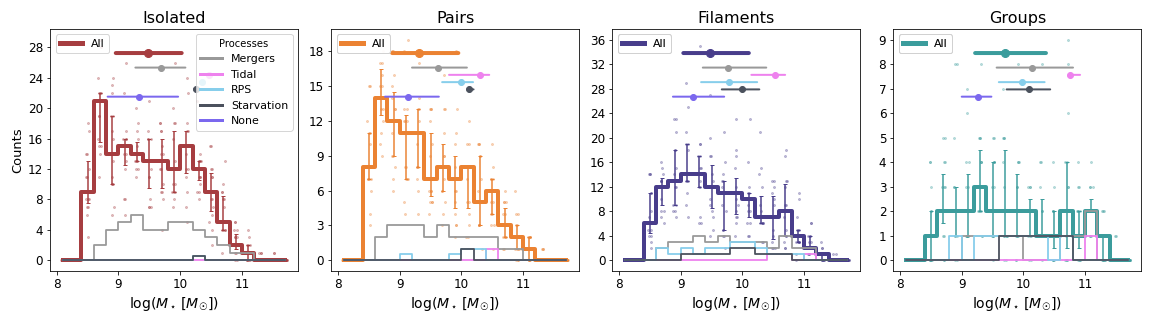}    \caption{Halo mass (top) and stellar mass (bottom) distributions  for galaxies in different environments and affected by different mechanisms. The 2D environment definition is adopted.  Panels, colors, lines and symbols are as in Figs. \ref{fig:halo_distr_halo} and \ref{fig:mass_distr_halo}, respectively.}
    \label{fig:distr_2D}
\end{figure*}

\begin{figure*}
    \centering
    \includegraphics[trim = 0 40 0 30, clip,width=0.9\linewidth]{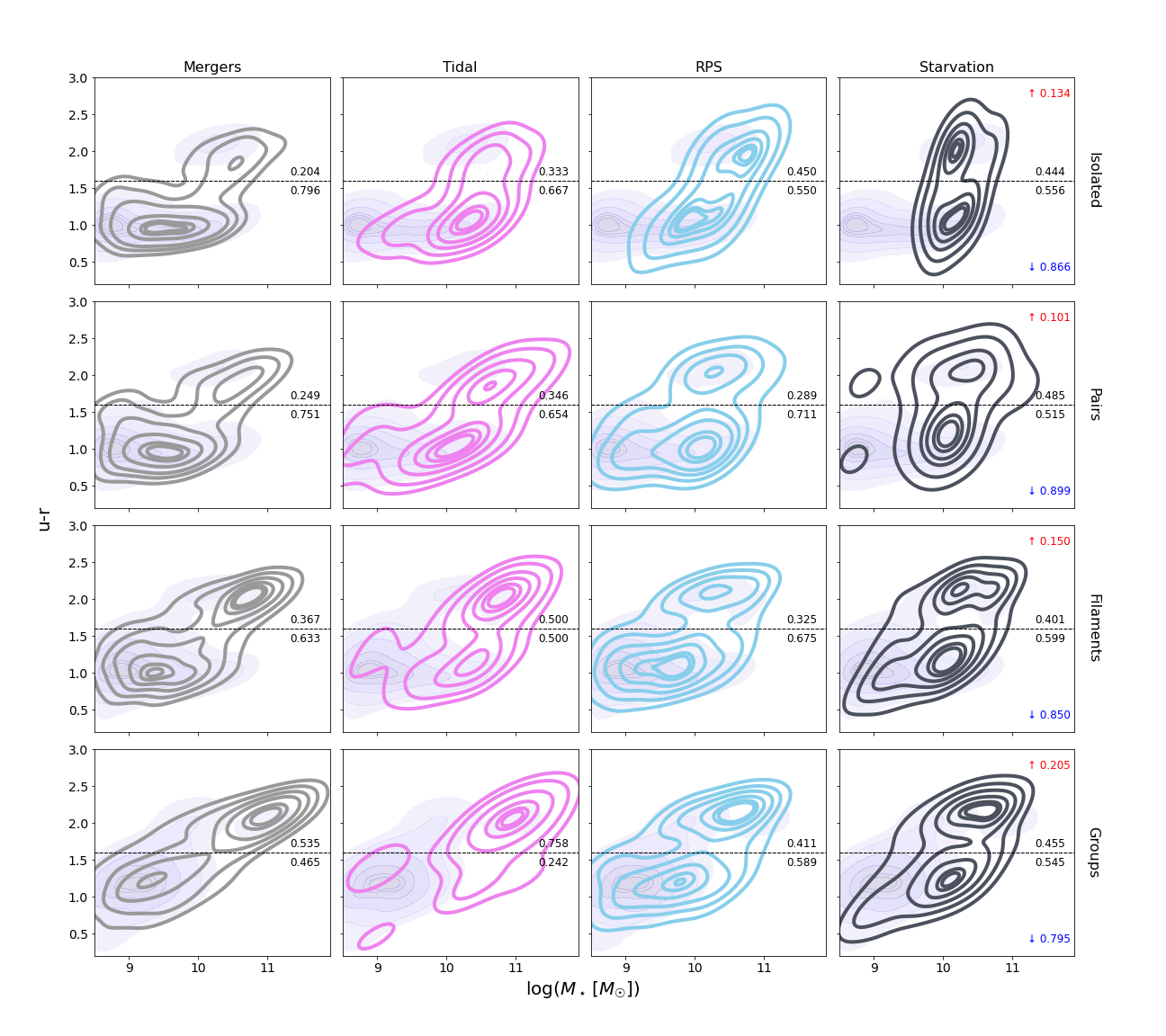}
    \caption{$u-r$ - stellar mass diagram for galaxies undergoing different mechanisms (columns) and in different Halo environments (rows).  The 2D environment definition is adopted.  Panels, colors, lines and symbols are as in Figs. \ref{fig:coloromass_halo}. }
    \label{fig:coloromass_2D}
\end{figure*}

\begin{figure*}
    \centering
    \includegraphics[width=0.85\linewidth]{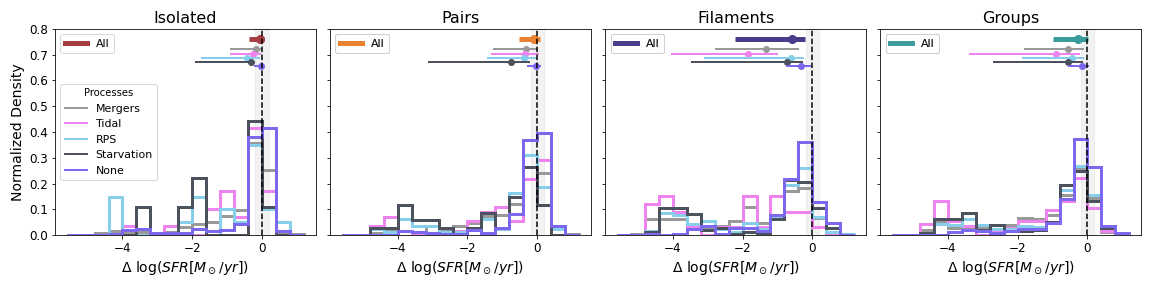}
    \includegraphics[width=0.85\linewidth]{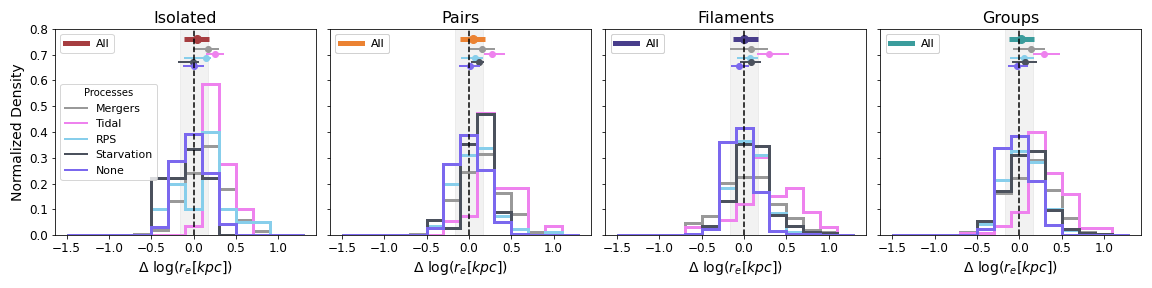}
   \includegraphics[width=0.85\linewidth]{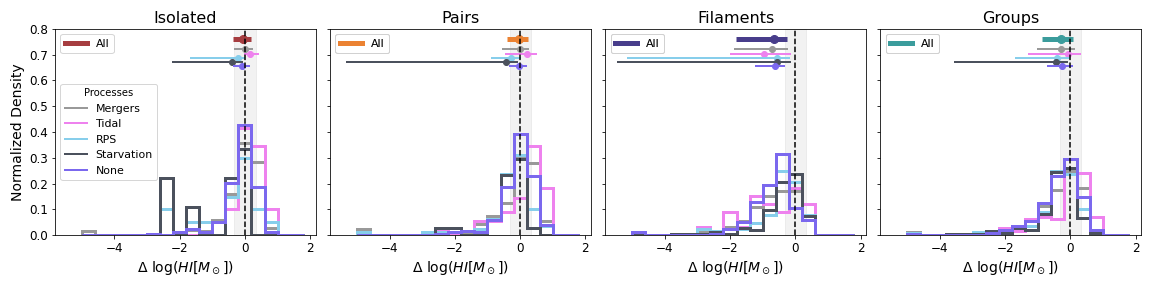}  
      \includegraphics[width=0.85\linewidth]{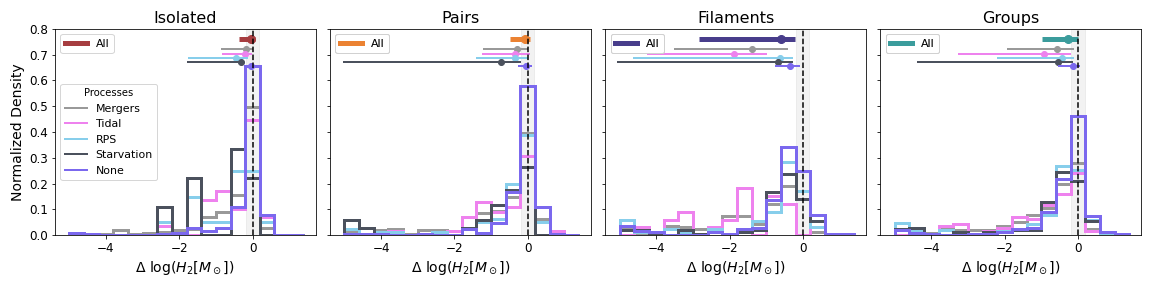}
    \caption{Normalized distribution of the difference between each galaxy SFR (top), size (second from the top), HI (second to bottom) and H$_2$ (bottom) and the expected corresponding value given its mass and the best fit. The 2D environment definition is adopted.  Panels, colors, lines and symbols are as in Fig. \ref{fig:delta_sfr}.}
    \label{fig:delta_2d}
\end{figure*}

\begin{figure*}
    \centering
\includegraphics[width=0.9\linewidth]{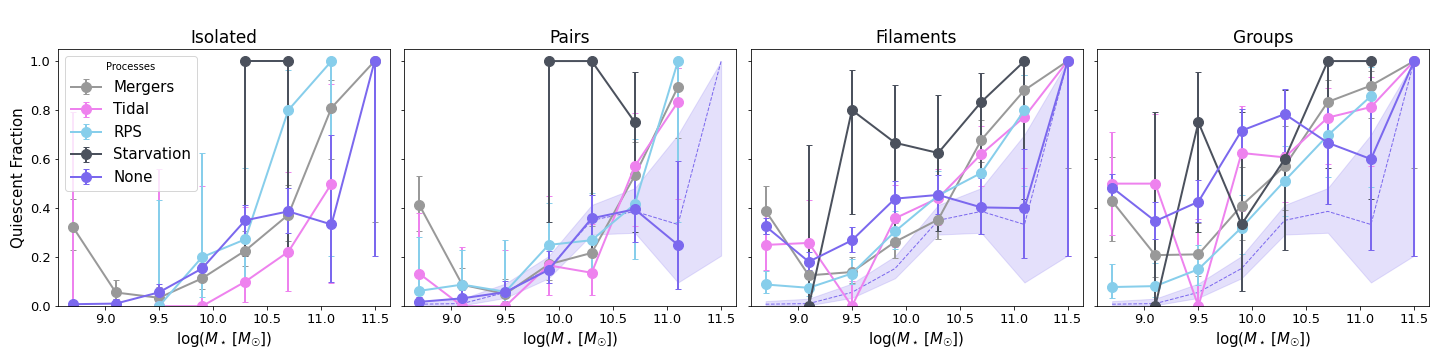}
    \caption{Quenched fraction as a function of stellar mass for galaxies in different categories and environments.  The 2D environment definition is adopted.  Panels, colors, lines and symbols are as in Fig. \ref{fig:quiefrac}.}
    \label{fig:quiefrac_2d}
\end{figure*}

\begin{figure*}
    \centering
    \includegraphics[width=0.9\linewidth]{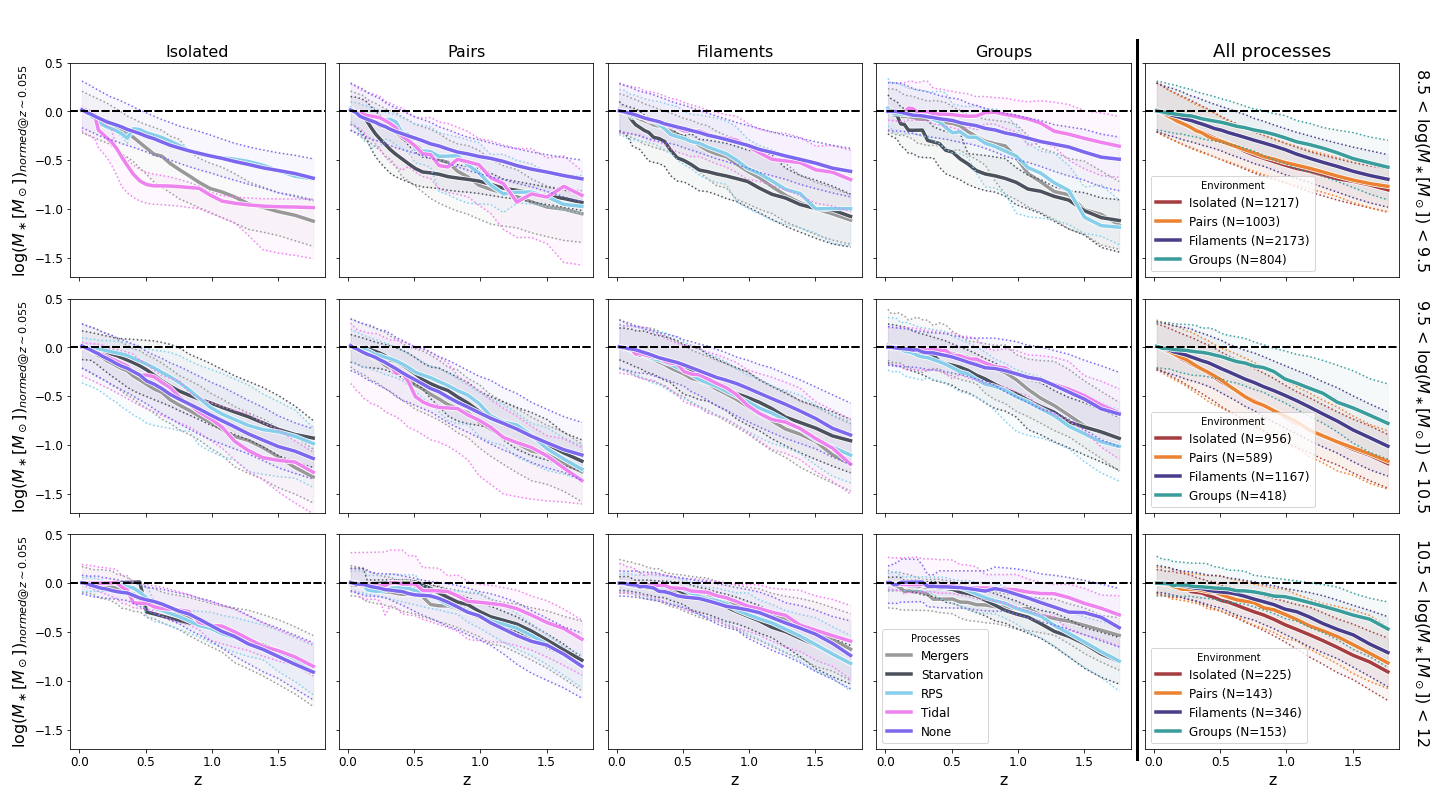}
    \includegraphics[width=0.9\linewidth]{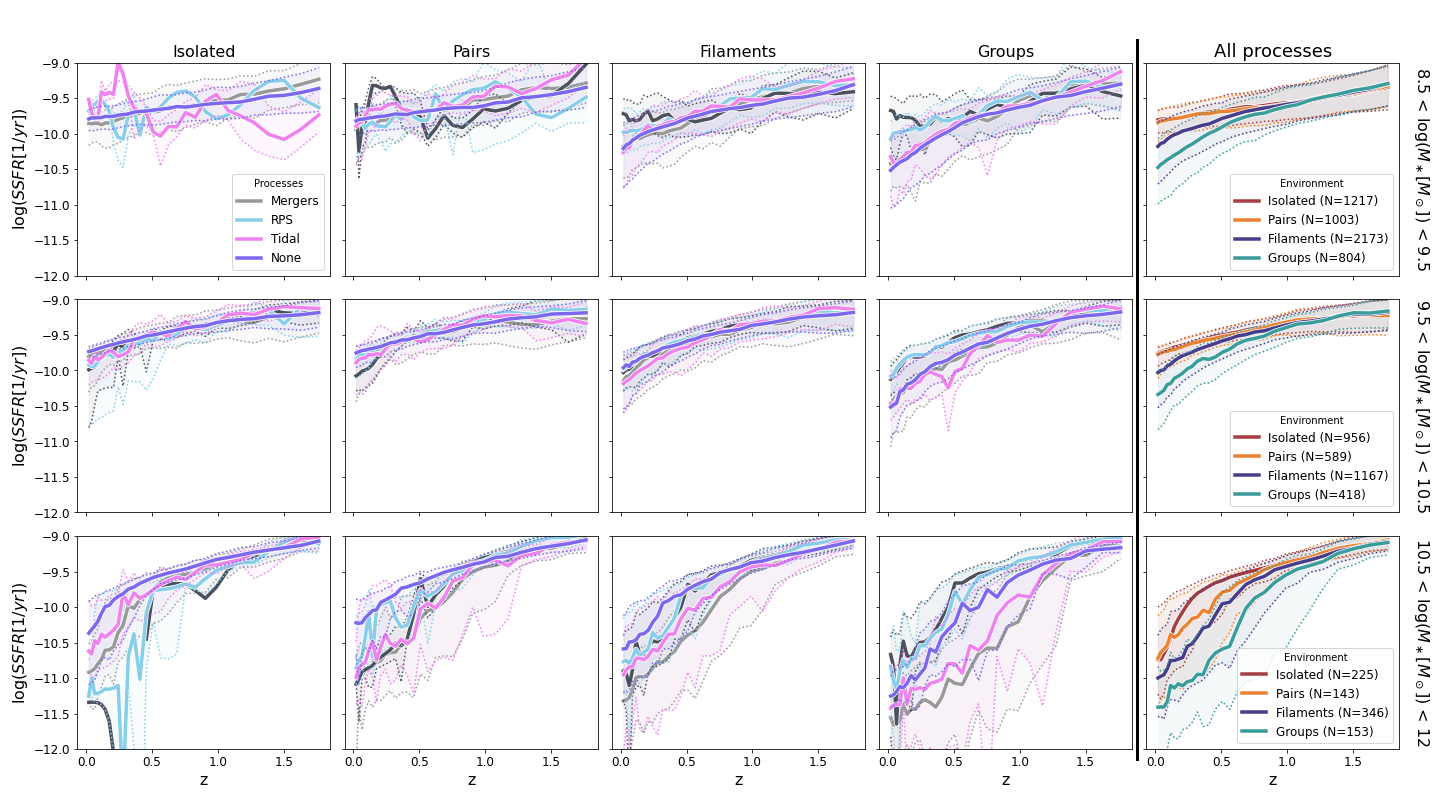}     \caption{Stellar mass growth (top) and SSFR evolution (bottom) as a function of time for  galaxies in different environments, undergoing different mechanisms and in three different stellar mass bins. The 2D environment definition is adopted. Panels, colors, lines and symbols are as in Fig. \ref{fig:mass_growth}.}
    \label{fig:mass_SFR_growth_2D}  
    \end{figure*}

    \begin{figure*}
    \centering
    \includegraphics[width=0.9\linewidth]{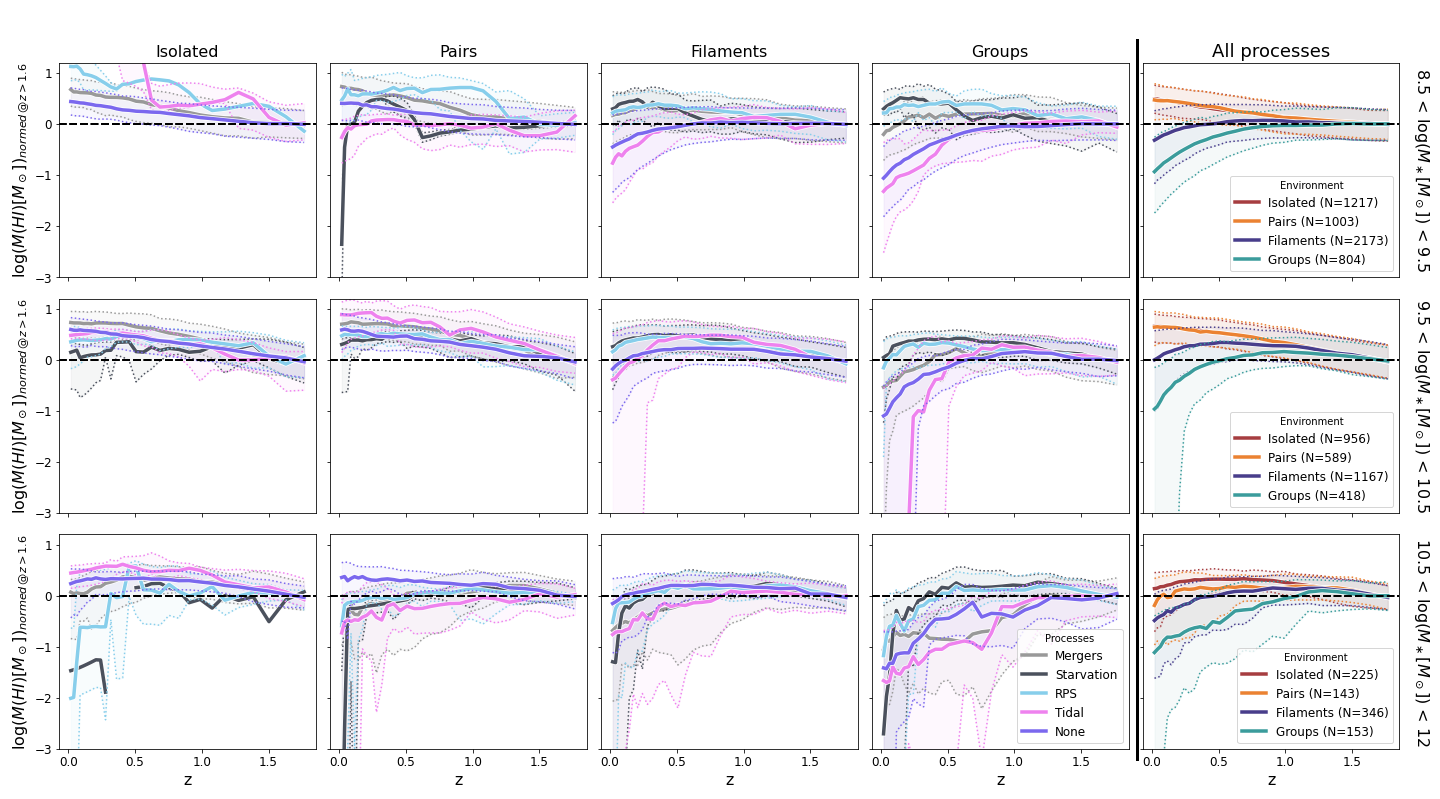}
    \includegraphics[width=0.9\linewidth]{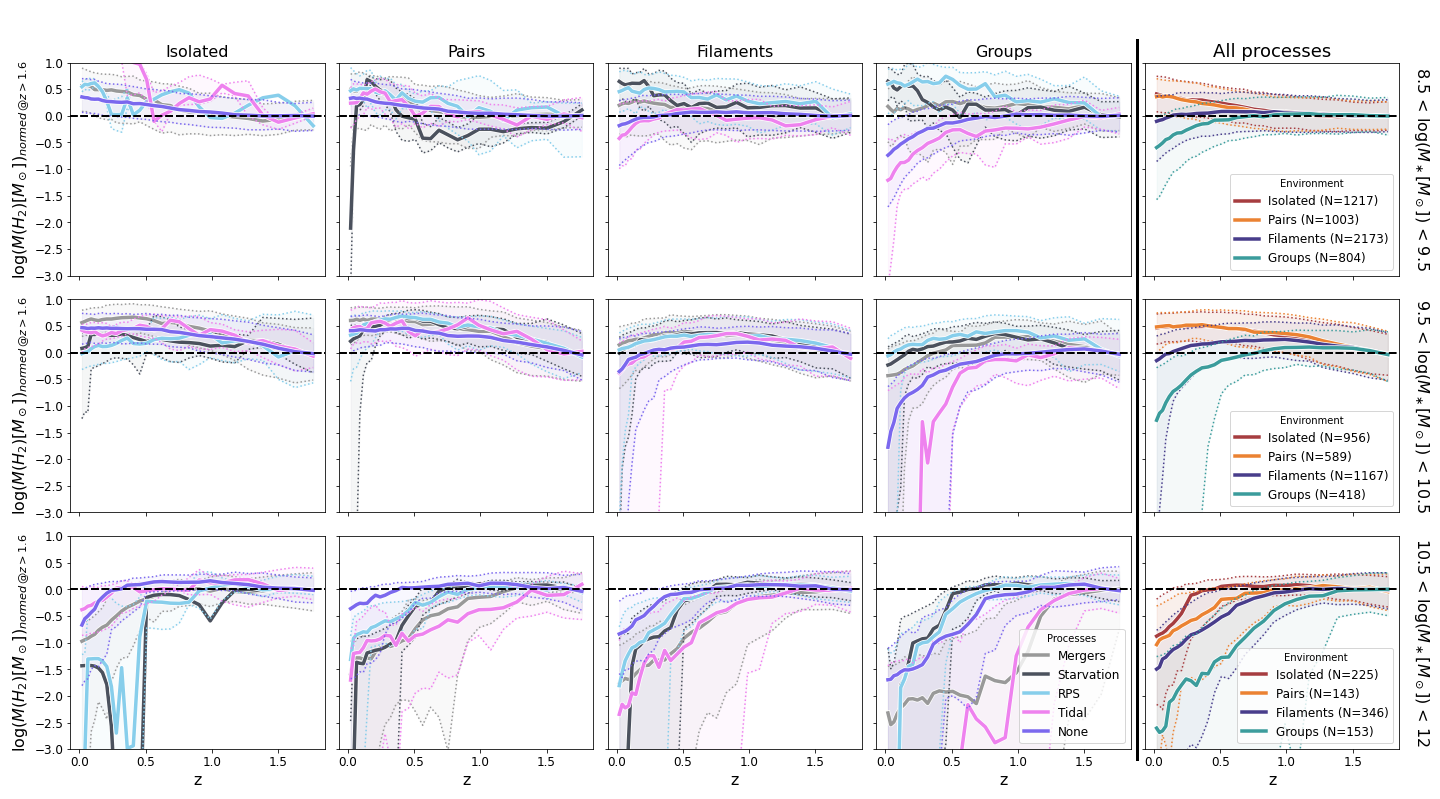}    \caption{Evolution of normalized HI (top) and H$_2$ (bottom) mass content as a function of time for  galaxies in different environments, undergoing different mechanisms and in three different stellar mass bins.  The 2D environment definition is adopted. Panels, colors, lines and symbols are as in Fig. \ref{fig:HI_growth}.}    \label{fig:HI_H2_growth_2D}  
    \end{figure*}

\begin{figure*}

\section{The impact of excluding orphan galaxies}
\label{app:orphans}

In the main analysis we considered only central and satellite galaxies. 
Here we assess the impact of this choice by comparing the fractions of galaxies 
experiencing different physical processes in various environments (Fig.~\ref{fig:distr-env-proc})
when orphan galaxies are either included or excluded from the statistics.  
For each stellar mass bin and classification scheme (Halo, 2D), 
we quantified the differences between the two samples by evaluating, 
for every environment–process combination, the deviation between the corresponding galaxy fractions and their uncertainties. 
Specifically, we computed
\begin{equation}
z = \frac{f_{\mathrm{data2}} - f_{\mathrm{data1}}}{\sqrt{\sigma_{\mathrm{data1}}^2 + \sigma_{\mathrm{data2}}^2}} \, ,
\end{equation}
where $f$ and $\sigma$ denote the measured fraction and its uncertainty, respectively. 
Here, data1 refers to the sample including only central and satellite galaxies, 
while data2 also includes orphans. 
This quantity expresses the difference between the two samples in units of the combined statistical error.
Figure~\ref{fig:zmap} displays the resulting $z$-values as a function of environment (x-axis) and process (y-axis) 
for each mass bin and classification scheme. 
The color scale reflects the sign and amplitude of the deviation: 
red regions correspond to higher fractions in data2 (i.e. when orphans are included), 
while blue regions indicate lower fractions. 
Thus, the color intensity directly traces the significance of the difference between the two datasets. In general, low- and intermediate-mass bins show moderate deviations, occasionally exceeding $2\sigma$ in specific environments, 
whereas the high-mass regime remains remarkably consistent, with differences well within the combined uncertainties. 
More specifically, across all stellar masses and for both classification schemes, the fraction of galaxies undergoing mergers 
systematically decreases when orphan galaxies are included, while the fraction affected by tidal interactions increases, 
particularly in groups at the highest masses. 
The fraction of galaxies experiencing ram-pressure stripping (RPS) decreases at all masses, with the strongest effect again seen in groups. 
At low masses, the fraction of galaxies undergoing starvation shows no significant change, but it increases progressively with stellar mass. 
Conversely, the fraction of galaxies unaffected by any physical mechanism rises—most notably at low masses. 
Overall, the largest differences are found in group environments, as expected, since these host the highest number of orphan galaxies.


\centering
\includegraphics[width=0.9\linewidth]{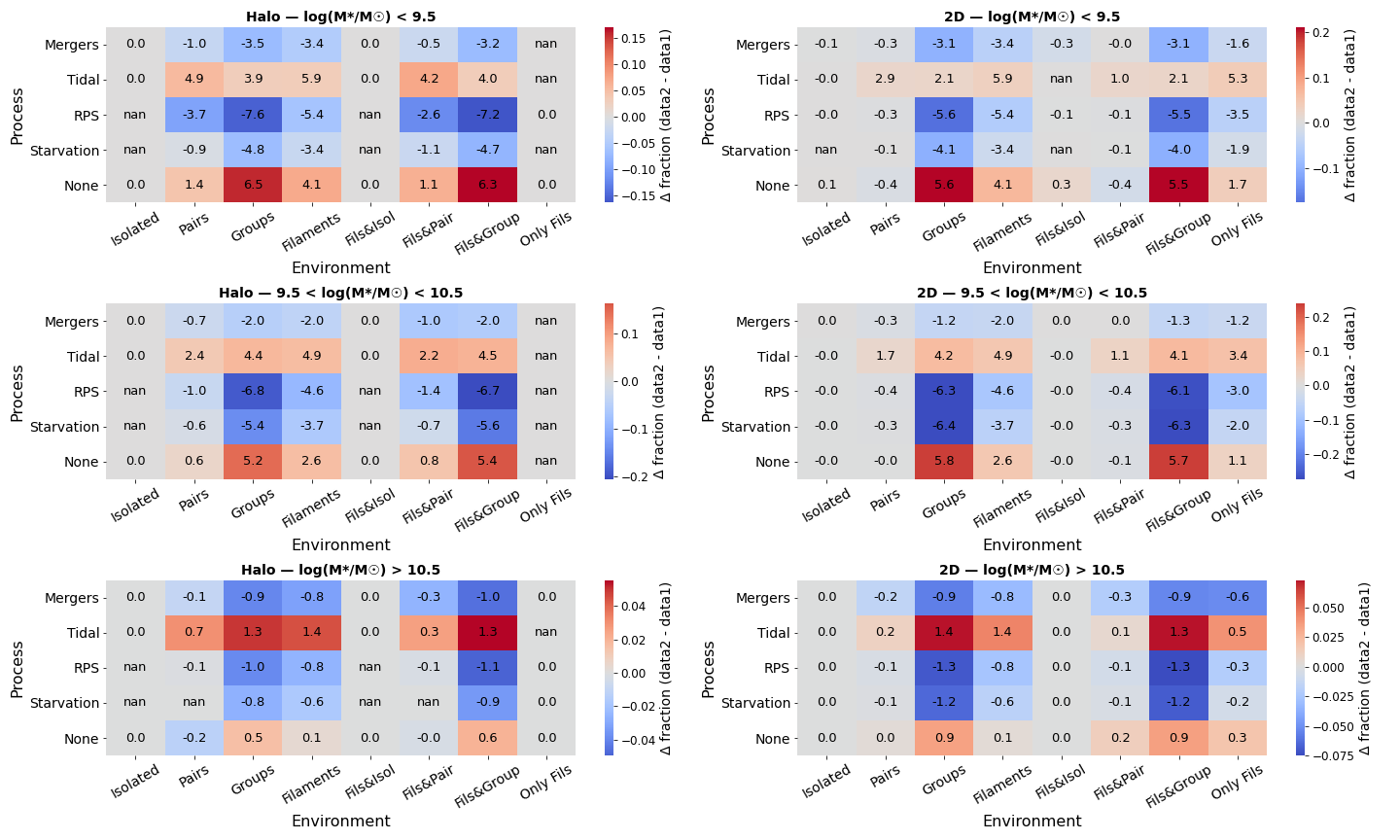}
\caption{
Heatmap representation of the fractional differences $\Delta f = f_{\mathrm{data2}} - f_{\mathrm{data1}}$ 
between the two datasets, for each stellar mass bin and classification scheme. 
Each panel shows the comparison as a function of environment (x-axis) and physical process (y-axis). 
Red cells denote higher fractions in data2 relative to  data1, 
while blue cells indicate lower fractions. 
The color intensity reflects the amplitude of the difference, with neutral tones corresponding to consistent results 
within uncertainties.}
\label{fig:zmap}
\end{figure*}


\end{appendix}

\end{document}